\documentclass[amsmath,amssymb,twocolumn,nofootinbib]{revtex4-2}
\usepackage{ragged2e}

\usepackage{color}
\usepackage[colorlinks=true,linkcolor=blue,urlcolor=blue,citecolor=blue]{hyperref}
\usepackage{graphicx}\usepackage{bm,color}

\usepackage{float}
\usepackage[caption = false]{subfig}

\definecolor{labelkey}{cmyk}{.4,.2,0,0}
\usepackage{times}

\newlength{\highlightlength}
\newlength{\highlightlengthbis}

\usepackage{tikz}
\usetikzlibrary{arrows,shapes,positioning}
\usetikzlibrary{decorations.markings}
\tikzstyle arrowstyle=[scale=1]
\tikzstyle directed=[postaction={decorate,decoration={markings,
    mark=at position .65 with {\arrow[arrowstyle]{stealth}}}}]
\tikzstyle endreversedirected=[postaction={decorate,decoration={markings,
    mark=at position 1.0 with {\arrow[arrowstyle]{stealth}}}}]
\tikzstyle enddirected=[postaction={decorate,decoration={markings,
    mark=at position 1.0 with {\arrow[arrowstyle]{stealth}}}}]

\tikzstyle reverse directed=[postaction={decorate,decoration={markings,
    mark=at position .65 with {\arrowreversed[arrowstyle]{stealth};}}}]

\newcommand{\Eq}[1]{Eq.~(\ref{#1})}
\newcommand{\Eqs}[1]{Eqs.~(\ref{#1})}
\newcommand{\eq}[1]{(\ref{#1})}
\newcommand{\ds}[1]{\displaystyle }

\newcommand{\half}{\frac12}

\newcommand{\bea}{\begin{eqnarray}}
\newcommand{\eea}{\end{eqnarray}}
\newcommand{\beq}{\begin{equation}}
\newcommand{\eeq}{\end{equation}}

\definecolor{mygrey}{gray}{0.5}

\definecolor{mygray}{gray}{0.4}
\newcommand{\gray}{\color{mygray}}

\newcommand{\rme}{\mathrm{e}}
\newcommand{\rmd}{\mathrm{d}}
\newcommand{\nn}{\nonumber}
\renewcommand{\epsilon}{\varepsilon}
\newcommand{\nott}[1]{}
\newcommand{\ca}[1]{{\cal #1}}
\newcommand{\be}{\begin{equation}}
\newcommand{\ee}{\end{equation}}
\newcommand{\Fig}[1]{\includegraphics[width=\columnwidth]{./figures/#1}} 
\newcommand{\fig}[2]{\includegraphics[width=#1\columnwidth]{./#2}}

\newlength{\bilderlength}

\graphicspath{{figures/},{}}

\renewcommand{\log}{\ln}

\renewcommand{\paragraph}{\subsubsection*}

\newlength{\thislength}

\newcommand{\MP}[2]{%
\begin{minipage}{#1\columnwidth}%
\centerline{#2}%
\end{minipage}}

\begin{document}

\title{Making complex CFTs real: The two-dimensional Potts model for $Q>4$ and complex $Q$}
\author{{\bf\normalsize Jesper Lykke Jacobsen{$^{1,2,3}$} and Kay J\"org Wiese{$^{1}$}}}
\affiliation{{$^{1}$}CNRS-Laboratoire de Physique de l'Ecole Normale Sup\'erieure, PSL Research University, Sorbonne Universit\'e, Universit\'e Paris Cit\'e, 24 rue Lhomond, 75005 Paris, France.\\
{$^{2}$}Sorbonne Universit\'e, Ecole Normale Sup\'erieure, CNRS, Laboratoire de Physique (LPENS).\\
{$^{3}$}Institut de Physique Th\'eorique, CEA, CNRS, Universit\'e Paris-Saclay.}

\begin{abstract}
 
The two-dimensional $Q$-state Potts model with real couplings has a first-order transition for $Q>4$. 
Starting from a triangular-lattice Potts model with two- and three-spin interactions, we study an equivalent
loop model in which $Q$ is a continuous parameter. By a combination of analytical and numerical arguments,
we show that this loop model allows for the collision of a critical and a tricritical fixed point at $Q=4$. These then
emerge as a pair of complex conformally invariant theories at $Q>4$, or even complex $Q$, for suitable
complex coupling constants. We conjecture that all conformal data (such as the central charge, critical exponents, and
three-point structure constants) can be obtained by analytic continuation of known exact results for the
loop model with $Q \le 4$. This conjecture is checked, both for real $Q>4$ and for $Q \in \mathbb{C}$,
by extensive transfer-matrix computations and comparison to previous studies for $Q=5$.

\end{abstract} 

\maketitle

\section{Introduction}

Second-order phase transitions in statistical-physics models are signaled by the divergence of thermodynamic quantities, such
as the free energy or the specific heat, when a control parameter (usually the temperature) is taken to a suitable critical value.
Accordingly, the correlation length associated with the two-point function of a local observable,
such as the spin or the energy, diverges at the transition point, and the generic exponential decay of
correlations changes to a power-law behavior characterized by universal, critical exponents.

From a more mathematical perspective, this is a manifestation of non-analytic behavior of the thermodynamical quantities with respect to the control parameter. In a classical paper, Lee and Yang \cite{LeeYang1952} initiated the fruitful idea of studying second-order phase transitions from the perspective of complex analysis. Specifically, they studied the free energy of the two-dimensional Ising
model in a complex magnetic field.  Fisher \cite{Fisher1965} later undertook a similar study at zero field, where the temperature takes complex values. In these works, the complex plane of the relevant parameter is partitioned into distinct regions, or phases. In each of these regions,  the dominant contribution to the free energy takes a different analytical expression.
Complex analysis then implies that the phase boundaries   take the form of loci across which these expressions cannot be continued analytically.  The interest of this approach is that knowledge about the physical model can ultimately be inferred by constraining to real, physical values of the relevant parameter.

This general scenario has been delineated for a multitude of different  models, thermodynamical quantities, and control parameters. In particular, in  a series of studies \cite{SalasSokal2001}  the phase transitions in the two-dimensional $Q$-state Potts model were investigated in the complex $Q$-plane. In the present paper we  also study the complex-$Q$ Potts model in two dimensions, but from a different perspective, where the phase transitions are followed by analytic continuation throughout the complex plane. We  pay special attention to
the role of conformal field theory (CFT) in the description of the second-order phase transitions, and study the properties of these CFTs by analytic continuation.

Before moving to field theory, let us start by recalling the lattice definition of the $Q$-state Potts model. For any (complex)
value of $Q$, its partition function can be reformulated as a Fortuin-Kasteleyn (FK) model \cite{FortuinKasteleyn1972}, in which each lattice bond is
considered occupied or empty with a probability related to the temperature. The   occupied bonds then form connected components,
known as FK clusters, each of which occur with a weight $Q$. An important special case is bond percolation, which arises in the limit of $Q \to 1$.
In contrast to the usual definition in terms of $Q$-component spins, the FK formulation permits us to study the Potts model for complex $Q$.

An additional advantage of the FK formulation, which has attracted much interest since the 1980s, stems from the fact that its correlation functions are related to probability measures in random geometry. Indeed, in the continuum limit, the FK clusters themselves, as well as related geometrical objects (hulls, external perimeters, backbones, shortest paths {\it etc.}),  form scale- and conformally invariant fractals at the critical temperature. In probability theory, starting around the year 2000, these fractals are studied in their own right using rigorous methods, complementary to CFT  (see \cite{Schramm2000,Sheffield2009,RohdeSchramm2011} and references therein).

Many models, including the Ising model referred to above, harbor a single or a finite number of second-order phase transitions (or universality classes), each one described by a CFT with a symmetry extending that of the discrete model ($\mathbb{Z}_2$ for the case of Ising). From that point of view the FK-Potts model is interesting, because its underlying symmetry is the permutation group $S_Q$, with $Q$ being allowed to take any complex value. This opens the perspective of a continuum limit whose CFT is characterized by the continuous parameter $Q$. And indeed, it is known that in dimension $d=2$, and for ferromagnetic interactions, criticality occurs in the range $0 \le Q \le Q_{\rm c}=4$ \cite{Baxter1973}, whereas the phase transition turns first-order when $Q>4$.
The situation in higher dimensions is similar, but for a $d$-dependent value of $Q_{\rm c}$.
For instance, $Q_{\rm c}=10/3$ for $d\to 6$ \cite{WieseJacobsen2024}. For any $d$ in the range between the lower and upper critical dimensions, $2 \le d \le 6$, first-order behavior sets in for $Q > Q_{\rm c}(d)$.

Intriguingly, it has been observed that   when $Q > Q_{\rm c}$, some kind of approximate conformal invariance continues to hold,
at least provided that $Q$ is not too large \cite{MaHe2019}. In fact,
it has been argued a long time ago \cite{CardyNauenbergScalapino1980} that such a scenario can happen when nearby a stable (critical) fixed point of the renormalization group annihilates with another unstable (tricritical) fixed point, upon varying a symmetry-related continuous parameter. For the Potts model this parameter is $Q$, and in $d=2$ the annihilation happens at $Q_{\rm c}=4$. This annihilation scenario epitomizes other similar situations, occurring for instance in deconfined quantum critical points \cite{WangNahumMetlitskiXuSenthil2017}, or in models of quantum impurity spins coupled to a bath \cite{Nahum2022}.

Based on the hypothesis that the CFT description valid for real $Q \le Q_{\rm c}$ could be analytically continued in $Q$, Gorbenko, Rychkov and Zan
\cite{GorbenkoRychkovZan2018b} proposed that approximate conformality in the first-order region $Q > Q_{\rm c}$ could result from the vicinity,
in the plane of complex $Q$, of a pair of complex conjugate fixed points. Furthermore it was argued \cite{GorbenkoRychkovZan2018b,WieseJacobsen2024}
that the Potts model could become truly conformal also for $Q > Q_{\rm c}$ upon introducing suitable complex coupling constants into the model.

The field-theoretical argument applies also to the closely related, but easier to handle $O(n)$ model. This model is originally defined as a lattice model
of $n$-component spins. In two dimensions, it can again be extended to complex values of $n$, by
using a diagrammatic reformulation in terms of lattice loops \cite{Nienhuis1982} each of which occurs with weight $n$. This model has, for any $0 \le n < n_{\rm c} = 2$, both a critical fixed point (governing the so-called dense phase of loops, corresponding to temperatures $T > T_{\rm c}$ for the vector spins) and a tricritical point (governing the dilute phase of loops, corresponding to the critical point $T = T_{\rm c}$ for the spins). The annihilation of fixed points can be   realized when $n \to n_{\rm c}$. It was in this model that approximate conformality (with real couplings) and true conformality (with suitable complex couplings) were first established in the range $n>n_{\rm c}$ \cite{HaldarTavakolMaScaffidi2023}, extending earlier and (by that time) rather puzzling results
\cite{GuoBloteWu2000}.

The question remained how to achieve something similar for the two-dimensional Potts model with $Q > Q_{\rm c}$.
An obvious candidate for the lattice model would be the usual FK model defined on a square lattice, but with a complex weight (temperature parameter)
for an occupied bond. However, this does not work, since this model realizes only the critical fixed point, but not the tricritical one that is required for the collision of fixed points. In our previous Letter \cite{JacobsenWiese2024} we showed how to overcome this obstacle with a more elaborate FK model defined on a triangular lattice. This model, defined  below, contains both two- and three-spin interactions, which   compete and  can  induce tricritical behavior. The collision of fixed points can be realized by this lattice model so as to generate conformal behavior for $Q > Q_{\rm c}$, and more generally for complex values of $Q$. A convenient formulation of this model is via completely packed loops on the triangular lattice, with specific discrete-rotational invariant interactions at the vertices.

The purpose of this paper is not only to give a more detailed account of our results than was possible in the Letter \cite{WieseJacobsen2024},
but also to extend them in several directions. In the meantime a number of related studies have been published.
Ref.~\cite{TangMaTangHeZhu2024} considers the quantum version of the 5-state Potts model, where the internal degrees of freedom  are explicitly realized. In contrast to our approach, this technique does not allow the authors to take $Q$ arbitrary. 
This is then extended to boundary critical phenomena \cite{TangLiuTangZhu2025}.
A common feature of these related studies is their use of the $Q$-state non-Hermitian Potts spin chain. Its representation in terms of $Q$-component
spins requires $Q$ to be a positive integer, mostly taken as $Q=5$ in order to be in the regime $Q > Q_{\rm c}$, while still keeping manageable
the dimension of the linear operators used in the numerical studies. We believe that our model has the decisive advantage of simply treating
$Q$ as a parameter, so that any value of $Q$, including complex ones, can be accessed in a framework where the dimension of the transfer matrix
(or the quantum Hamiltonian) is independent of $Q$.
This confers to the lattice model of loops the same flexibility as that of the corresponding field theory.

To expand on this motivation, we round off this Introduction by a brief overview of the present analytical understanding of the continuum limit
of the loop-model version of the FK-Potts model. The loop-model representation itself is due to Temperley and Lieb \cite{TemperleyLieb1971}. The first exact results were obtained by Baxter. Using a mapping to the six-vertex model, valid at the critical temperature, he obtained the partition function and some critical exponents
for the square-lattice model using integrability techniques \cite{Baxter1973}. As we have just explained, this model is not sufficient for the
purpose of the present investigation. The required triangular-lattice model with competing interactions was introduced by Wu and Lin \cite{WuLin1980},
but only the case of vanishing three-spin interactions was found to be integrable \cite{BaxterTemperleyAshley1978}.

Strong results about the continuum limit of this family of models arrived with the field-theoretical advances of the early 1980s. The loops,
which are the contours of the FK clusters \cite{BaxterKellandWu1976}, were found to become, in the continuum limit, the level lines of a
compactified bosonic field, to which the methods of Coulomb Gas (CG) \cite{Nienhuis1984} and Conformal Field Theory (CFT)
\cite{BelavinPolyakovZamolodchikov1984} can be applied. The study of the model on a torus \cite{DiFrancescoSaleurZuber1987}
fixes the set of critical exponents, sometimes known as the {\it spectrum} of the theory. However, the CG approach is only one part of the field-theoretical description of loop models.

In the terminology currently in use \cite{JacobsenRibaultSaleur2023,NivesvivatRibaultJacobsen2024}
operators in loop models are of three types: (1) degenerate operators (local energy densities), (2) diagonal operators (which modify the weights
of loops encircling the insertion point), and (3) non-diagonal operators (which insert defects in the form of open loop segments, or paths).
From heuristic arguments, there are indications of a relation to Liouville CFT (LCFT) \cite{Kondev1997}, an exactly solvable interacting CFT.
However, LCFT has a continuous spectrum and is unitary for central charge $c \in (1,\infty)$, facts which are in tension with CG
results for the Potts loop model for generic $Q \in [0,4]$, since these establish that the corresponding CFT is non-unitary, has $c \le 1$, and a discrete spectrum \cite{DiFrancescoSaleurZuber1987}.

The last 10 years of research on the loop model have aimed at solving this conundrum. In this endeavor the theoretical physicists
have been joined by probability theorists, who use methods such as Schramm-Loewner Evolution (SLE)
\cite{Cardy2005,Lawler2005,Lawler2007}, the Conformal Loop Ensemble (CLE) \cite{Sheffield2009} and the probabilistic construction of
the LCFT path integral \cite{GuillarmouRhodesVargas2019,GuillarmouKupiainenRhodesVargas2024}.
One might wonder whether the loop model could be obtained from LCFT by analytic continuation through complex values of $Q$; this idea can, however,  be quickly dismissed. Indeed,
in LCFT the structure constants of three-point correlation functions are given by the so-called DOZZ formula \cite{DornOtto1994,ZamolodchikovZamolodchikov1996}, but
analytic continuation of the latter to $c \in (-\infty,1)$ is impossible. On the other hand, there is a variant of LCFT, called ${\rm LCFT}_{c \le 1}$
or sometimes imaginary Liouville CFT, for which it has been established \cite{PiccoSantachiaraVitiDelfino2013,IkhlefJacobsenSaleur2016}
that the DOZZ-type formula for ${\rm LCFT}_{c \le 1}$ \cite{Zamolodchikov2005} correctly predicts three-point structure constants of diagonal operators
in the Potts loop model. Very recently, this result on the three-point functions has been promoted to a theorem by probability theorists 
Ang, Cai, Sun and Wu \cite{AngCaiSunWu2024}, while physicists have established a more general result for three-point structure constants
involving any of the three types of operators \cite{JacobsenNivesvivatRibaultRoux2025} (one special case of the latter result has been proved \cite{AngCaiSunWu2024}). The results in  \cite{JacobsenNivesvivatRibaultRoux2025,AngCaiSunWu2024} go beyond both the CG and LCFT scenarios.

From this perspective, the present paper establishes numerically that the Potts model outside of the interval $Q \in [0,4]$ can be made
conformal in a large portion of the complex $Q$-plane, and that the corresponding universality class coincides with an analytic 
continuation of the CFT discussed above. More precisely, we provide precise numerical evidence that the conformal data
(the central charge, the spectrum of critical exponents, and the three-point structure constants) of
the two complex CFTs that result from the collision of the real critical and tricritical CFTs coincide with
the appropriate analytic continuation of the results in \cite{DiFrancescoSaleurZuber1987,JacobsenRibaultSaleur2023,NivesvivatRibaultJacobsen2024}
for the $c \le 1$ loop models.
Based on this link, we are confident in conjecturing that recent results on three- and four-point correlation functions \cite{HeJacobsenSaleur2020,GransSamuelssonJacobsenNivesvivatRibaultSaleur2023,NivesvivatRibaultJacobsen2024,RouxJacobsenNivesvivatRibault2025}
and the symmetries of the space of states \cite{JacobsenRibaultSaleur2023} of the loop models
 carry over to the complex CFT by the same analytic continuation.

Let us outline the content of this paper,
for which a table of contents can be found on page \pageref{table-of-contents}:
In Section \ref{s:Theory}, the model is defined (\ref{s:Model}), together with its transfer matrix formulation (\ref{sec:TM}) and its duality and symmetry structure (\ref{Duality and symmetries}). The self-dual manifold is parametrized by a free parameter $z$. The phase diagram is then analyzed both in the complex $z$-plane (\ref{Phase diagram as a function of z}) and as a function of $Q$ (\ref{Phase diagram as a function of Q}), with section \ref{Analytic continuation from minimal models} describing the analytic continuation of the critical theory to complex $Q$ and $Q>5$. The operator content and conformal dimensions are given in Sec.~\ref{sec:op-content}, and the role of the global $S_Q$ symmetry (\ref{sec:global-sym}) and the question how to calculate  three-point structure constants (\ref{Three-point structure constants}) are clarified.

We then proceed to our numerical results: 
Section \ref{Fisher zeroes in the model with nearest-neighbor  interactions} briefly studies a simpler model with only nearest-neighbor interactions, showing that it cannot access the complex CFT. Returning then to the model of interest,
in Section \ref{Numerics for c and omega on the real line for Q<4}, numerical results for central charge $c$ and scaling corrections $\omega$ are presented for $Q<4$, along with the RG flows between tricritical and critical theories.
This sets the stage for section 
\ref{s:Q=5} on $Q=5$, analyzing the central charge in the complex $z$-plane in section \ref{Central charge in the complex z-plane}, with improvements in precision discussed in section \ref{Improving the precision}, while Section \ref{Other values of  Q} extends the analysis to large real values of $Q$, and complex $Q$.
Section \ref{Spectrum} discusses the spectrum, starting again with a case with $Q<4$ (\ref{s:B_8}), before proceeding to $Q=5$ (\ref{s:Q=5 spectrum}).

Finally, Section \ref{Conclusions} summarizes the results and their implications for complex CFTs in statistical physics.

Appendices \ref{s:Q=10} to \ref{s:Q=8+i} provide additional data for large and complex $Q$ values ($Q=10,20,40,5+2i,8+i$). We also examine whether the location of a complex CFT can be inferred from data for real couplings only (section \ref{Can one find zc(Q=5) from data on the real line only?}). 
Finally we provide alternate derivations of some OPE coefficients in the Dotsenko–Fateev framework, complementing the general formula in section \ref{Three-point structure constants}.

\section{Theory}
\label{s:Theory}
\subsection{Model}
\label{s:Model}

We use a variant of the $Q$-state Potts model introduced by Wu and Lin \cite{WuLin1980}. It is defined on a triangular
lattice $\mathbb{T}_0$, which we orient so that one half of the faces form up-pointing triangles, while the other half are down-pointing.
Let $\Delta_{ijk}$ denote an up-pointing triangle with vertices $i$, $j$, $k$ hosting the three spins  $\sigma_i$, $\sigma_j$, $\sigma_k$,
each of which takes values in $\{1,2,\ldots,Q\}$:
$$
\begin{tikzpicture}[rotate=0]
 \draw (0.0,0.0) -- (1.0,0.0) -- (0.5,0.866) -- cycle;
 \draw [fill] (0,0) circle (.5ex);
 \draw [fill] (1.0,0.0) circle (.5ex);
 \draw [fill] (0.5,0.866) circle (.5ex);
 \node at (-0.2,-0.2) {$i$};
 \node at (0.5,1.2) {$j$};
 \node at (1.2,-0.2) {$k$};
\end{tikzpicture} \quad
$$
The interactions within $\Delta_{ijk}$ are given by the local Hamiltonian ${\cal H}_{ijk}$, where
\begin{equation}
 \label{Hijk}
 {\cal H}_{ijk} = -K( \delta_{\sigma_i,\sigma_j} + \delta_{\sigma_j,\sigma_k} + \delta_{\sigma_k,\sigma_i}) + K_3 \delta_{\sigma_i,\sigma_j,\sigma_k} \,.
\end{equation}
The Kronecker symbol $\delta_{x,y} = 1$ if $x=y$, and $0$ if $x \neq y$; we have set $\delta_{x,y,z} = \delta_{x,y} \delta_{y,z}$ for convenience.
The Hamiltonian for the whole lattice is  
\begin{equation}
 \label{Hlatt}
 {\cal H} = \sum_{\Delta_{ijk}} {\cal H}_{ijk} \,,
\end{equation}
and the sum runs over all up-pointing triangles.
There are no interactions in the down-pointing triangles.
Notice that each bond belongs to precisely one up-pointing triangle, so there is a two-spin interaction $K$ along each bond, but a three-spin
interaction $K_3$ only in the up-pointing triangles.

If we absorb $k_{\rm B} T$ into the coupling constants $K$ and $K_3$, the local Boltzmann weights read
\begin{equation}
 {\rm e}^{-{\cal H}_{ijk}} = 1 + c_2 (\delta_{\sigma_i,\sigma_j} + \delta_{\sigma_j,\sigma_k} + \delta_{\sigma_k,\sigma_i}) + c_3 \delta_{\sigma_i,\sigma_j,\sigma_k} \,, \label{lbw}
\end{equation}
and the parameters $c_2$ and $c_3$ are related to $K$, $K_3$ by
\begin{eqnarray}
 \label{eKL}
 {\rm e}^{K} &=& 1 + c_2 \,, \\
 {\rm e}^{K_3 + 3 K} &=& 1 + 3 c_2 + c_3 \,. \label{K3K} \nonumber
\end{eqnarray}
Each term on the right-hand side of \Eq{lbw} can be represented graphically as
\begin{equation} \label{FKdiag}
\begin{tikzpicture}[rotate=0]
 \draw (0.0,0.0) -- (1.0,0.0) -- (0.5,0.866) -- cycle;
 \draw [fill] (0,0) circle (.5ex);
 \draw [fill] (1.0,0.0) circle (.5ex);
 \draw [fill] (0.5,0.866) circle (.5ex);
 \draw (0.5,-0.3) node {$1$};
\end{tikzpicture} \quad
\begin{tikzpicture}[rotate=0]
 \draw (0.0,0.0) -- (1.0,0.0) -- (0.5,0.866) -- cycle;
 \draw[blue,ultra thick] (0,0) -- (0.5,0.289) -- (0.5,0.866);
 \draw [fill] (0,0) circle (.5ex);
 \draw [fill] (1.0,0.0) circle (.5ex);
 \draw [fill] (0.5,0.866) circle (.5ex);
 \draw (0.5,-0.3) node {$c_2$};
\end{tikzpicture} \quad
\begin{tikzpicture}[rotate=0]
 \draw (0.0,0.0) -- (1.0,0.0) -- (0.5,0.866) -- cycle;
 \draw[blue,ultra thick] (1,0) -- (0.5,0.289) -- (0.5,0.866);
 \draw [fill] (0,0) circle (.5ex);
 \draw [fill] (1.0,0.0) circle (.5ex);
 \draw [fill] (0.5,0.866) circle (.5ex);
 \draw (0.5,-0.3) node {$c_2$};
\end{tikzpicture} \quad
\begin{tikzpicture}[rotate=0]
 \draw (0.0,0.0) -- (1.0,0.0) -- (0.5,0.866) -- cycle;
 \draw[blue,ultra thick] (0,0) -- (0.5,0.289) -- (1.0,0.0);
 \draw [fill] (0,0) circle (.5ex);
 \draw [fill] (1.0,0.0) circle (.5ex);
 \draw [fill] (0.5,0.866) circle (.5ex);
 \draw (0.5,-0.3) node {$c_2$};
\end{tikzpicture} \quad
\begin{tikzpicture}[rotate=0]
 \draw (0.0,0.0) -- (1.0,0.0) -- (0.5,0.866) -- cycle;
 \draw[blue,ultra thick] (0,0) -- (0.5,0.289) -- (0.5,0.866);
 \draw[blue,ultra thick] (0.5,0.289) -- (1.0,0.0);
 \draw [fill] (0,0) circle (.5ex);
 \draw [fill] (1.0,0.0) circle (.5ex);
 \draw [fill] (0.5,0.866) circle (.5ex);
 \draw (0.5,-0.3) node {$c_3$};
\end{tikzpicture}
\end{equation}
Each connected component of thick blue segments is called a Fortuin-Kasteleyn (FK) cluster.
The partition function corresponding to \eqref{Hlatt} is 
\begin{equation}
 Z = \sum_{\{\sigma\}} {\rm e}^{-{\cal H}} = \sum_{\Delta} Q^C c_2^{N_2} c_3^{N_3} \,. \label{ZWuLin}
\end{equation}
Here, the first sum runs over the possible $Q$ states of each spin on the lattice.
The second sum is over the five possible configurations \eqref{lbw} of each up-pointing triangle, with
$N_2$ and $N_3$ denoting the number of diagrams of weight $c_2$ and $c_3$, respectively, and
$C$ is the number of FK clusters.

In this last writing $Q$ appears only as a parameter, so we can {\em define} the FK model for
any (complex) $Q$ by the partition function \eqref{ZWuLin}, forgetting that $Q$ had to be a positive integer
in the original Potts model \eqref{Hlatt}.

In a final step, we trade the clusters for the loops bouncing off their boundaries. To this aim, replace the diagrams \eqref{FKdiag} by
\begin{equation} \label{loopdiag}
\begin{tikzpicture}[rotate=0]
 \draw (0.0,0.0) -- (1.0,0.0) -- (0.5,0.866) -- cycle;
 \draw [fill] (0,0) circle (.5ex);
 \draw [fill] (1.0,0.0) circle (.5ex);
 \draw [fill] (0.5,0.866) circle (.5ex);
 \draw [thick,red,rounded corners=2mm] (-0.166,0.289) -- (0.5,0.289) -- (0.167,-0.289);
 \draw [thick,red,rounded corners=2mm] (0.167,0.866) -- (0.5,0.289) -- (0.833,0.866);
 \draw [thick,red,rounded corners=2mm] (1.166,0.289) -- (0.5,0.289) -- (0.833,-0.289);
 \draw (0.5,-0.5) node {$1$};
\end{tikzpicture} \quad
\begin{tikzpicture}[rotate=0]
 \draw (0.0,0.0) -- (1.0,0.0) -- (0.5,0.866) -- cycle;
 \draw [fill] (0,0) circle (.5ex);
 \draw [fill] (1.0,0.0) circle (.5ex);
 \draw [fill] (0.5,0.866) circle (.5ex);
 \draw [thick,red,rounded corners=2mm] (-0.166,0.289) -- (0.5,0.289) -- (0.167,0.866);
 \draw [thick,red,rounded corners=2mm] (0.167,-0.289) -- (0.833,0.866);
 \draw [thick,red,rounded corners=2mm] (0.833,-0.289) -- (0.5,0.289) -- (1.166,0.289);
 \draw (0.5,-0.5) node {$z$};
\end{tikzpicture} \quad
\begin{tikzpicture}[rotate=0]
 \draw (0.0,0.0) -- (1.0,0.0) -- (0.5,0.866) -- cycle;
 \draw [fill] (0,0) circle (.5ex);
 \draw [fill] (1.0,0.0) circle (.5ex);
 \draw [fill] (0.5,0.866) circle (.5ex);
 \draw [thick,red,rounded corners=2mm] (1.166,0.289) -- (0.5,0.289) -- (0.833,0.866);
 \draw [thick,red,rounded corners=2mm] (0.167,0.866) -- (0.833,-0.289);
 \draw [thick,red,rounded corners=2mm] (0.167,-0.289) -- (0.5,0.289) -- (-0.166,0.289);
 \draw (0.5,-0.5) node {$z$};
\end{tikzpicture} \quad
\begin{tikzpicture}[rotate=0]
 \draw (0.0,0.0) -- (1.0,0.0) -- (0.5,0.866) -- cycle;
 \draw [fill] (0,0) circle (.5ex);
 \draw [fill] (1.0,0.0) circle (.5ex);
 \draw [fill] (0.5,0.866) circle (.5ex);
 \draw [thick,red,rounded corners=2mm] (-0.166,0.289) -- (1.166,0.289);
 \draw [thick,red,rounded corners=2mm] (0.833,-0.289) -- (0.5,0.289) -- (0.167,-0.289);
 \draw [thick,red,rounded corners=2mm] (0.167,0.866) -- (0.5,0.289) -- (0.833,0.866);   
 \draw (0.5,-0.5) node {$z$};
\end{tikzpicture} \quad
\begin{tikzpicture}[rotate=0]
 \draw (0.0,0.0) -- (1.0,0.0) -- (0.5,0.866) -- cycle;
 \draw [fill] (0,0) circle (.5ex);
 \draw [fill] (1.0,0.0) circle (.5ex);
 \draw [fill] (0.5,0.866) circle (.5ex);
 \draw [thick,red,rounded corners=2mm] (0.167,-0.289) -- (0.5,0.289) -- (0.833,-0.289);
 \draw [thick,red,rounded corners=2mm] (0.833,0.866) -- (0.5,0.289) -- (1.166,0.289);
 \draw [thick,red,rounded corners=2mm] (-0.166,0.289) -- (0.5,0.289) -- (0.167,0.866);
 \draw (0.5,-0.5) node {$w$};
\end{tikzpicture}
\end{equation}
The red curves   go straight across the down-pointing triangles. Applying the Euler relation we have,
up to an unimportant global factor,
\begin{equation} \label{Zloop}
 Z = \sum_{\Delta} Q^{\ell/2} z^{N_2} w^{N_3} \,,
\end{equation}
where $\ell$ denotes the number of closed loops formed by the red curves, while $N_2$ and  $N_3$ are defined as before.
We call this representation the {\em loop model}. The local weights $z$ and $w$ are related to the previous ones by
\begin{eqnarray}
 \label{zw}
 z &=& Q^{-1/2} c_2 \,, \\
 w &=& Q^{-1} c_3 \,. \nonumber
\end{eqnarray}
The diagrams \eqref{loopdiag} are such that the loops live on another, shifted triangular lattice $\mathbb{T}$. At each
vertex of $\mathbb{T}$ (corresponding to the barycenter of a face in the original lattice $\mathbb{T}_0$) the loop
segments split so as to avoid   crossing one another. We call the loops {\em completely packed}, since each edge of $\mathbb{T}$ is covered by a loop segment. Each of the five possible splittings \eqref{loopdiag} at the vertices of $\mathbb{T}$ defines what is sometimes called a non-intersecting transition system. The loop model on $\mathbb{T}$ is thus completely packed, and has configurations given by the sum over all possible non-intersecting transition systems.

\subsection{Transfer matrix}
\label{sec:TM}

In the numerical part of this paper we will study the model \eqref{Zloop} via the transfer matrix $T_L$ which builds a row of $L$ triangles, with
periodic boundary conditions. Thus $T_L$ propagates $\sqrt{3}/2$ lattice spacings upwards and $1/2$ to  the right.
Consider now the operator $\check{R}_k$ that propagates through triangle number $k$ of $\mathbb{T}$ within a given row:
\begin{equation}
\label{labelled-triangle}
\begin{tikzpicture}[rotate=0]
 \draw (0.0,0.0) -- (1.0,0.0) -- (0.5,0.866) -- cycle;
 \draw [fill] (0,0) circle (.5ex);
 \draw [fill] (1.0,0.0) circle (.5ex);
 \draw [fill] (0.5,0.866) circle (.5ex);
 \draw [thick,red] (-0.166,0.289) -- (1.166,0.289);
 \draw [thick,red] (0.167,-0.289) -- (0.833,0.866);
 \draw [thick,red] (0.167,0.866) -- (0.833,-0.289);
 \draw (-0.366,0.339) node {$k$};
 \draw (-0.033,-0.489) node {$k+1$};
 \draw (1.033,-0.489) node {$k+2$};
 \draw (0.067,1.096) node {$k'$};
 \draw (1.133,1.066) node {$(k+1)'$};
 \draw (1.766,0.289) node {$(k+2)'$};
\end{tikzpicture}
\end{equation}
The vertex of $\mathbb{T}$ where the red lines cross   implies a sum over the five splittings, as
in \eqref{loopdiag}, with the corresponding local Boltzmann weight for each term. We have labelled the
three lines as respectively $\{k,k+1,k+2\}$ and $\{k,k+1,k+2\}'$ before and after the action by $\check{R}_k$; the
operator $\check{R}_k$ does nothing to the remainder of the system. Note that with this labelling,
the imaginary time flows in the North-East direction, as usual in the transfer matrix formalism.

We now explain that the transfer matrix $T_L$
that adds another row of $L$ triangles to the lattice can be expressed as follows:
\begin{equation}
 \label{TM_R}
 T_L = u^{-1} \text{qTr} \check{R}_{2L-2} \cdots \check{R}_2 \check{R}_0 \,.
\end{equation}
Before the action by $T_L$, we suppose that the red lines sticking out of the previously completed row are labelled $1,2,\ldots,2L-1,2L$. 
At the beginning of the new row (i.e., before the action by the
first factor $\check{R}_0$) the horizontal space (also called `auxiliary space') is inserted at site $0$. A further site labelled $2L+1$, initially
identified with site $0$, is also inserted in order to manage the periodic boundary conditions along the horizontal direction.
Acting with the successive factors $\check{R}_0, \check{R}_2,\ldots$ in \eqref{TM_R} the auxiliary space is constantly relabelled (as $0 \to 2 \to 4 \to \cdots$),
and finally comes out with label $2L$ after the action by the last factor $\check{R}_{2L-2}$. 
The operation $\text{qTr}$ in \eqref{TM_R}
denotes the `quantum trace' over the horizontal space, which amounts to the initial insertion of the extra sites $0$ and $2L+1$, as
well as the final identification of sites $2L$ with $2L+1$ followed by their removal. In this way the periodic boundary conditions have been
imposed and a row of the lattice is completed. After the $\text{qTr}$ operation, the sites on
the top have the new labels $0', 1',2',\ldots(2L-1)'$, when read from left to right, which we can relabel (working modulo $2L$, and removing the primes)
as $2L,1,2,\ldots,2L-1$. The final application of the left-shift operator $u^{-1}$ in \eqref{TM_R} --- where by convention $u$ denotes the operator
that shifts the sites cyclically towards the right --- brings the labelling back in order as $1,2,\ldots,2L-1,2L$, and we are ready for the next
application of $T_L$.

The periodic Temperley-Lieb algebra on $N=2L$ sites is generated by the identity $1$ and the $e_i$ with $i=1,2,\ldots,N$,
satisfying the algebraic relations
\begin{eqnarray}
 e_i^2 &=& n e_i \,, \label{e_i^2} \\
 e_i e_{i \pm 1} e_i &=& e_i \,, \\
 e_i e_j &=& e_j e_i \quad \mbox{for } |i-j| > 1 \,,
\end{eqnarray}
where indices $i,j$ are considered modulo $N$. The affine Temperley-Lieb algebra is generated by $e_i$, $u$ and $u^{-1}$ and satisfies
the additional relations
\begin{eqnarray}
 u e_i u^{-1} &=& e_{i+1} \,, \\
 u u^{-1} &=& u^{-1} u = 1 \,, \\
 u^{-2} e_1 &=& e_{N-1} e_{N-2} \cdots e_2 e_1 \,.
\end{eqnarray}
These algebras have a faithful diagrammatic representation in
which $e_i$ acts non-trivially on sites $i$ and $i+1$, as
\begin{equation}
 e_i = \begin{tikzpicture}[scale=0.5,baseline=5]
\draw[thick,red] (-3,0) -- (-3,1);
\draw[thick,red] (-2,0) -- (-2,1);
\draw (-1,0.5) node {$\cdots$};
\draw[thick,red] (0,1) .. controls (0.5,0.5) .. (1,1); 
\draw[thick,red] (0,0) .. controls (0.5,0.5) .. (1,0); 
\draw (2,0.5) node {$\cdots$};
\draw[thick,red] (3,0) -- (3,1);
\draw[thick,red] (4,0) -- (4,1);
\draw (-0.2,-0.4) node {$i$};
\draw (1.4,-0.42) node {$i+1$};
\draw (-0.2,1.45) node {$i'$};
\draw (1.4,1.4) node {$(i+1)'$};
\end{tikzpicture} \,.
\end{equation}
By inspecting \eqref{loopdiag} and comparing with \eqref{labelled-triangle} we can therefore write
\begin{equation}
 \label{R_check} 
 \check{R}_k = e_k + z \big( 1 + e_{k+1} e_k + e_k e_{k+1} \big) + w e_{k+1} \,,
\end{equation}
where $e_i$ are generators of the periodic Temperley-Lieb algebra \cite{JacobsenRibaultSaleur2023} on $N=2L$ sites.
Each $\check{R}_k$ acts like the identity operator on the $2L-3$ sites different from $\{k,k+1,k+2\}$. The formulae
\eqref{R_check} and \eqref{TM_R} define the transfer matrix as an element of the $N$-site affine Temperley-Lieb algebra.

Notice, however, that the affine Temperley-Lieb algebra is infinite-dimensional, because non-contractible loops may
pile up and lines may spiral around the periodic direction.
To obtain a manageable problem involving a finite-dimensional linear operator, we therefore let $T_L$ act
in the space of link patterns, sometimes called standard modules ${\cal W}_{j,\tilde{z}}$, with $2j$ defect lines (FK cluster boundaries)
propagating from bottom to top, and (pseudo) momentum $\tilde{z}$ describing the winding of lines with respect to the periodic boundary
condition. We refer to \cite{JacobsenRibaultSaleur2023} for  details.

\subsection{Duality and symmetries}
\label{Duality and symmetries}

We are interested in placing the model \eqref{Zloop} at its critical point, hoping for conformal invariance in the continuum limit.
Since this implies continuous rotational invariance, it seems that the least one may ask for in the lattice model is 6-fold rather
than 3-fold discrete rotational invariance of the diagrams \eqref{loopdiag}. From this consideration, 
Wu and Lin \cite{WuLin1980} conjectured that the critical manifold is given by
\begin{equation} \label{w=1}
 w = 1 \,.
\end{equation}
It will be a huge advantage for us that the model with \eqref{w=1} is now defined in terms of only one local weight $z$.
In what follows we shall take $z \in \mathbb{C}$. We sometimes split it into real and imaginary parts,
\begin{equation}
 z=x+i y \qquad (x,y \in \mathbb{R}) \,.
\end{equation}
We briefly mention  another way of thinking about the constraint \eqref{w=1}. In ordinary graph theory, an edge connects precisely two vertices,
and the usual FK model (mentioned in the Introduction) consists in summing each edge over an occupied and an empty state.
More generally, one may define a hypergraph by allowing for edges to connect more than two vertices. From that point of view, the
diagrams \eqref{FKdiag} correspond to summing each hyperedge in the regular three-uniform hypergraph $\mathbb{T}_0$ over
its possible states of occupancy. The dual graph $\mathbb{T}_0^*$ is obtained by replacing each up-pointing triangle with a down-pointing triangle,
here shown as a dashed line  superimposed on the original triangle:
\begin{equation}
\begin{tikzpicture}[rotate=0]
 \draw (0.0,0.0) -- (1.0,0.0) -- (0.5,0.866) -- cycle;
 \draw [fill] (0,0) circle (.5ex);
 \draw [fill] (1.0,0.0) circle (.5ex);
 \draw [fill] (0.5,0.866) circle (.5ex);
 \draw [dashed] (0.0,0.578) -- (1.0,0.578) -- (0.5,-0.288) -- cycle;
 \draw [thick] (0,0.578) circle (.5ex);
 \draw [thick] (1.0,0.578) circle (.5ex);
 \draw [thick] (0.5,-0.288) circle (.5ex);
 \node at (-0.2,-0.2) {$i$};
 \node at (0.5,1.2) {$j$};
 \node at (1.2,-0.2) {$k$};
 \node at (-0.2,0.778) {$k'$};
 \node at (1.2,0.778) {$i'$};
 \node at (0.5,-0.588) {$j'$};
\end{tikzpicture}
\end{equation}
The duality transformation amounts to replacing each hyperedge on $\mathbb{T}_0$ with a hyperedge on $\mathbb{T}_0^*$
which is in the opposite state of occupancy: $i'$ is connected to $j'$ iff $k$ is disconnected from $i$ and $j$, and so on by
cyclic permutation of the labels $(i,j,k)$. In particular, the first and fifth diagrams in \eqref{FKdiag} or \eqref{loopdiag} get
interchanged (up to rotation). From this perspective, the constraint \eqref{w=1} simply means that we place the model at its
{\em selfdual point} (in the sense of hyperduality).

Let us return to the loop model defined by \eqref{Zloop} with configurations \eqref{loopdiag}. Superficially this looks like
an $O(n)$ model, with $n = Q^{1/2}$. However,  the global symmetry is larger than $O(n)$. Consider endowing the underlying
lattice $\mathbb{T}$ with the following {\em fixed} orientations of its edges:
\begin{equation}
\begin{tikzpicture}[rotate=0]
 \draw [thick,red] (-0.166,0.289) -- (1.166,0.289);
 \draw [thick,red] (0.167,-0.289) -- (0.833,0.866);
 \draw [thick,red] (0.167,0.866) -- (0.833,-0.289);
 \draw [thick,red,-stealth] (-0.033,0.289) -- (0.034,0.289);
 \draw [thick,red,-stealth] (0.900,0.289) -- (1.099,0.289);
 \draw [thick,red,-stealth] (0.766,0.751) -- (0.733,0.693);
 \draw [thick,red,-stealth] (0.300,-0.058) -- (0.200,-0.231);
 \draw [thick,red,-stealth] (0.766,-0.174) -- (0.733,-0.116);
 \draw [thick,red,-stealth] (0.300,0.635) -- (0.200,0.808);
 \draw (0.0,0.0) -- (1.0,0.0) -- (0.5,0.866) -- cycle;
 \draw [fill] (0,0) circle (.5ex);
 \draw [fill] (1.0,0.0) circle (.5ex);
 \draw [fill] (0.5,0.866) circle (.5ex);
\end{tikzpicture}
\end{equation}
Each horizontal line is oriented from left to right, and the other lines are obtained from this by rotations through $\pm \tfrac{2\pi}{3}$. Then  each of the transition systems \eqref{loopdiag} is such, that an observer moving along any one of its loop segments   sees consistently oriented arrows. By extension, each loop is consistently oriented with respect to this fixed orientation of the lattice.
This implies that the loop model has in fact the global symmetry $PSU(n)$. The simplest square-lattice loop model with this symmetry was studied extensively in \cite{RouxJacobsenRibaultSaleur2025}. The symmetry $PSU(n)$ is an example of a Deligne category, which is well-defined for any $n$ (and hence for any $Q \in \mathbb{C}$). The CFT properties of the loop model are enhanced by the properties of this global symmetry, as explained in \cite{RouxJacobsenRibaultSaleur2025} to which we refer for further details.

\subsection{Phase diagram as a function of $z$}
\label{Phase diagram as a function of z}
The ordinary nearest-neighbor Potts model on the triangular lattice is a special case of \eqref{Hijk} with $K_3=0$.
Defining $c_2 = {\rm e}^K - 1$ as above, it is known to be integrable along the curve \cite{BaxterTemperleyAshley1978} 
\begin{equation}
 \label{cubic}
 (c_2)^3 + 3 (c_2)^2 = Q \,.
\end{equation}
From \eqref{eKL} we see that this requires $c_3 = Q$, so from \eqref{zw} this means $w=1$ in agreement with \eqref{w=1}.
In other words, the Wu-Lin model on its critical manifold \eqref{w=1} is a deformation of the integrable nearest-neighbor
model, in which the free parameter $z$ determines how $K_3$ should be adjusted against $K$ to maintain criticality.

A useful parameterisation of the cubic equation \eqref{cubic} is obtained by setting \cite{JacobsenSalasScullard2017}
\begin{equation}
 c_2 = -1 + 2 \cos \left( \frac{2\pi(g-1)}{3} \right) \,, \quad Q = 4 \cos^2(\pi g) \,,
\end{equation}
with $-\tfrac12 \le g \le 1$. We have then
\begin{equation}
 z = \frac{c_2}{\sqrt{Q}} = \frac{-1 + 2 \cos \left( \frac{2\pi(g-1)}{3} \right)}{2 \cos(\pi g)} \,.
\end{equation}
Notice that for each $0 < Q < 4$ there are three distinct solutions for $g$, one in each of the intervals $(-\tfrac12,0)$, $(0,\tfrac12)$ and $(\tfrac12,1)$.
Only the latter corresponds to $z > 0$.

It was established numerically \cite{JacobsenSaleur2006} that the interval $0 < g \le 1$ corresponds to a CFT with central charge
\begin{equation}
 \label{cc_CG}
 c = 1 - \frac{6 (1-g)^2}{g} \,.
\end{equation}
This is easily recognised as the usual Coulomb gas expression. It is this CFT whose analytic continuation shall be our main interest in this paper.
Meanwhile, the interval $-\tfrac12 < g < 0$ is described by a quite different CFT with central charge
\begin{equation}
 c = 2 + 6g
\end{equation}
and unusual properties, such as a continuous spectrum of critical exponents. It can be identified by the black-hole sigma model,
whose square-lattice counterpart has been extensively studied \cite{Baxter1982,JacobsenSaleur2006,IkhlefJacobsenSaleur2008,IkhlefJacobsenSaleur2012,Bazhanov2019}.

\subsection{Analytic continuation}\nopagebreak
\label{Analytic continuation from minimal models}

We henceforth concentrate on the CFT with central charge \eqref{cc_CG}.
To discuss its analytic continuation away from the real interval $c \in [-\infty,1]$, or equivalently away from a Potts model with
real $Q \in [0,4]$, we find it useful to introduce the parameter
\begin{equation}
 b = \frac{1}{1-g} \,.
\end{equation}
The corresponding CFT, denoted $\ca B_b$, has
\bea\label{Q-of-b}
Q(b) &=& 4 \cos\Big(\frac{\pi}b\Big)^2 \,, \\
\label{c-of-b}
c(b) &=& 1-\frac{6}{b(b-1)} \,,
\eea 
so that
\be
b(Q)=\frac{\pi }{\mbox{arccos} (\frac{\sqrt{Q}}{2} )}. 
\ee
The choice of branch is such that $b>0$ corresponds to $Q<4$.

Notice that for integer $b = 3,4,\ldots$, \eqref{c-of-b} coincides formally with the series of unitary minimal models $\ca M_{m+1,m}$ \cite{DiFrancescoMathieuSenechal}
via the identification
\be
\ca M_{b,b-1} \equiv \ca B_b.
\ee
Within these models it is customary to specify the dimension of the holomorphic part of the operator $\phi_{r,s}(z)$ in terms of the Kac table
\be
\label{Kac-table}
h_{r,s}=\frac{[b (r-s)+s]^2-1}{4 (b-1) b} \,.
\ee
The dimension of the full operator $\Phi(z,\bar z): =  \phi_{r,s}(z) \bar\phi_{\bar r,\bar s}(\bar z)$ and its spin are   given by
\bea
\Delta_\Phi &=&  h_{r,s} + h_{\bar r,\bar s}, \\
s_\Phi &=&  h_{r,s} - h_{\bar r,\bar s}.
\eea
In this paper we are   interested in the continuum limit of the loop model which is neither unitary, nor minimal.
Minimality in \Eq{Kac-table} would require $1 \le r \le b-2$ and $1 \le s \le b-1$, but the operator content of loop models is
not confined to this range, as will be discussed below. We shall    consider \eqref{Kac-table}   a convenient parametrization, and allow  $b \in \mathbb{C}$ to take complex values.

\begin{figure}[t]
\Fig{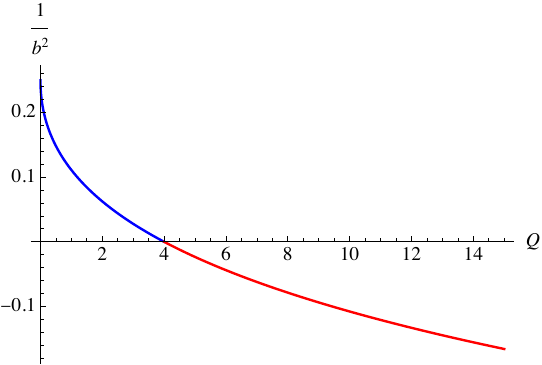}
\caption{$1/b^2$ as a function of $Q$. For $Q\le 4$, $b$ is real, while for $Q>4$, $b$ is purely imaginary.}
\label{1ob2ofQ}
\end{figure}

While $\ca B_{b}$ is a critical model for $b$ integer, its tricritical counterpart
identifies with the next minimal model $\ca B_{b+1}$, which according to the above can also be written as $\ca B_{-b}$. 
This means that the collision between critical and tricritical theories occurs when $b \to \infty$.
The desired analytic continuation must be performed around this value, for which  $\frac1 {b(Q)^2}$
has a regular Taylor series around $Q=4$, see Fig.~\ref{1ob2ofQ},

\begin{figure*}[t]
\centerline{\begin{minipage}{\columnwidth}
{\bf Upper path}
\end{minipage}
\hfill
\begin{minipage}{\columnwidth}
{\bf Lower path}
\end{minipage}}
\centerline{\begin{minipage}{\columnwidth}
\Fig{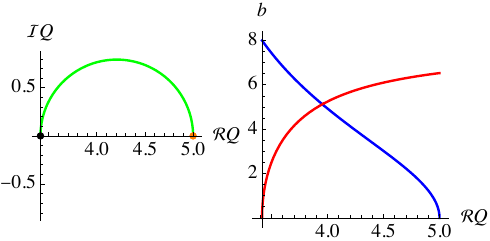}
\end{minipage}
\hfill
\begin{minipage}{\columnwidth}
\Fig{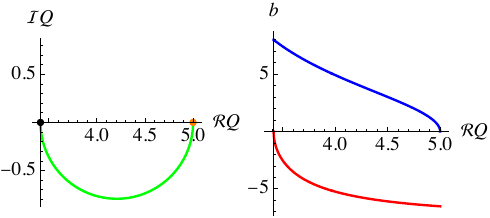}
\end{minipage}}
\begin{minipage}{\columnwidth}
\Fig{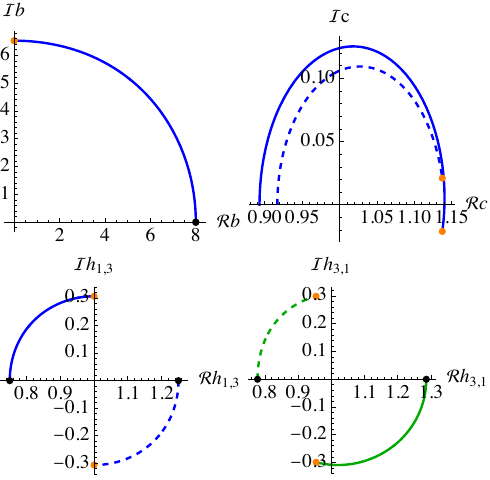}
\end{minipage}
\hfill
\begin{minipage}{\columnwidth}
\Fig{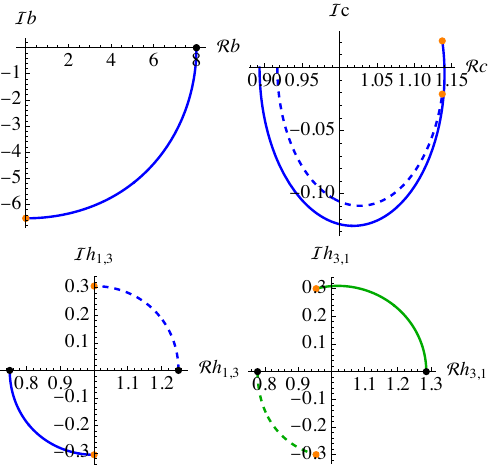}
\end{minipage}
\caption{Upper row: paths taken in the analytic continuation, starting at $Q=3.41421$ ($b=8$, black dot) and ending at $Q=5$ ($b=\pm 6.5285 i$, orange dot [gray in B\&W]).
In the left part of the figure, the path taken is a semicircle in the upper half-plane in $Q$ (``upper path''), while in the right part it is a semicircle in the lower half-plane (``lower path'').
The remaining plots show the central charge $c$, as well as the dimensions $h_{1,3}$ and $h_{3,1}$, for the corresponding continuations of the critical point $\ca B_b$ (solid lines) and the tricritical point $\ca B_{-b}$ (dashed lines).
Once we arrive at $Q\to 5$ (with $b \in i \mathbb R$), shown as an orange dot, the two continuations of $c$, $h_{1,3}$ and $h_{3,1}$, starting from either $\ca B_b$ or $\ca B_{-b}$, are complex conjugates of each other. The same is true if we compare the two continuations along the upper and lower paths, with the same starting point.}
\label{f:paths}
\end{figure*}
\bea
\frac{4\pi^2} {b(Q)^2} &=&(4-Q)+\frac{(Q-4)^2}{12} +\frac{(4-Q)^3}{90} +\frac{(Q-4)^4}{560} \nn\\
&&   +\frac{(Q-4)^5}{3150}+\ca O (4-Q)^6 \,.
\eea
With our choice of $b>0$ for $0<Q<4$, this uniquely defines the analytic continuation in $Q$. As shown in Fig.~\ref{f:paths}, 
there are two distinct paths one can take, passing above or below $Q=4$ in the complex plane.

For $Q=5$ we find\footnote{There are two more solutions, $b=\pm t$, $t= 0.977075 + 0.149663 i$. They have central charge $c=13. + 38.252 i$ (for $t=b$) and $c=-1.98251 + 0.690625 i$ for $t=-b$. We do not know what these branches mean. What makes them unlikely to meet our criteria is the fact that they are not complex conjugate of each other, as we expect for real $Q$.}
\be
b(Q=5) = \mp 6.5285 i.
\ee
The upper sign, corresponding to analytic continuation along the lower path in Fig.~\ref{f:paths},
is what will be necessary for the critical coupling $z_{\rm c}$ to belong to the upper half plane. 
We then use this value for analytic continuation, e.g.
\be\label{c-Q=5}
c(Q=5)= 1.13755 \pm 0.0210687 i
\ee
There are also the complex conjugate solutions $\bar b$ and $\bar c$. As we show below, 
 for  $\Im z_{\rm c}>0$  we see the solutions for $b$ and $c$   with the upper sign. 

\begin{figure}[b]
\Fig{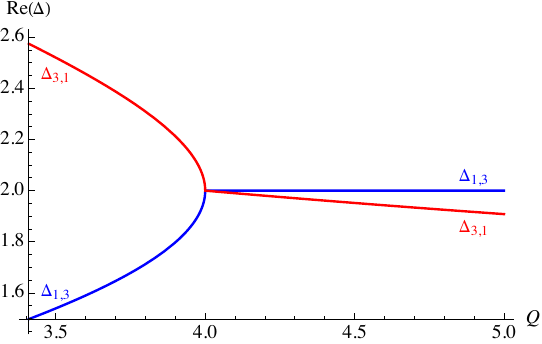}
\caption{$\Delta_{1,3}$ (unphysical branch)  and $\Delta_{3,1}$ physical branch.}
\label{Delta13+31}
\end{figure}

\subsection{Phase diagram as a function of $Q$}
\label{Phase diagram as a function of Q}

In this section, we determine the stability of the critical and tricritical fixed points in the complex $Q$-plane by analyzing
the subleading scalar operator \(\epsilon'\), which governs the perturbation of the corresponding fixed points,
given that we have tuned the amplitude of the leading scalar operator $\epsilon$ to 0 by restricting to the
critical manifold \eqref{w=1}. Readers familiar with \cite{Zamolodchikov1987} will agree that the most natural
candidates for these operators are $\epsilon = \phi_{2,1} \times \bar \phi_{2,1}$ or
$\phi_{1,2} \times \bar \phi_{1,2}$, and $\epsilon' = \phi_{3,1} \times \bar \phi_{3,1}$ or
$\phi_{1,3} \times \bar \phi_{1,3}$. Pending the detailed discussion of the operator content of the Potts loop model
(section~\ref{sec:op-content-Potts}) and a detailed numerical investigation (section~\ref{sec:flowB8})
we shall admit here that the correct identification is
\begin{eqnarray}
 \epsilon &=& \phi_{2,1} \times \bar \phi_{2,1} := V^d_{(2,1)} \,, \\
 \epsilon' &=& \phi_{3,1}\times \bar \phi_{3,1} := V^d_{(3,1)}
\end{eqnarray}
for both the critical point $\ca B_{b}$ and the tricritical point $\ca B_{-b}\equiv \ca B_{b+1}$, and indeed
also for the complex CFT obtained by making both fixed points collide.

In order to decide whether a fixed point is stable or unstable towards the perturbation by $\epsilon'$,
we thus have to check whether $\Re h_{3,1}$ is smaller or larger than $1$.
To simplify our considerations, we  note the curious relation 
\be
\label{curious}
h_{3,1}({\ca B}_b) h_{3,1}({\ca B}_{-b}) = 1 \,.
\ee
For $0 \le Q < 4$, when $b$ and hence $h_{3,1}$ are real, this implies that one of the fixed points is attractive
(critical) and the other repulsive (tricritical), in agreement with the discussion above.
Having two repulsive fixed points is possible, but only if both $h_{3,1}({\ca B}_b)$ and $ h_{3,1}({\ca B}_{-b})$
are complex.

For future reference we note the numerical values
\bea
 2 h_{1,3}({\ca B}_8) &=& \frac32 , \label{2h13=32} \\
 2 h_{3,1}({\ca B}_8) &=& \frac{18}{7} = 2.57143. \\
2 h_{1,3}(Q=5) &=& 2 - 0.612698 i,\\
\label{h31}
2 h_{3,1}(Q=5) &=& 1.9083 + 0.598652 i.
\eea
Curiously, for $Q=5$ the first is marginal, while the second is slightly relevant. This is illustrated in Fig.~\ref{Delta13+31}. We will come back to this  below.

\begin{figure}[t]
\Fig{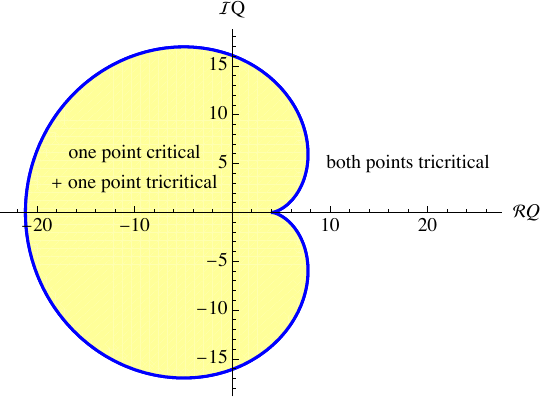}
\caption{The critical domain as a function of $Q$.}
\label{Reh31=1-v2}
\end{figure}%
The separatrix in the complex $Q$-plane between the two situations is given by the curve
$\Re h_{3,1}=1$.
It can be recast in the parametric form \cite{BohmJacobsenJiangZhang2022}
\be\label{parametric-curve}
Q(\varphi)=2+2 \cos \left(\frac{\pi }{\frac{1}{2}+i \varphi}\right)\,, \quad \varphi \in [-\tfrac12,\tfrac12]
\ee
and is plotted in Fig.~\ref{Reh31=1-v2}. On the inside of the curve (shown in yellow) we
have one critical and one tricritical point, while in its complement there is no 
stable fixed point (both points are tricritical).

The cusp of the separatrix is at $Q=4$. The maximum of $\Re Q$ is $7.746\,347\ldots$ and is attained
for $\varphi = \pm 0.193\,437\ldots$. The maximum of $|\Im Q|$ is $16.949\,180\ldots$ and is attained
for $\varphi = \pm 0.336\,521\ldots$. Finally, the minimum of $\Re Q$ is $2 - 2 \cosh \pi = -21.183\,906\ldots$
and is attained for $\varphi = \pm \frac12$.

In Fig.~\ref{h31prodPlot} we show the same plot as a function of the conformal weight $h_{3,1}({\ca B}_b)$.
The yellow domain with $\Re h_{3,1}({\ca B}_b) > 1$ corresponds   to ${\ca B}_b$ being attractive
(i.e.\ critical) while  ${\cal B}_{-b}$ to being repulsive (i.e.\ tricritical). The other yellow domain corresponds to
the opposite situation, where ${\ca B}_b$ is tricritical and ${\cal B}_{-b}$ is critical.
By \eqref{curious} it is delimited by the separatrix $\frac{1}{1 + i \mathbb{R}}$,   a fancy rewriting of a circle of radius $1/2$ centered at $1/2$.

\subsection{Operator content and conformal dimensions}
\label{sec:op-content}

The non-unitary CFTs describing the continuum limit of the $Q$-state Potts-FK model with $0 \le Q \le 4$ and the $O(n)$ loop model
with $-2 \le n \le 2$, both with real central charge $c \le 1$, were studied in \cite{DiFrancescoSaleurZuber1987} and a recent series of papers including
\cite{JacobsenSaleur2019,HeJacobsenSaleur2020,JacobsenRibaultSaleur2023,GransSamuelssonJacobsenNivesvivatRibaultSaleur2023,NivesvivatRibaultJacobsen2024,RouxJacobsenNivesvivatRibault2025,JacobsenNivesvivatRibaultRoux2025}. We can formulate the main result of the present paper in the form of a pair of conjectures:

\begin{figure}[t]
\Fig{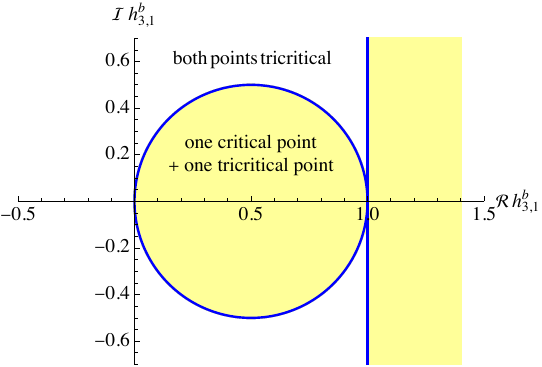}
\caption{In yellow we show the domains in which one fixed point is stable while the other is unstable.}
\label{h31prodPlot}
\end{figure}%

\medskip
\noindent
{\bf Conjecture 1}. For any $Q \in \mathbb{C}$ inside the critical domain,
the loop model \eqref{Zloop} at its selfdual point \eqref{w=1} and at the critical value $z_{\rm c}$ of the parameter $z$ \eqref{zw}
has a continuum limit that is a complex CFT.

\medskip
\noindent
{\bf Conjecture 2}. For any $Q \in \mathbb{C}$, this complex CFT is the analytic continuation,   as prescribed in Section~\ref{Analytic continuation from minimal models},
of the non-unitary CFT studied in the preceding list of references.

\medskip
We now unroll the consequences of Conjecture 2 in terms of the conformal data for the critical loop models.
We wish to make contact with the results and notations contained in the above list of references which treat in a unified way loop models of both $O(n)$ and Potts type.
In the latter case, the loop weight $n = Q^{1/2}$. Although the present paper is concerned with the Potts-type loop model \eqref{Zloop},
we can give results that cover also the $O(n)$ loop model at almost no extra cost.
In other words, although Conjectures~1--2 are stated for the Potts loop model, which is our main focus here,
the equivalent statements should also hold true for the analytic continuation of the $O(n)$ loop model
to $n \in \mathbb{C}$, that was studied in \cite{HaldarTavakolMaScaffidi2023}.

That being said, there are some notational differences and small subtleties that distinguish the Potts and $O(n)$
models, and which need to be stated clearly. To avoid any undue confusion, we split the remainder of this
section into two parts, one for each model.

\subsubsection{$O(n)$ loop model}
\label{sec:op-content-On}

We start with the $O(n)$ model, because most of the results in
\cite{JacobsenSaleur2019,HeJacobsenSaleur2020,JacobsenRibaultSaleur2023,GransSamuelssonJacobsenNivesvivatRibaultSaleur2023,NivesvivatRibaultJacobsen2024,RouxJacobsenNivesvivatRibault2025,JacobsenNivesvivatRibaultRoux2025}
are formulated in a way that gives precedence to that model.
To make contact with those references, we parametrize the loop weight $n$ by $\beta^2$, as
\begin{align}
 n = -2\cos\left(\pi\beta^2\right) \,,
 \label{n}
\end{align}
with $\Re \beta^2 > 0$.
We also denote the left and right conformal dimension by $(\Delta,\bar{\Delta})$. The dimension
and the central charge will be parametrized by
\begin{eqnarray}
 c &=& 1 - 6 (\beta - \beta^{-1})^2 \,, \\
 \Delta_{(r,s)} &=& \frac14\left(\beta r-\beta^{-1}s\right)^2 -\frac14\left(\beta-\beta^{-1}\right)^2\ . \label{Delta_rs}
\end{eqnarray}
It is also useful to rewrite a general conformal dimension $\Delta$ and its Kac-table parametrization $\Delta_{(r,s)}$
in terms of momentum variables $P$ and $P_{(r,s)}$ as follows:
\begin{eqnarray}
 \Delta &=& P^2 - P_{(1,1)}^2 \,, \\ 
 \Delta_{(r,s)} &=& P^2_{(r,s)} - P_{(1,1)}^2  \,, \label{Kac_On} \\
 P_{(r,s)} &=& \tfrac12\left(-\beta r +\beta^{-1}s\right) \,. \label{P-momentum}
\end{eqnarray}
It was then shown in \cite{DiFrancescoSaleurZuber1987,JacobsenSaleur2019} that the operator content (or {\em spectrum}) of the $O(n)$ loop-model CFT is:
\begin{align}
 \renewcommand{\arraystretch}{1.6}
 \begin{array}{|l|l|l|l|}
 \hline 
 \multicolumn{4}{|c|}{\text{Primary fields in the $O(n)$ loop model CFT}} \\
 \hline
  \text{Name} & \text{Notation} & \text{Parameters} & (\Delta,\bar\Delta)
  \\
  \hline 
  \text{Degenerate} & V^d_{(r,s)} & r{=}1; s{\in}2\mathbb{N}+1 & \left(\Delta_{(r,s)},\Delta_{(r,s)}\right) 
  \\
  \text{Diagonal} & V_P & P\in\mathbb{C} & \left( \Delta_P ,\Delta_P \right)
  \\
  \text{Non-diagonal} & V_{(r,s)} & r{\in}\frac12\mathbb{N}^*;s{\in}\frac{1}{r}\mathbb{Z} & \left(\Delta_{(r,s)},\Delta_{(-r,s)}\right) 
  \\
  \hline 
 \end{array}
\nonumber 
\end{align}
where $\mathbb{N}^* = \mathbb{N} \setminus \{0\}$.

This spectrum involves three different types of operators. The {\em degenerate operators} $V^d_{(r,s)}$ encompass the continuum limit of the identity operator,
the local energy operator $\epsilon = V^d_{(1,3)}$,
and the whole series of higher energy operators. These operators have one singular vector at level $s$, so their highest-weight representations are Kac modules. 
The label $s \in \mathbb{N}^*$ can take any parity for the Potts model, but must be odd for the $O(n)$ model. 

The {\em diagonal operators} $V_P$ correspond to changing the weight of a loop surrounding the insertion point of one such operator from $n$ to the modified value 
\begin{equation}
 w(P) = 2\cos\left(2\pi\beta P\right),
 \label{wp}
\end{equation}
determined by a momentum $P$, which can be an arbitrary complex number.
Such operators are non-degenerate and correspond to a full Verma module.

Finally, the {\em non-diagonal operators} $V_{(r,s)}$ result from inserting a defect of $2r$ open loop segments, each of which   acquires a complex phase ${\rm e}^{i \pi s}$
upon making a full turn around the insertion point.   
These open segments are not allowed to contract among themselves, but must end on another non-diagonal
operator.%
\footnote{Open segments can also end on a boundary, if the system has one, and if the boundary conditions are appropriate (non-diagonal).
In this paper we consider the bulk CFT, so there are no boundaries.}
Their conformal spin $r s$ is integer. The representations of non-diagonal operators are again Verma modules; when $s > 0$  they are 
logarithmic with a rank-two Jordan cell at level $r s$. We are here   interested in the spinless case $s=0$ in which no phase factors appear.
These non-diagonal operators correspond to the Temperley-Lieb standard modules ${\cal W}_{j,\tilde{z}^2}$ discussed in 
Section~\ref{sec:TM}, setting $j = r$ and $\tilde{z} = {\rm e}^{i \pi s}$ there. We  refer to \cite{JacobsenRibaultSaleur2023} for details.

To establish the link between the parametrizations \eqref{n}--\eqref{Delta_rs} and those used above --- in particular
in the the algebraic relation \eqref{e_i^2} and the parametrizations \eqref{Q-of-b}--\eqref{c-of-b} --- we should set
\begin{equation}
 n = 2 \cos \left( \frac{\pi}{b} \right) \ \Longrightarrow\ \beta^2= g = \frac{b-1}{b} \,.
\end{equation}
We then retrieve \eqref{Q-of-b} with $Q(b) = n^2$. Moreover, it is then straightforward to show that the relation to the standard Kac table \eqref{Kac-table} is
\begin{equation}
 \Delta_{(r,s)} = h_{s,r} \,, \qquad \text{[$O(n)$ model]} \,.
\end{equation}
i.e., the indices $r,s$ are {\em interchanged} on the right-hand side.

\subsubsection{Potts loop model}
\label{sec:op-content-Potts}

The results for the Potts loop model are almost the same, except for this interchange of indices.
We have the following operator content:%
\footnote{The modular invariant torus partition function \cite{DiFrancescoSaleurZuber1987} contains the degenerate,
order parameter and non-diagonal fields, except non-diagonal fields $V_{(r,s)}$ with $s=1$. The latter $V_{(r,1)}$, as well
as the most general diagonal fields $V_P$, can  be realized in the representation theory of the underlying
affine Temperley-Lieb algebra, which is the basis of our transfer-matrix treatment, so they are included in the table.}
\begin{align}
 \renewcommand{\arraystretch}{1.6}
 \begin{array}{|l|l|l|l|}
 \hline 
 \multicolumn{4}{|c|}{\text{Primary fields in the Potts loop model CFT}} \\
 \hline
  \text{Name} & \text{Notation} & \text{Parameters} & (\Delta,\bar\Delta)
  \\
  \hline 
  \text{Degenerate} & V^d_{(r,s)} & s{=}1; r{\in}\mathbb{N}^* & \left(\Delta_{(r,s)},\Delta_{(r,s)}\right) 
  \\
  \text{Diagonal} & V_P & P\in\mathbb{C} & \left( \Delta_P ,\Delta_P \right)
  \\
  \text{Order parameter} & V_{(r,0)} & r \in \mathbb{N}+\tfrac12 & \left(\Delta_{(r,0)},\Delta_{(r,0)}\right) 
  \\
  \text{Non-diagonal} & V_{(r,s)} & s{\in}\mathbb{N}^* ;r{\in}\frac{1}{s}\mathbb{Z} & \left(\Delta_{(r,s)},\Delta_{(-r,s)}\right) 
  \\
  \hline 
 \end{array}
\nonumber 
\end{align}
where now the notation $\Delta_{(r,s)}$ coincides with that of the Kac table \eqref{Kac-table}:
\begin{equation}
 \Delta_{(r,s)} = h_{r,s} \,, \qquad \text{[Potts model]} \,.
\end{equation}
Accordingly, in the column ``Parameters'' the labels $r,s$ are reversed with respect to the corresponding table
for the $O(n)$ model. For the same reason \eqref{Kac_On}--\eqref{P-momentum} are replaced by
\begin{eqnarray}
 \Delta_{(r,s)} &=& P^2_{(r,s)} - P_{(1,1)}^2  \,, \\
 P_{(r,s)} &=& \tfrac12\left(-\beta s +\beta^{-1}r\right) \,. \label{P-momentum-Potts}
\end{eqnarray}
But there are a few more subtle differences, which we now highlight.

First, in the list of degenerate operators, the non-trivial index $r$ can now take both even and odd values.
In particular, the local energy operator is $\epsilon = V^d_{(2,1)}$.

Second, the non-diagonal operators are now interpreted as inserting a defect of $s$ Potts cluster boundaries.
Each such boundary can be thought of as a marked point that inserts a pair of open loop segments, so we
have $2s$ open loop segments as before, but now $s$ must be an integer rather than half an integer.
Another way to see this is that odd values of $2s$ simply do not make sense in the microscopic formulation
of the Potts loop model, because of the completely-packing constraint.

Third, the non-diagonal operator with $s=1$ actually drops out of the spectrum (it has zero multiplicity in the modular invariant torus partition function).
It is replaced by the {\em order-parameter operators} $V_{(r,0)}$, where $r \in \mathbb{N}+\tfrac12$.
The first member of this series is $\sigma = V_{(1/2,0)}$, which can be interpreted in the spin formulation as the 
local spin operator, and in the cluster formulation as an operator, that inserts one cluster (not a cluster boundary).
The order-parameter operators are in fact just special cases of diagonal operators. To see why, notice that
for any $r \in \mathbb{N}+\frac12$, inserting $P=P_{(r,0)}$ into \eqref{wp} we find $w(P) = 0$. This means that
the two-point function $\langle V_P V_P \rangle$ does not admit any loop that separates one of the operator
insertion points from the other. Therefore they will have to belong to the same FK cluster, which is the defining
property obeyed by the two-point function $\langle V_{(r,0)} V_{(r,0)} \rangle$ of an order-parameter operator.

We remark in passing that comparing \eqref{Delta_rs} with \eqref{Kac_On} shows that the dimension $\Delta_{(r,s)}$
of any operator can be evaluated in this way by passing through the equivalent momentum $P = P_{(r,s)}$.

\subsection{Global $S_Q$ symmetry}
\label{sec:global-sym}
Instead of using the loop representation \eqref{Zloop} of the Potts model, it can also be studied in the spin representation for integer $Q$.
When $Q=1$ the resulting model is trivial (all Potts spins take the same value). $Q=2$ and $Q=3$  give the minimal models
with $b=4$ and $b=6$, as discussed in Section~\ref{Analytic continuation from minimal models};  $Q=4$ similarly arises as the
limit $b \to \infty$. In those cases the spin representation results from taking a quotient of the underlying Temperley-Lieb algebra. This quotient renders the CFT unitary and, for $Q=2$ and $Q=3$  minimal. If the quotient is not taken, the non-local operators make the CFT
logarithmic: the same would be true for taking $Q \to 1$ as a limit in the loop model. 
Our   interest instead is   to take {\em generic} values of the loop weight $n = \sqrt Q$ --- namely those where $n = q + q^{-1}$, and $q$ is not   a root of unity.
The diagram algebra of Section~\ref{sec:TM} then has a  semisimple representation,  corresponding to the non-logarithmic CFT
with operator content described in Section~\ref{sec:op-content}.

Taking $Q>4$ integer and using the spin representation is another generic choice. This generic representation   allows us to make contact with 
the works \cite{TangMaTangHeZhu2024,TangLiuTangZhu2025} mentioned in the introduction, where the complex $Q$-state Potts CFT
was studied as the scaling limit of a $Q=5$ state non-Hermitian spin chain.
To clarify this link we need to discuss the global $S_Q$ symmetry of the model, following \cite{JacobsenRibaultSaleur2023}.
As explained there, operators can be classified in terms of $S_Q$ representations in the generic case,
where $Q$ may take non-integer or even complex values. Recall that finite-dimensional irreducible representations of $S_Q$, that we  
first think of for $Q \in \mathbb{N}^*$,
are parametrized by Young diagrams with $Q$ boxes. To extend this to a more general setting, we pick an integer partition $\lambda = [\lambda_1,\lambda_2,\lambda_3\cdots]$
with integer $\lambda_1 \ge \lambda_2 \ge \lambda_3 \ge \cdots$ that we can visualize as row lengths of a diagram. Let $|\lambda| = \sum_i \lambda_i$ be the size of this
diagram. Then, for $Q \ge |\lambda| + \lambda_1$ integer, $\widetilde{\lambda} = [\lambda_0,\lambda]$ is a Young diagram with an extra zeroth row of length
$\lambda_0 = Q-|\lambda|$ and size $|\lambda_0| = Q$.
It turns out that $\widetilde{\lambda}$ is a valid $S_Q$ irreducible representation, in the sense of Deligne categories, even when $Q \in \mathbb{C}$. In particular we can compute
its dimension (not necessarily a positive integer) and write tensor products $\widetilde{\lambda}_1 \times \widetilde{\lambda}_2$ as a sum of irreducibles \cite{JacobsenRibaultSaleur2023}.

For the Potts model, the trivial, one-dimensional (scalar) representation is written $\lambda = [\,]$; a $Q$-dimensional vector in which the Potts indices
are permuted under $S_Q$ decomposes on the irreducibles $[1] + [\,]$; and generally any operator acting on $M$ spins can be decomposed on irreducibles $\lambda$
with $|\lambda| \le M$ \cite{CouvreurJacobsenVasseur2017}.

The state space $\mathcal{S}$ of the $Q$-state Potts CFT is a representation of ${\cal C}_c \times S_Q$, where ${\cal C}_c$ denotes the conformal algebra for the CFT with
central charge $c$, and $S_Q$ is the global symmetry of the symmetric group in the Deligne sense. Decomposing this on irreducibles one finds \cite{JacobsenRibaultSaleur2023} that 
\begin{eqnarray}
 \mathcal{S} &=& \bigoplus_{r\in \mathbb{N}^*} \mathcal{R}_{\langle r,1\rangle}\otimes [\,] \ \ \oplus\bigoplus_{r\in \mathbb{N}+\frac12} \mathcal{W}_{(r,0)}\otimes [1] \nonumber \\
 & & \oplus \bigoplus_{s\in\mathbb{N}+2}\bigoplus_{r\in\frac{1}{s}\mathbb{Z}}  \mathcal{W}_{(r,s)}\otimes \Xi_{(r,s)}\ .
 \label{spotts}
\end{eqnarray}
Here the ${\cal C}_c$ irreducibles $\mathcal{R}_{\langle r,1\rangle}$, $\mathcal{W}_{(r,0)}$ and $\mathcal{W}_{(r,s)}$ correspond  to the set of degenerate
(energy-type) operators $V^d_{(r,1)}$, the order-parameter (spin-type) operators $V_{(r,0)}$, and the non-diagonal (defect-type) operators $V_{(r,s)}$ acting on $2s$ TL lines
or $s$ FK clusters.  They are exactly the operators that were discussed in
Section~\ref{sec:op-content-Potts}.
On the right-hand side of the tensor products we find the $S_Q$ representations which are, respectively, for the three types of operators: 
the trivial representation $[\,]$, the standard representation $[1]$, and another family of representations $\Xi_{(r,s)}$.
This latter family of representations satisfy $\Xi_{(r,s)} = \Xi_{(-r,s)} = \Xi_{(r+1,s)}$, so the representations of ${\cal C}_c$ can   be regrouped into larger
representations of a so-called interchiral algebra, obtained from ${\cal C}_c$ by acting with the degenerate energy operator $V^d_{(2,1)}$.
This regrouping   amounts to constraining the $r$-index to the interval $0 \le r \le 1/2$, keeping the spin $r s$   integer.
Finally, $\Xi_{(r,s)}$ can be decomposed into $S_Q$ irreducibles $\lambda$. For the first few cases one finds \cite{JacobsenRibaultSaleur2023}
\begin{subequations}
\label{x-all}
\begin{align}
 \Xi_{(0,2)} &= [2] \ , 
 \\
 \Xi_{(\frac12,2)} &= [1^2] \ ,
 \\
 \Xi_{(0,3)} &= [3]+[1^3]  \ , 
 \\
 \Xi_{(\frac13,3)} &= [21]  \ ,
\end{align}
\end{subequations}
where $[1^k]$ is a short-hand for $[1,1,\ldots]$ with an index repeated $k$ times.

After this review, we are   ready to discuss the link between the operator content of Section~\ref{sec:op-content-Potts} and that exposed
in \cite{TangMaTangHeZhu2024}. Table I of that reference lists 11 low-lying primary operators in the $Q=5$ state complex Potts CFT.
Here is a table where we compare them to the above discussion:
\begin{align}
 \renewcommand{\arraystretch}{1.3}
 \begin{array}{|l|l|l|l|l|}
 \hline 
  \text{\cite{TangMaTangHeZhu2024}} & \text{This work} & \lambda & \text{Dimension} & \text{Spin}
  \\
  \hline 
  \epsilon & V^d_{(2,1)} & [\,] & 2 h_{2,1} = 0.466 + 0.255 i & 0
  \\
  \epsilon' & V^d_{(3,1)} & [\,] & 2 h_{3,1} = 1.908 + 0.599 i & 0
  \\
  \epsilon'' & V^d_{(4,1)} & [\,] & 2 h_{4,1} = 4.328 + 1.123 i & 0
  \\
  \sigma & V_{(\frac12,0)} & [1] & 2 h_{\frac12,0} = 0.134 + 0.021 i & 0
  \\
  \sigma' & V_{(\frac32,0)} & [1] & 2 h_{\frac32,0} = 1.111 + 0.170 i & 0
  \\
  \sigma'' & V_{(\frac52,0)} & [1] & 2 h_{\frac52,0} = 3.065 + 0.469 i & 0
  \\
  Z & V_{(0,2)} & [2] & 2 h_{0,2} = 2.011 - 0.305 i & 0
  \\
  Z' & V_{(0,3)} & [1^3] & 2 h_{0,3} = 4.511 - 0.688 i & 0
  \\
  X & V_{(\frac12,2)} & [1^2] & h_{\frac12,2} + h_{-\frac12,2} = 2.134 - 0.286 i & 1
  \\
  Y & V_{(1,2)} & [2] & h_{1,2} + h_{-1,2} = 2.500 - 0.230 i & 2
  \\
  Z & V_{(\frac32,2)} & [1^2] & h_{\frac32,2} + h_{-\frac32,2} = 3.111 - 0.136 i & 3
  \\
  \hline 
 \end{array}
 \label{tab:Q5ops}
\nonumber 
\end{align}
\label{page10}
Our dimensions agree with those of \cite{TangMaTangHeZhu2024} up to complex conjugation\footnote{For $Q>4$ and real there is  a pair of complex conjugate theories, transforming into each other under complex conjugation.}.
The $S_Q$ irreducibles $\lambda$ also agree with the $Q$-box Young diagrams $\widetilde{\lambda} = [Q-|\lambda|,\lambda]$ given
in \cite{TangMaTangHeZhu2024}, once the decompositions \eqref{x-all} of $\Xi_{(r,s)}$ are taken into account. In all but one of the 11 cases
this is manifestly simple, because the operator is of the energy or spin type (with $\lambda = [\,]$ or $[1]$ respectively),
or because the right-hand side of \eqref{x-all} contains only one term.

The remaining case of the operator called $Z'$ in \cite{TangMaTangHeZhu2024} requires a bit more discussion. In our setup, $Z'$ is contained
in the representation $\Xi_{(0,3)}$ which by \eqref{x-all} decomposes into the $S_Q$ irreducibles $\widetilde{\lambda} = [Q-3,3]$ and $[Q-3,1^3]$.
The first of those does not exist for $Q=5$ (since $[2,3]$ is not a valid Young diagram), but the other one $[2,1^3]$ does. In Table I of \cite{TangMaTangHeZhu2024},
$Z'$ is however identified with the $S_5$ irreducible $[4,1]$. Notice that $[4,1]$ and $[2,1^3]$ are conjugate tableaux, related
by interchanging rows and columns, so they have in particular the same dimension. It is possible that, in the context of the methods
employed in \cite{TangMaTangHeZhu2024} and the issues being addressed there, it is permissible to use one or the other. From our point of view the two diagrams do not correspond to the same symmetry.

In summary, we are in agreement with the operator content and (up to the minor point just mentioned) with its $S_Q$ symmetry classification
reported in \cite{TangMaTangHeZhu2024} for $Q=5$;
 our construction is more general, since it makes sense for all $Q>4$, even non-integer, as well as for generic $Q \in \mathbb{C}$.

\subsection{Three-point structure constants}
\label{Three-point structure constants}

The identification of the $Q$-state Potts CFT by analytic continuation of the results of 
\cite{JacobsenSaleur2019,HeJacobsenSaleur2020,JacobsenRibaultSaleur2023,GransSamuelssonJacobsenNivesvivatRibaultSaleur2023,NivesvivatRibaultJacobsen2024,RouxJacobsenNivesvivatRibault2025,JacobsenNivesvivatRibaultRoux2025}
can be taken beyond the operator content (Section~\ref{sec:op-content}) and the global symmetry (Section~\ref{sec:global-sym}).
Indeed, by Conjecture 2 it applies to all features of the CFT;   we now discuss   the structure constants entering into the definition of three-point correlation functions.%
\footnote{More generallly we would expect the identification to hold for any correlation function that can be expressed in terms of the loop model.
The cited references contain much progress on the four-point functions on the Riemann sphere, albeit still falling short of full and explicit analytical expressions.}

A recent piece of work \cite{JacobsenNivesvivatRibaultRoux2025} (inspired by \cite{NivesvivatRibaultJacobsen2024}) 
proposed a general formula for the structure constants in the loop model
that covers all types of operators discussed in Section~\ref{sec:op-content}. It contains as special cases the structure constants of degenerate
operators which are covered by the work of Dotsenko and Fateev \cite{DotsenkoFateev1984} or can be found by analytic continuation thereof
(see Appendix \ref{OPE-coefficients following Dotsenko-Fateev}). It also contains the result for diagonal operators, first found in \cite{IkhlefJacobsenSaleur2016} and then established rigorously
 \cite{AngCaiSunWu2024}.

As mentioned in Section~\ref{sec:op-content}, the notations used in \cite{JacobsenNivesvivatRibaultRoux2025}
apply to the $O(n)$ model, but we report below the results for the Potts model --- the object of interest in this
paper --- obtained after swapping the $r,s$ labels with respect to \cite{JacobsenNivesvivatRibaultRoux2025},
as discussed in Section~\ref{sec:op-content-Potts}.
Using the parameter $\beta$ of \eqref{n} and setting 
\be
 Q=\beta +\beta ^{-1} \,,
\ee 
the Barnes double Gamma function $\Gamma_\beta(w)$ is defined from its integral representation, convergent for   $\Re w>0$,
\bea\label{66}
\log\Gamma_\beta (w) &=& \int\limits_0^\infty\frac{\rmd t}{t}\bigg[\frac{\rme^{-wt}-\rme^{-\frac{Q}{2}t}}{(1-\rme^{-\beta t})(1-\rme^{-\beta ^{-1}t})} \nn\\
&& \qquad -\frac{\left(\frac{Q}{2}-w\right)^2}{2}\rme^{-t} -\frac{\frac{Q}{2}-w}{t}\bigg]. \qquad
\eea
It satisfies the shift relations
\bea 
\label{65a}
\Gamma_\beta(w+\beta) &=& \sqrt{2\pi}\frac{\beta^{\beta w-\frac12}}{\Gamma(\beta w)}\Gamma_\beta (w) , \\
\label{65b}
\Gamma_\beta (w+\beta ^{-1}) &=& \sqrt{2\pi}\frac{\beta ^{-\beta ^{-1}w+\frac12}}{\Gamma(\beta ^{-1}w)} \Gamma_\beta (w)\ . 
\eea
The so-called reference structure constant for three non-diagonal operators, $V_{(r_i,s_i)}$ with $i=1,2,3$, is defined as \cite{JacobsenNivesvivatRibaultRoux2025} 
\begin{align}
 \label{cref}
 \lefteqn{C_{(r_1,s_1)(r_2,s_2)(r_3,s_3)}} \\
&&=\prod_{\epsilon_1,\epsilon_2,\epsilon_3=\pm} \Gamma_\beta^{-1} \left(\tfrac{\beta+\beta^{-1}}{2} + \tfrac{\beta}{2}\left|\textstyle{\sum_i} \epsilon_i s_i\right| + \tfrac{\beta^{-1}}{2}\textstyle{\sum_i} \epsilon_i r_i\right)\ , \nn
\end{align}
where $\Gamma_\beta^{-1}=1/{\Gamma_\beta}$.
The result for diagonal and degenerate operators is also given by \Eq{cref}, upon using the indices $(r,s) = (2\beta P,0)$ with $P$ being the momentum
variable defined in \eqref{P-momentum-Potts}. For instance, the identity operator has momentum $P = P_{(1,1)}$ and can be represented by using $(r,s) = (1-\beta^2,0)$
in \eqref{cref}.
The three-point structure constant, normalized by two- and zero-point constants in such a way that the result is invariant under field rescalings $V_{(r,s)}\to \lambda_{(r,s)}V_{(r,s)}$, is finally given by
\begin{align}
  \omega_{123} = C_{123} \sqrt{\frac{C_{000}}{C_{011}C_{022}C_{033}}} \ ,
  \label{wott}
\end{align}
where subscripts $1,2,3$ are short-hands for the pair $(r_i,s_i)$ with $i=1,2,3$, whereas
$0$ refers to the identity field with $(r,s) = (1-\beta^2,0)$.
This expression can be numerically evaluated using the {\sc Julia} package  BarnesDoubleGamma.jl~\cite{Roux2024}.

We have checked that the 8 structure constants given in Table II of \cite{TangMaTangHeZhu2024} for the $Q=5$ state complex Potts CFT
agree with \eqref{wott}, up to complex conjugation as noticed above.
These structure constant involve the degenerate energy operators $\epsilon$, $\epsilon'$, $\epsilon''$ and the non-diagonal spin operators $\sigma$, $\sigma'$, $\sigma''$.
For instance we find, for the structure constant denoted $C_{\sigma \sigma'' \epsilon'}$ in \cite{TangMaTangHeZhu2024}
\begin{equation}
 \omega_{(\frac12,0)(\frac52,0)(3,1)} = 0.6710 + 0.1143i \,,
\end{equation}
with $\sigma = V_{(\frac12,0)}$, $\sigma'' = V_{(\frac52,0)}$ and $\epsilon' = V^d_{(3,1)}$ being identified from the table in Section~\ref{sec:op-content-Potts}.

In the same way we can produce results not covered by Table II of \cite{TangMaTangHeZhu2024}. For instance
\begin{equation}
 C_{\sigma \sigma Z}  = \omega_{(\frac12,0)(\frac12,0)(0,2)} = 0.05193 + 0.03344i \,.
\end{equation}
The authors of \cite{TangMaTangHeZhu2024} have kindly provided their unpublished numerical result for this structure constant \cite{TangMaTangHeZhuUnpublished},
which is $C_{\sigma \sigma Z} = 0.0581-0.0352i$ with unspecified error bars. The agreement looks rather convincing, up to the usual
complex conjugation.

\begin{figure}[t]
\Fig{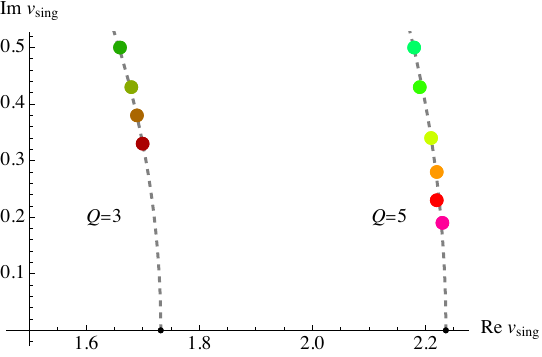}
\caption{The location of the nearest Fisher singularity in the complex $v_{\rm c}$ plane, for $Q=3$ (left) and $Q=5$ (right), evaluated on a grid of size $\delta v=0.01$ for both the real and imaginary part. The different colors indicate different system sizes: for $Q=3$ (darker) there are  $L=6$ (green, top) to $L=9$ (red, bottom); for $Q=5$ (brighter)  there are  $L=5$ (blue-green, top) to $L=9$ (magenta, bottom). The gray dashed lines are the curves $|v_{\rm c}|^2=Q$. Small grid artifacts are visible.}
\label{f:singularity-location-Q=3+5}
\end{figure}
\begin{figure}[t]
\Fig{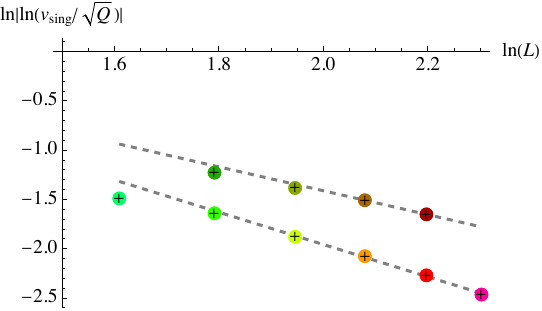}
\caption{Scaling of the distance to the critical point $v_{\rm c}=\sqrt Q$ for $Q=3$ and $Q=5$. 
The plotted  slopes are $-1.2(2)$ for $Q=3$ and $-1.6(2)$ for $Q=5$, fitting to the last  points. The crosses use ${\rm Im}(\ln v_{\rm c}/\sqrt{Q})$  instead of $ |\ln v_{\rm c}/\sqrt{Q}|$. We interpret the seemingly increasing slope for large $L$ as a sign for the first-order transition.}
\label{f:singularity-location-scaling3-Q=3+5}
\end{figure}
\begin{figure}
\Fig{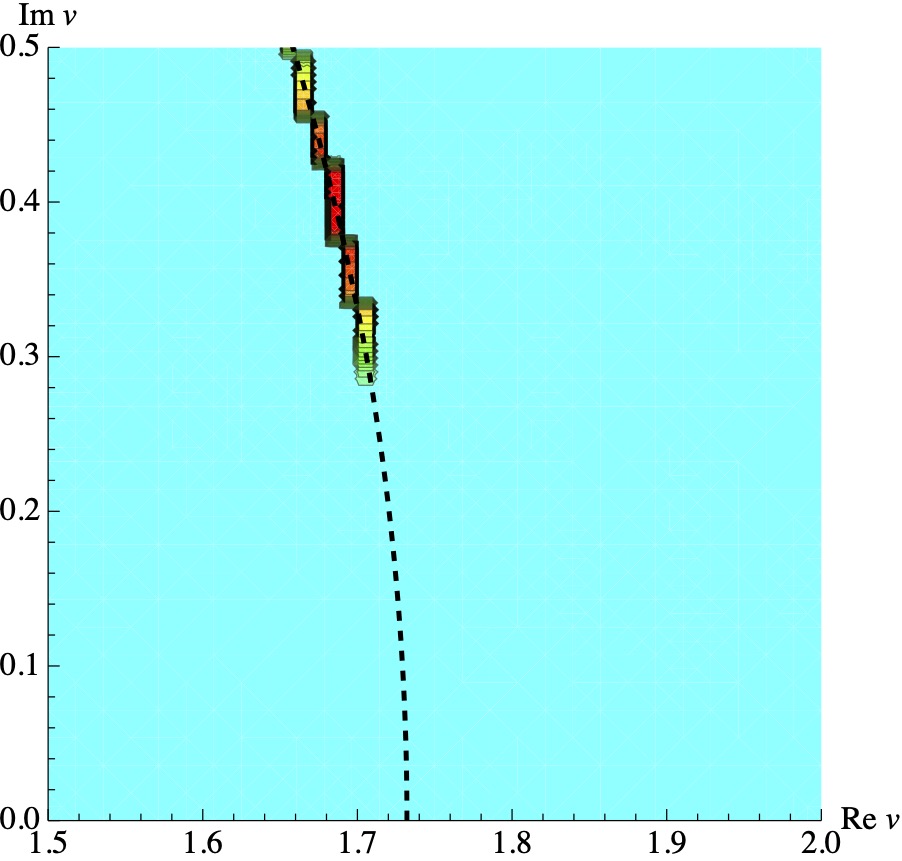}
\Fig{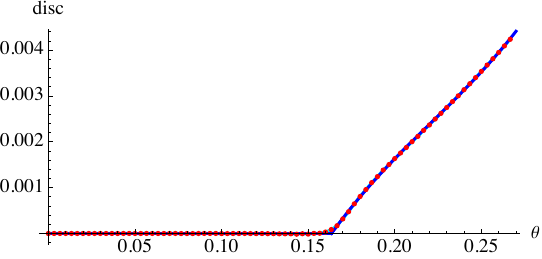}
\caption{Top: Singularity of $f$ for $Q=3$, $L=10$. Bottom: Discontinuity of $\mbox{Re}(f)/|f|$ across the cut.}
\label{singularity-obs=f-Q=3-L=10-jump}
\end{figure}

\begin{figure}
\Fig{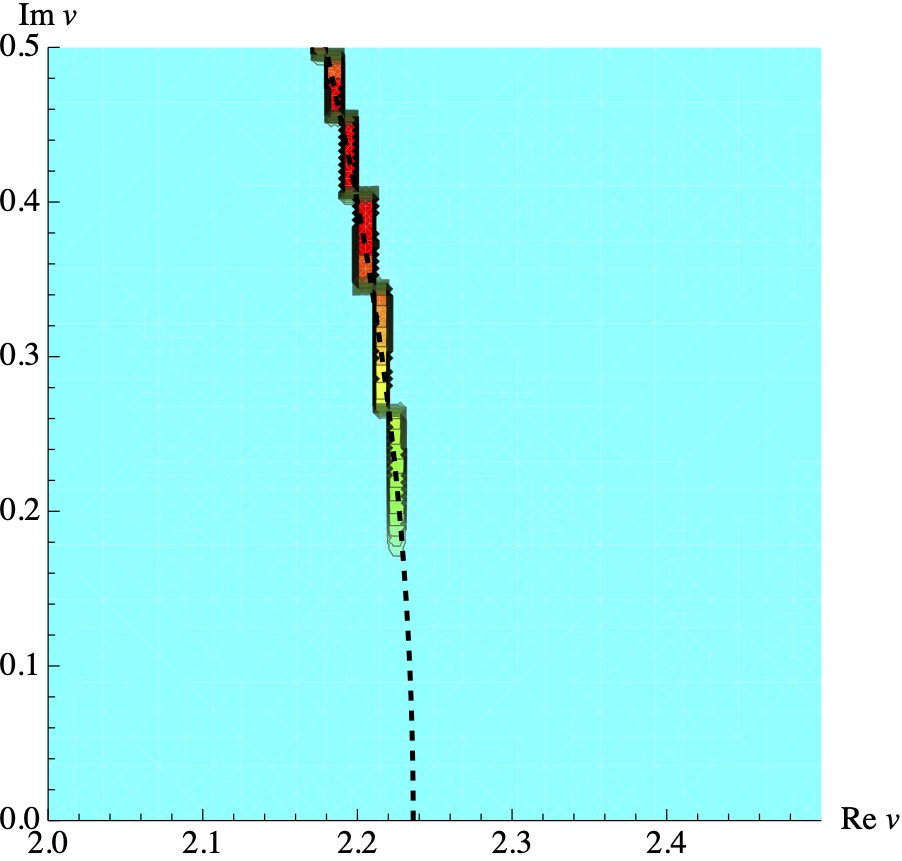}
\Fig{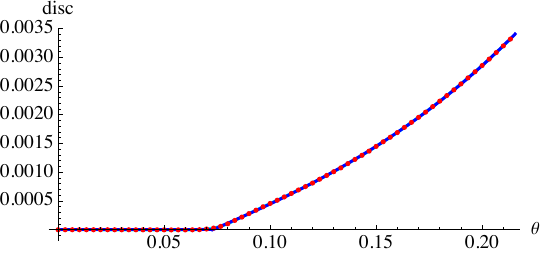}
\caption{Top: Singularity in $Z$ for  $Q=5$, $L=11$. Bottom: Discontinuity of $\mbox{Re}(f)/|f|$ across the cut. In red numerical points, in blue fit to a polynomial for $\theta > \theta_{\rm c}$.}
\label{f:singularity-Q=5-L=10bis}
\end{figure}

\section{Fisher zeroes in the model with nearest-neighbor  interactions}
\label{Fisher zeroes in the model with nearest-neighbor  interactions}

The bulk of this paper addresses the complex criticality of the model \eqref{Hlatt} defined on triangular lattice, with local
interactions \eqref{Hijk} comprising two-spin interactions $K$ between nearest neighbors and
three-spin interactions $K_3$ in up-pointing triangles. In this section we show that the usual Potts model,
with only nearest-neighbor interactions, is insufficient to access the complex fixed point.

Consider therefore the square-lattice Potts model
\begin{equation}
 \label{Hsq}
 {\cal H}_{\rm sq} = -J \sum_{\langle ij \rangle} \delta_{\sigma_i,\sigma_j} \,,
\end{equation}
where $E = \langle ij \rangle$ denotes the set of nearest neighbors. The corresponding FK cluster model
[cf.~\eqref{ZWuLin}] takes the simple form
\begin{equation}
 \label{Zsq}
 Z_{\rm sq} = \sum_{\{\sigma\}} {\rm e}^{{\cal H}_{\rm sq}} = \sum_{A \subseteq E} Q^C v^{|A|} \,,
\end{equation}
where $v = {\rm e}^J - 1$, and $C$ is the number of connected components in the subgraph having
only the subset of edges $A$.

\begin{figure*}
{\setlength{\unitlength}{\columnwidth}\begin{picture}(2.07,0.65)
\put(0,0){\Fig{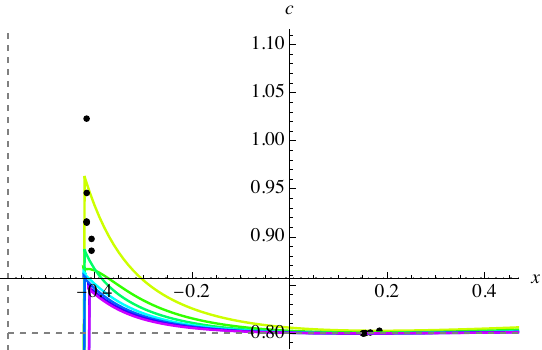}}
\put(1.07,0){\Fig{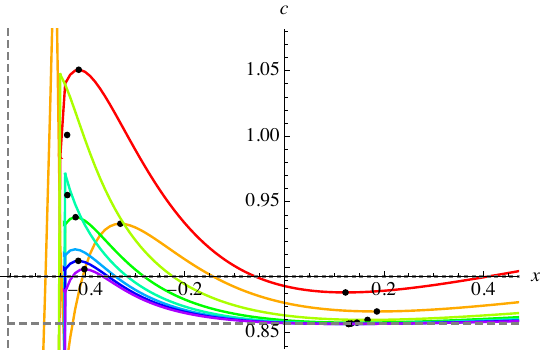}}
\put(0.66,0.62){$Q=3$, $b= 6$}
\put(1.76,0.62){$Q=3.24698$, $b=7$}
\end{picture}}
\caption{Left: $c(x)$ for $\ca B_6$, i.e.~$Q=3$, extracted from \Eq{c-eff-L} from $L=2$ and $3$ (red) over green, blue to violet  for $L=10$ and $11$.
The dots are at the maximum and minimum of $c(z)$. The dashed horizontal line represents the central charge for $B_6$, the dotted line (on the axis) the central charge for $\ca B_7\equiv \ca B_{-6}$. The left dashed vertical line marks the location of the singularity at $x=-1/\sqrt{Q}$.
Right: The same for $\ca B_7$, i.e.\ $Q=3.24698$.
On both plots, one clearly sees a minimum of the effective central charge $c$ at $x\approx 0.17$, and a maximum at $x\approx =0.4$. For $Q=6$, the latter is obscured by the singularity  at $x=-1/\sqrt Q$. It becomes visible at $Q=7$, but only for large values of $L$.}
\label{f:B6+B7}
\end{figure*}

To inquire into the analytical structure of \eqref{Zsq}, we examine its Fisher zeros in the complex $v$-plane.
This is of course a well-studied subject, and since our sole purpose is to elucidate the question stated above,
we shall be brief.

Previous studies of Fisher zeros include
\cite{ItzyksonPearsonZuber1983,ChenHuWu1996,Kim2005} for the Ising model, and \cite{ChangSalasShrock2002}
for the Potts model. While Refs.~\cite{Kim2005,ChangSalasShrock2002} consider lattices of size $L\times L$
with free boundary conditions, our study is via the transfer matrix for a cylinder of circumference $L$, i.e.\ an $L \times \infty$ lattice.
From its dominant eigenvalue $\Lambda_0$ (the largest in norm), we obtain $f = -\frac{1}{L} \ln \Lambda_0$,
the free energy per unit area.

The critical coupling $J_{\rm c}$ is given by
 \be\label{64}
\rme^{J_{\rm c}} -1=  v_{\rm c} =\sqrt{Q} \,. 
 \ee
Fig.~\ref{f:singularity-location-Q=3+5} shows that both for $Q=3$ and $Q=5$ 
there is a singularity close to $v_{\rm c}$, at
\be
v_{\rm sing} \approx v_{\rm c} \, \rme^{i a} ,\quad a \in \mathbb R,
\ee
i.e.\ on a circle of radius $v_{\rm c}$, slightly off from the real axis.
The parameter $a$ decreases  with system size $L$. Numerically we find in Fig.~\ref{f:singularity-location-scaling3-Q=3+5}
\bea\label{72}
a &\sim& L^{-1.2(2)}  \mbox{ for }Q=3, \\
\label{73}
a &\sim& L^{-1.6 (2)} \mbox{ for }   Q=5.
\eea
This is in reasonable agreement with the theoretical expectation  \cite{ItzyksonPearsonZuber1983}
\be
a\sim L^{-\frac1\nu}.
\ee 
Indeed, the CFT gives  the exponent as $\frac1\nu =  2-2h_{2,1}$. For $Q=3$ and $Q=5$ this evaluates to
\bea
\label{75}
\frac1\nu\Big|_{Q=3}  &=&  2-2h_{2,1}\Big|_{Q=3} =  \frac65 , \\
\frac1\nu\Big|_{Q=5}  &=&   2-2h_{2,1}\Big|_{Q=5} = 1.53439 \pm 0.224495 i . \label{74}
\eea
\Eq{75} is in agreement with \Eq{72} for $Q=3$. We however know from exact results \cite{Baxter1973} that
the phase transition in the square-lattice model \eqref{Hsq} is first order for $Q > 4$. Thus, for $Q=5$ the argument of the singularity $a$ should not tend
to zero as $L \to \infty$, but rather converge to a very small, but finite value $a_\infty > 0$. 
So in a log-log plot like Fig.~\ref{f:singularity-location-scaling3-Q=3+5} we should eventually see the $Q=5$ curve level
off to a finite value, if only we could access $L$ large enough.
The correlation length $\xi$ at $v_{\rm c}$ can be
estimated from its analytical expansion around $Q=4$ \cite{BuffenoirWallon1993}.
Evaluated at $Q=5$ this gives $\xi \approx 10^3$, so the sizes employed in our study satisfy $L \ll \xi$.
Therefore the first-order nature of the transition is not visible to us, and the scaling looks critical, being
dominated by the two nearby complex fixed points and a very slow running of the effective coupling
constant, a phenomenon known as walking
\cite{GorbenkoRychkovZan2018a,GorbenkoRychkovZan2018b}.
From the point of view of effective scaling, in the regime $L \ll \xi$ the apparent slope in Fig.~\ref{f:singularity-location-scaling3-Q=3+5}
should thus be $\Re (\nu^{-1}) \approx 1.53$, which is in reasonable agreement with \Eq{73}. 

The point is now that even though the model \eqref{Hsq} can detect the proximity of the complex fixed
point, its parameter space is not large enough to enter its basin of attraction.
In particular, we have checked that there is no value of $v \in \mathbb{C}$ for which the finite-size scaling of $f$ reveals the
central charge of the complex fixed point.

Figures \ref{singularity-obs=f-Q=3-L=10-jump} (for $Q=3$) and \ref{f:singularity-Q=5-L=10bis} (for $Q=5$) show that there is a branch cut  of  the free energy $f$  per site, visible in the color plot at the top;
this branch cut sits on a  circle of radius $\sqrt Q$, i.e.~at $v = \sqrt{Q} \rme^{i\theta}$ (marked by a dashed line). 
The bottom plots explore the   discontinuity across the cut as a function of the angle $\theta$. The discontinuity   grows approximately  linearly from the cut's endpoint.

\section{Numerics for $c$ and $\omega$ on the real line ($Q<4$, $b\in \mathbb R$)}
\label{Numerics for c and omega on the real line for Q<4}

\subsection{Outline}
Let us return to the model with nearest-neighbor and plaquet interactions defined in section \ref{s:Model}.
In this section, we extract the central charge $c$ and the correction-to-scaling exponent $\omega$ from transfer-matrix calculations of the Potts model on strips of width $L$. Before addressing the more challenging regime of complex couplings and $Q>4$, we first study the case of real $b$ (corresponding to $Q<4$). In this regime the critical behavior is well understood from conformal field theory. This allows us to understand the 
role of extremas of the effective central charge in locating the critical point. Once these tools are validated, we apply them to the more intricate situations discussed later.

\subsection{Extrapolation on the lattice for $L\to \infty$}
\label{Extrapolation on the lattice for L to infty}
Transfer-matrix calculations  give us the free energy density $f_L$ of a system of size $L$.
Following \cite{BloteCardyNightingale1986,Affleck1986,Cardy1986}, we use this to extract the effective central charge $c$, 
\be
f_L = f_\infty - \frac{\pi c }{6 L^2}+ ...
\ee
Solving for two consecutive sizes, we find
\be\label{c-eff-L}
c_{\rm eff}(L) \equiv  \frac{6 L^2 (L+1)^2  (f_{L+1}-f_L  )}{\pi (2L+1)}.
\ee
This is the effective central charge which we   plot as a function of $x$, or $z=x+iy $.  
Note that while $c_{\rm eff}(L)$ is indexed by $L$, we need system size $L+1$ to evaluate it.
The same holds later for the correction-to-scaling exponent $\omega$, which needs system sizes $L-1$, $L$ and $L+1$ to evaluate it.

Before attacking $Q=5$, we test our transfer-matrix calculations. 
We start our exploration for $Q<4$, which corresponds to real $b$. 
A special role is played by the minimal models for $b\in \mathbb N$, with $b=6$ for the 3-state Potts model.  The limit of $b\to \infty$ corresponds to $Q=4$.

\begin{figure*}[t]
\centerline{\MP{1}{\bf Left  fixed point: $\ca B_{-8}$ (tricritical)}
\hfill
\MP{1}{\bf Right  fixed point: $\ca B_{8}$ (critical)}}
\MP{0.5}{\Fig{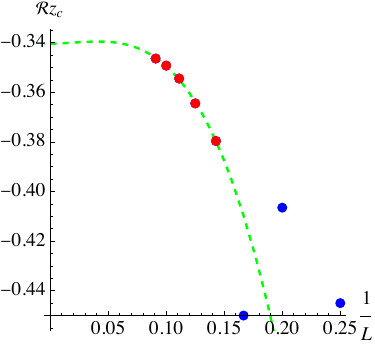}}~\MP{0.5}{\Fig{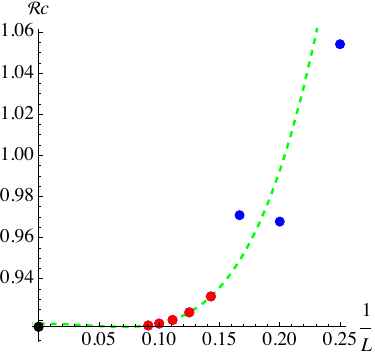}}
\hfill
\MP{0.5}{\Fig{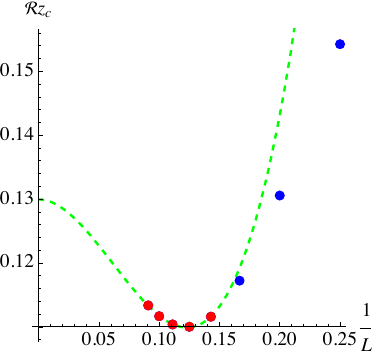}}~\MP{0.5}{\Fig{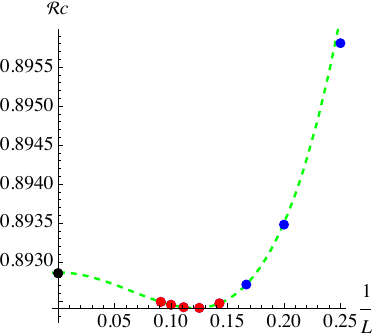}}
\MP{0.5}{\small $z_{\rm c}(L\to \infty  )= -0.34036 $}~\MP{0.5}{\small $c(L\to \infty  )= 0.918224 $}
\hfill
\MP{0.5}{\small $z_{\rm c}(L\to \infty  )= 0.129947 $}~\MP{0.5}{\small $c(L\to \infty  )= 0.892867 $}
\MP{0.5}{\Fig{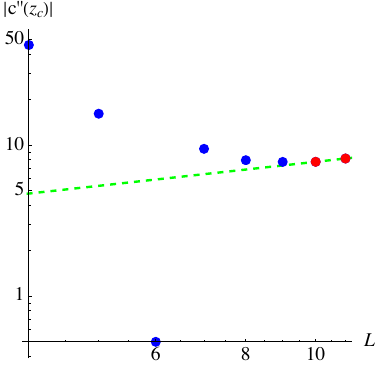}}~\MP{0.5}{\Fig{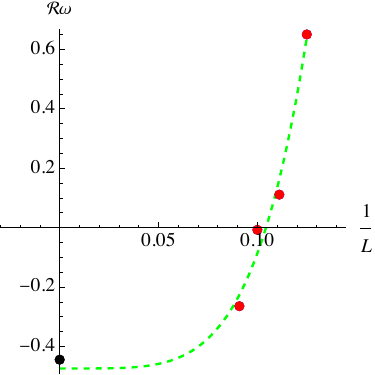}}
\hfill
\MP{0.5}{\Fig{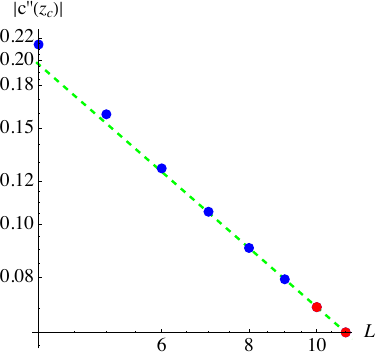}}~\MP{0.5}{\Fig{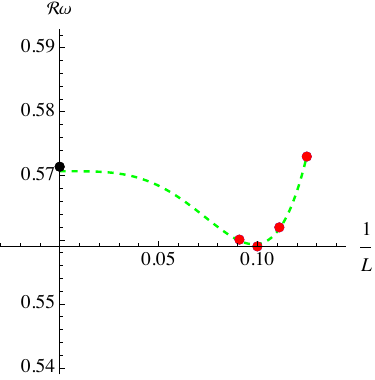}}
\MP{0.5}{\small slope:   $ \omega=-0.26441$ }~\MP{0.5}{\small   $\omega(L\to \infty)=-0.474152$}
\hfill
\MP{0.5}{\smallskip  slope:   $  \omega=0.563151$}~\MP{0.5}{\small  $\omega(L\to \infty)=0.57143$}
\caption{Left half of plot: Scaling analysis of $c(z)$ (see Fig.~\ref {f:B8}) for the model $\ca B_8$ ($Q=3.41421$), for  the left fixed point: The first plot shows the convergence of $x\equiv \Re z$ towards its critical value. (All imaginary parts vanish.)  The next plot shows the convergence of $c= \Re c$, followed by an estimate of $\omega$ from $c''(z)$ from the two largest system sizes, and finally an extrapolation to $L=\infty$ for $\omega$.  Right half of plot: {\em idem}  for the right fixed point.}
\label{B8-analysis}
\end{figure*}

\begin{figure}[b]
{\setlength{\unitlength}{1mm}\begin{picture}(86,56)
\put(0,0){\Fig{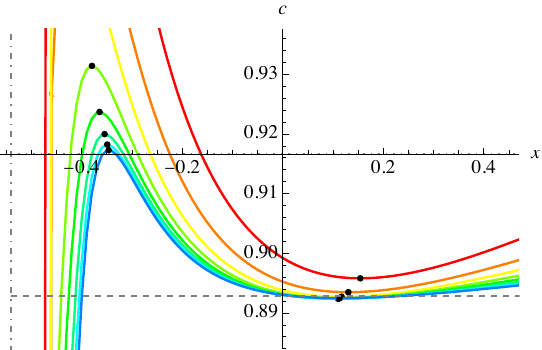}}
\put(48,1){\gray critical point}
\put(54.2,3.6){\gray\vector(0,1){4}} 
\put(15,1){\gray tricritical point}
\put(17.5,3.6){\gray\vector(0,1){26}} 
\put(58,53){$Q=3.41421$, $b= 8$}
\end{picture}}
\caption{$c_L(x)$ for $b=8$, $Q \simeq 3.41421$, using sizes $L=4$ (red) to $12$ (blue). The discontinuity near $x=-1/\sqrt{Q}$
(vertical dot-dashed line) is related to level crossings for $K_2 \to -\infty$.
These prevent us from using this approach for smaller $Q$.
CFT predictions are shown for $B_8$ (dashed line) and $B_{-8}\equiv B_9$ (axis, and gray dots). 
}
\label{f:B8}
\end{figure}
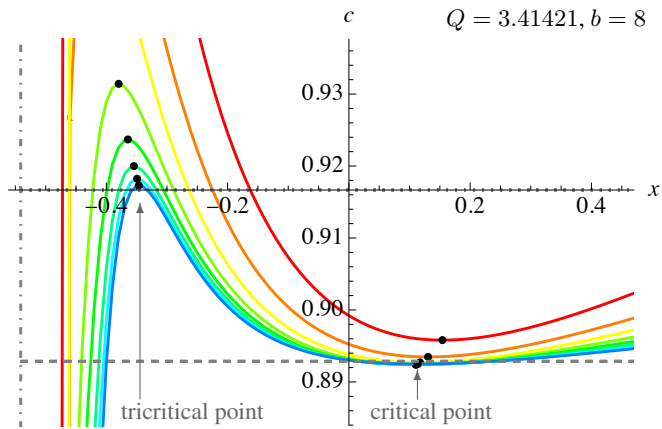

\subsection{$\ca B_6$, $\ca  B_7$ and $\ca B_8$}
\label{sec:flowB8}

We start our exploration with the models $\ca B_6$ and $\ca B_7$, see figure \ref{f:B6+B7}. 
On both plots, one clearly sees a minimum of the effective central charge $c$ at $x\approx 0.17$, and a maximum at $x\approx -0.4$.
For $b=6$, the latter is obscured by the singularity  at $x=-1/\sqrt Q$.%
\footnote{The limit $z \to -1/\sqrt{Q}$, or $c_2 \to -1$ from \Eq{zw}, corresponds to $K \to -\infty$ by
\Eq{eKL}, hence an infinitely strong antiferromagnetic two-spin interaction. Since $c_3 = Q$ at criticality
by \Eq{w=1}, we must have $K_3 \to \infty$ in \Eq{K3K} in such a way that keeps ${\rm e}^{K_3 + 3K} = Q-2$
fixed, implying an infinitely strong ferromagnetic three-spin interaction when $Q > 2$.}
It becomes visible at $b=7$, but only for large values of $L$. 

Things get clearer when we proceed to $\ca B_8$, see Fig.~\ref{f:B8}.
We see a clear minimum of the effective central charge $c_L(x)$ for $x\approx 0.12$, and a maximum at $x\approx -0.34$. 
Fig.~\ref{f:B8}    shows that upon increasing $L$, the minimum of  $c_L(z)$ becomes shallower, and the maximum   more pointed.
We associate this with a critical point for  the minimum: increasing $L$,  the measured effective central charge $c_L(z)$ converges to a limiting value, independent of the initial value of $z$, as long as the latter is close enough to the minimum. In contrast, the value at the maximum becomes more and more sensitive to the value of $x$, and we associate the deviation from the critical $x$ as a relevant perturbation, characteristic of a tricritical point.

In other words, thanks to the self-duality condition $w=1$ [see \eqref{w=1}] we are already on a critical manifold for any $x$.
Upon ajusting $x$ to a particular value we can reach a tricritical, repulsive fixed point $P_{\rm t}$ at $x_{\rm t}$.
The part $x > x_{\rm t}$ of the $w=1$ manifold is then governed by the critical, attractive fixed point $P_{\rm c}$ at $x_{\rm c}$.
There is an RG flow from $P_{\rm t}$ to $P_{\rm c}$ along which the central charge decreases, in agreement
with Zamolodchikov's $c$-theorem \cite{Zamolodchikov1986}.

A quantitative analysis is shown in Fig.~\ref{B8-analysis}. One finds for the left fixed point
\bea
z_{\rm c}(L\to \infty  ) &=&  -0.34036, \nn\\
   c(L\to \infty  ) &=& 0.918224, \nn\\
    c(\ca B_{-8}) &=& 0.916667.
\eea
The same analysis for the right fixed point yields
\bea
z_{\rm c} (L\to \infty  ) &=& 0.129947, \nn \\
  c(L\to \infty  ) &=& 0.892867, \nn \\
  c(\ca B_{8}) &=& 0.892857.
\eea
We see that for both fixed points, the central charge can be extracted, and agrees rather precisely with the prediction \eq{c-of-b} from the minimal model. 

Let us next return to our discussion of critical versus tricritical behavior. 
 This can be made quantitative through a finite-size analysis of the curvature $c_L''(z)$. It allows us to extract the correction-to-scaling exponent $\omega$ via 
\be\label{59}
c_L''(z_{\rm c}) \sim L^{-2 \omega}. 
\ee%
\begin{figure*}[t]
{\setlength{\unitlength}{\columnwidth}\begin{picture}(2.07,0.65)
\put(0,0){\Fig{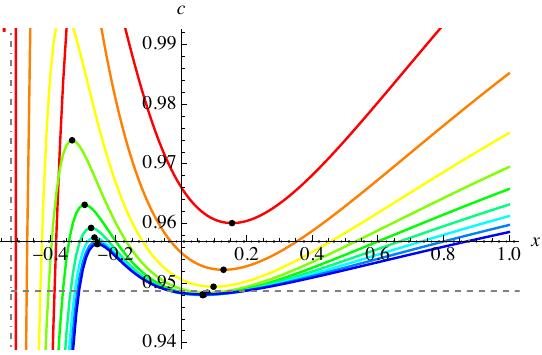}}
\put(1.07,0){\Fig{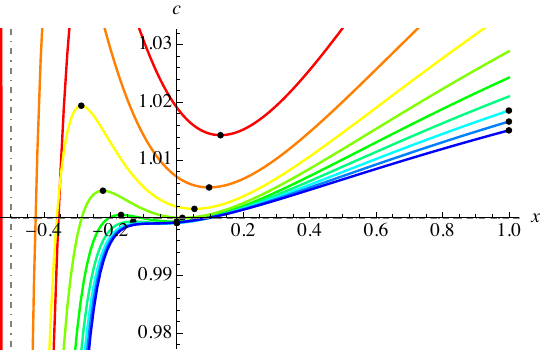}}
\put(0.4,0.62){$Q=3.7$, $b= 11.3249$}
\put(1.5,0.62){$Q=4$, $b\to \infty$}
\end{picture}}
\caption{Left: $c(x)$ for $b= 11.3249$, i.e.~$Q=3.7$, extracted from \Eq{c-eff-L} from $L=3$ over green, blue to violet  for $L=11$ and $12$; see Fig.~\ref{f:B6+B7} for details on what is plotted. 
Right: {\em idem} for $Q=4$ ($b\to \infty$). We see that  fro $Q\to 4$, the two fixed points approach, and merge at $Q=4$, at $z_{\rm c}\approx -0.02\pm 0.02$. }
\label{f:Q=3.7to4}
\end{figure*}%
{\em Derivation}: Since the critical points are stationary points (maxima or minima in $z$), 
\be\label{60}
c_L(z) = c(z_{\rm c}) + \half \partial_z^2 c(z) (z-z_{\rm c})^2 + ... 
\ee
Here, $z-z_{\rm c}$ is the deviation from the critical point in terms of the bare coupling $z$. In terms of the renormalized coupling $z^{\rm R}$, this reads 
\be\label{61}
(z^{\rm R}-z_{\rm c})L^{\omega} = z-z_{\rm c}
\ee
Inserting \Eq{61} into \Eq{60} yields \Eq{59}.

A more precise way to extract $\omega$ is to write 
\be
\frac{c''(L)}{c''(L+1)} =\frac{L^{-2\omega}}{(L+1)^{-2 \omega}}.
\ee
Taking the logarithm yields
\be
\ln \frac{c''(L)}{c''(L+1)} =  2{\omega} \ln \Big(1+\frac1L \Big) .
\ee
We define
\be\label{omega-L}
 \omega_L := \frac12 \frac{\ln \big( \frac{c''(L)}{c''(L+1)} \big) }{ \ln \big(1+\frac1L \big).}
\ee
This is plotted on the lower left plot in Fig.~\ref{B8-analysis}. 
Finally, we can   extrapolate these values to $L=\infty$, as shown in the following plot of  Fig.~\ref{B8-analysis}.

The exponent $\omega$  is related to the dimension $\Delta$ of the perturbing operators as
$\Delta= 2 + \omega $. (The 2 comes from the spatial integral over the perturbing operator.)
Using what we found in Fig.~\ref{B8-analysis}, we get (with c = critical and t = tricritical)
\bea\label4
\Delta^{\rm c} = 2+\omega^{\rm c} \approx 2.57 \,, &\qquad& \Delta^{\rm t} = 2+\omega^{\rm t} \approx  1.53 \,, \\
 2h_{3,1}^{b=8} = \frac{18}{7}= 2.571 \,, &\quad & 2h_{3,1}^{b=-8} = \frac{14}{9} =1.556 \,.
\label5
\eea
We find good agreement of measurements \eq{4} with the predictions of CFT \eq5.

At the critical point $P_{\rm c}$, according to section~\ref{sec:op-content-Potts}   the leading
energy operator is $\epsilon = V^d_{(2,1)}$. It governs the perturbation away from the critical
manifold $w=1$ and the latter is obtained, in the continuum limit, by tuning the amplitude of 
$\epsilon$ to zero. The sub-leading energy operator $\epsilon' = V^d_{(3,1)}$ is irrelevant
and governs the corrections to scaling upon flowing into $P_{\rm c}$ along the critical manifold.
At the tricritical point $P_{\rm t}$ the leading perturbation is again by $V^d_{(3,1)}$ and it governs
the flow towards $P_{\rm c}$.

This result was subsumed in section~\ref{Phase diagram as a function of Q} and used there to
determine the phase diagram as a function of $Q \in \mathbb{C}$.
It appears to be in tension with a classical paper by Zamolodchikov \cite{Zamolodchikov1987},
in which he establishes a massless flow from the minimal model ${\cal M}_p$ to ${\cal M}_{p-1}$
induced by the perturbation by $\phi_{1,3}$ (not $\phi_{3,1}$). However, there needs not be any
contradiction. The models ${\cal B}_b$ considered here are loop models, not minimal models,
even though we have often taken $b$ integer for practical purposes (e.g.\ having rational exponents).
In particular, the loop models have the spectrum given in section~\ref{sec:op-content-Potts} which
does simply not contain the field $V^d_{(1,3)}$. Indeed, if this field was present in our CFT all along
the RG trajectory, it would lead to a contradiction with the flow from $P_{\rm t}$ to $P_{\rm c}$ observed in Fig.~\ref{f:B8}, 
since according to \Eq{2h13=32} the field $V^d_{(1,3)}$ is relevant. So   though they look similar, our flow from
${\cal B}_{b+1}$ to ${\cal B}_b$ is not the same as the one studied by Zamolodchikov \cite{Zamolodchikov1987}.

To summarize this section, we are   able to identify a critical or tricritical point by   $c'(z)=0$, and   use a finite-size analysis on $c_L''(z_{\rm c})$ to decide whether it is attractive (critical) or repulsive (tricritical) once $T_c$ is fixed.\footnote{Such a prescription is missing in related work on the ${\rm O}(n)$ model, where the location of the critical point is known analytically \cite{HaldarTavakolMaScaffidi2023}.}

\subsection{The limit of $Q\to 4$}
In Fig.~\ref{f:Q=3.7to4} we study what happens in the limit of $Q\to 4$. One sees that maximum and minimum values get closer together, and become equal at $Q=4$ ($b\to \infty$). This phenomenology is accounted for by the following toy model inspired by conformal perturbation theory \cite{GorbenkoRychkovZan2018a,GorbenkoRychkovZan2018b}
\be\label{model-for-c(z)}
c(z) =1+  3  {z^3}  +\frac{3(Q-4)}{(2\pi)^2} (2+3 z).  
\ee
The coefficients are chosen s.t.\ they are analytic in $z$ and $Q$; especially there is  no singularity  at $Q=4$. 
The maximum and minimum   obtained from  $c'(z)=0$ are at 
\bea
z_{\pm} &=&\pm  \frac{\sqrt{4-Q} }{2\pi} \\
c(z_{\rm \pm}) &=& 1-\frac{3 (4-Q)}{2 \pi ^2}\pm \frac{3 (4-Q)^{3/2}}{4 \pi ^3}
\label{c-pert}
\eea
The chosen numerical values in the model \eq{model-for-c(z)} ensure that $c(z_+) = c(b(Q)) + \ca O(4-Q)^2$, and $c(z_-) = c(-b(Q)) + \ca O(4-Q)^2$.
These values are unchanged under smooth deformations, as e.g.\ $z\to a(Q) z-z_{\rm c}(Q)$, thus the  model \eq{model-for-c(z)} is not unique.
A very similar approach was used in \cite{GorbenkoRychkovZan2018a,GorbenkoRychkovZan2018b}, which gives 
 \Eq{c-pert}, evaluating to 
\be
c(Q=5) \approx 1.152 \pm 0.0242 i .
\ee
This should be compared to \Eq{c-Q=5}.


\section{$Q=5$}
\label{s:Q=5}
\subsection{Central charge in the complex $z$-plane}
\label{Central charge in the complex z-plane}

We now proceed to $Q=5$, for which we   run the transfer-matrix calculations on a discretization of the complex  $z=x+i y$ plane.
We chose the range $-0.5\le x \le 0.5$, and $0\le y \le 1$, with a step size of $ \delta z=0.01$. For future reference we call this numerical scheme ``grid''.

\begin{figure}[t]
\centerline{\fig{0.5}{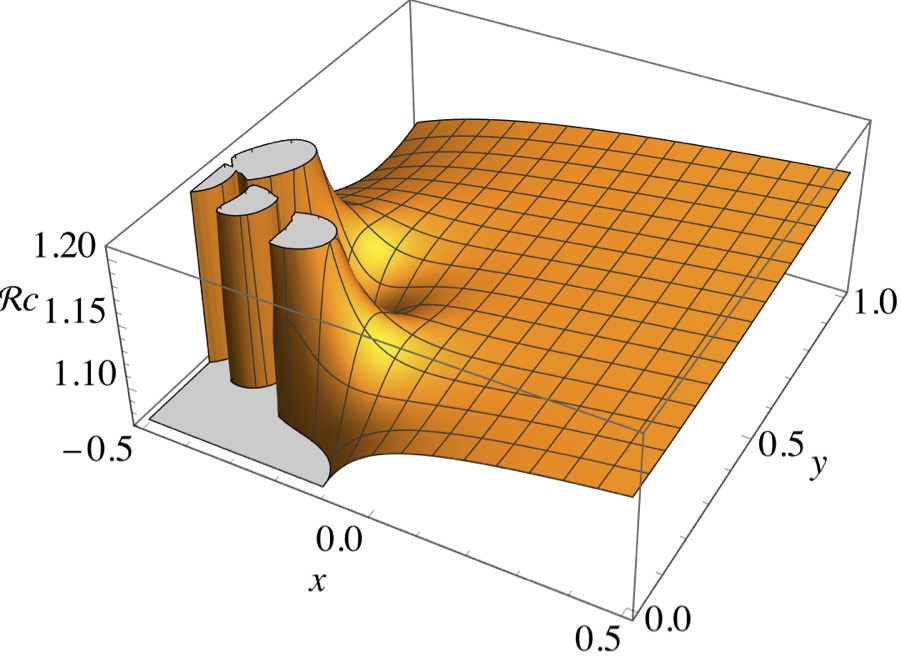}~\fig{0.5}{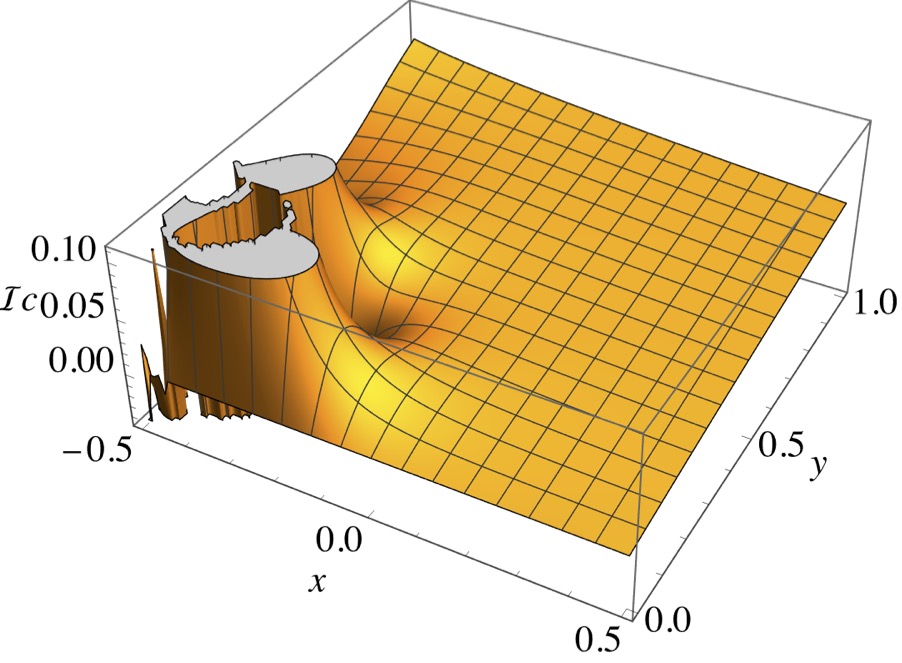}}
\caption{$Q=5$, $L=10$. Data for $\Re c$ (left) and $\Im c$ right, as a function of $z=x+iy$.}
\label{c-Q=5-top}
\centerline{\fig{0.5}{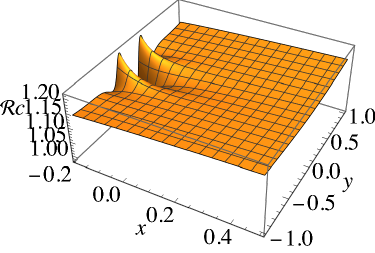}~\fig{0.5}{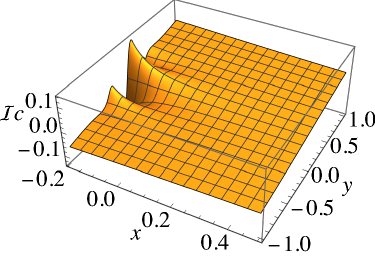}}
\caption{Data from Fig.~\ref{c-Q=5-top}, after restriction to a smaller range in $z$, and doubling the data using that $ c(\bar z) = \overline{c(z)}$. (Note the different scales between Figs.~\ref{c-Q=5-top} and \ref{c-Q=5-middle}.)
}
\label{c-Q=5-middle}
\end{figure}%
\begin{figure}
\centerline{\fig{0.5}{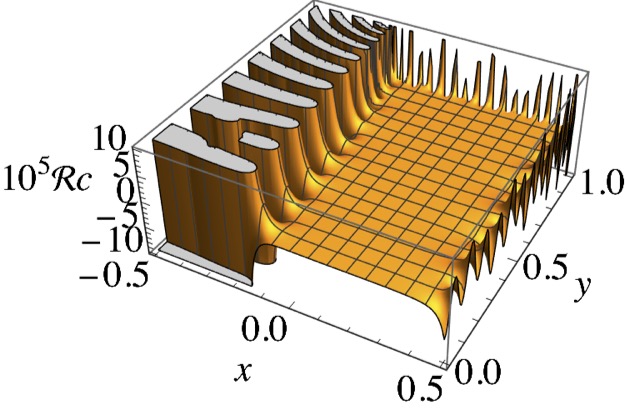}~\fig{0.5}{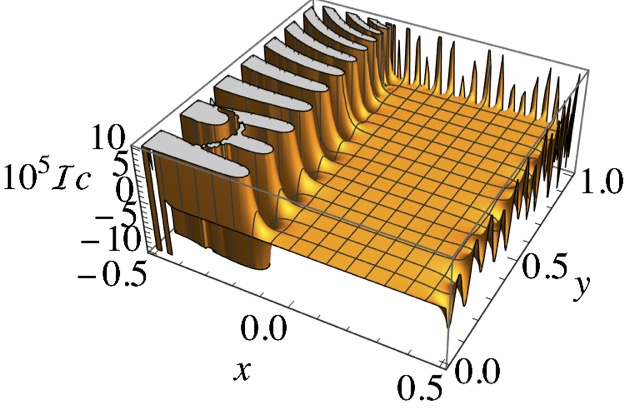}}
\caption{Error made in the interpolation for the real and imaginary parts of $c$ in Fig.~\ref{c-Q=5-middle}, $L=10$. Note multiplication by $10^5$.}
\label{c-Q=5-bottom}
\end{figure}
\begin{figure}[t]
\Fig{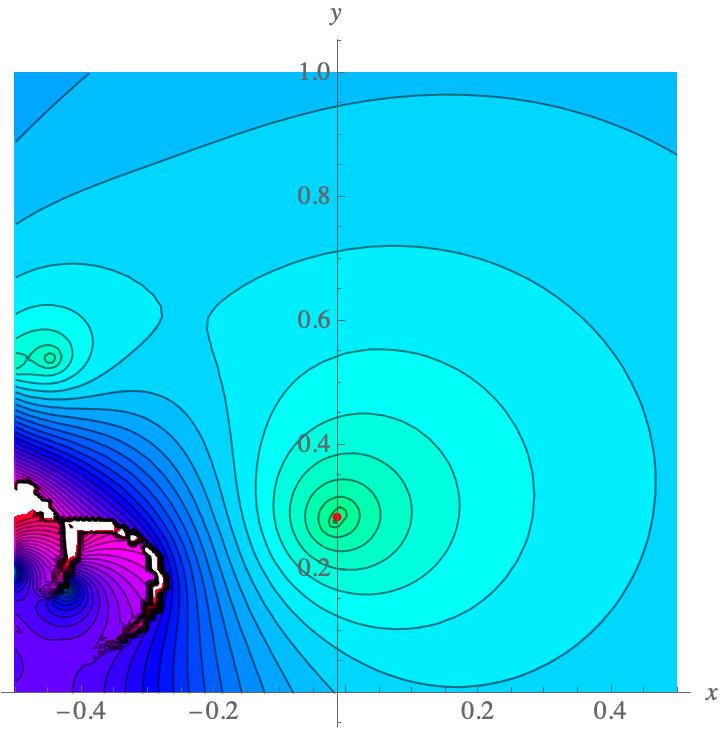}
\caption{$|c_L(z)-c_L(z_{\rm c})|^{0.25}$  for $Q=5$, $L=10$. To the lower left we see singularities at the level-crossing lines.}
\label{|c_L(z)-c_L(zc)|^0.25}
\end{figure}
\begin{figure*}
 \centerline{\fig{0.6}{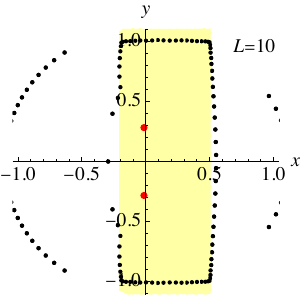}\hfill\fig{0.6}{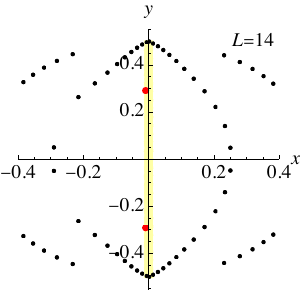}\hfill\fig{0.6}{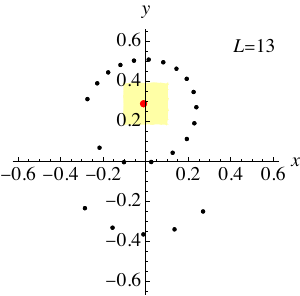}}
\caption{Zeros of $c'(z)$  using measurements on a rectangular grid of points indicated by the yellow shaded region. Left: data in the plane with $\delta z=0.01$, for $L=10$, with $10^4$ data points, plus their complex conjugate images. Middle: $L=14$ on the imaginary line, $\delta z =0.01$, 50 data points, plus their complex conjugate images. Right $L=13$ on the mini-grid, $\delta z=0.05$, 25 data points. We always display the largest available system size.}
\label{f:zerosQ=5}
\end{figure*}

\begin{figure*}[t]
\newcommand{\locsize}{0.33}
\centerline{\fig{\locsize}{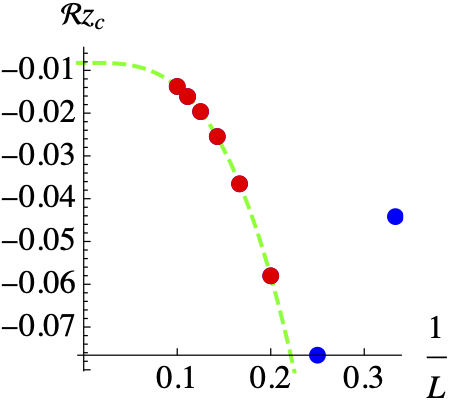}\fig{\locsize}{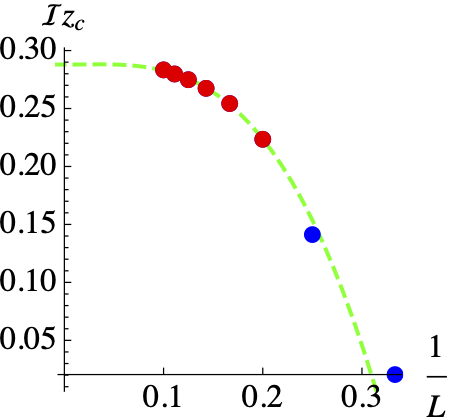}~\hfill~\fig{\locsize}{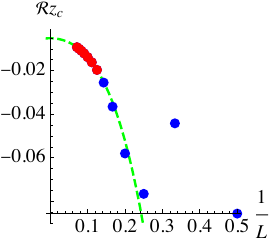}\fig{\locsize}{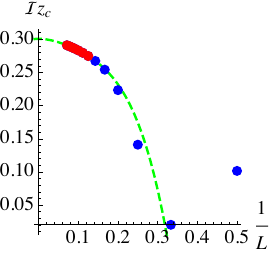}~\hfill~\fig{\locsize}{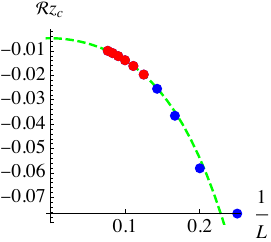}\fig{\locsize}{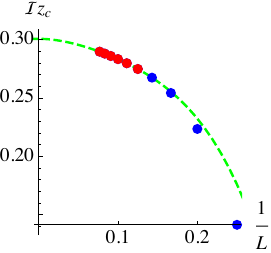}}
\centerline{\parbox{\locsize\textwidth}{\centerline{$z_{\rm c}^{L\to \infty} = -0.00826+0.2879 i$}}\hfill
\parbox{\locsize\textwidth}{\centerline{$z_{\rm c}^{L\to \infty} = -0.00510+0.30077 i$}}\hfill
\parbox{\locsize\textwidth}{\centerline{$z_{\rm c}^{L\to \infty }=-0.00478+0.30073i$}}}
\centerline{\fig{\locsize}{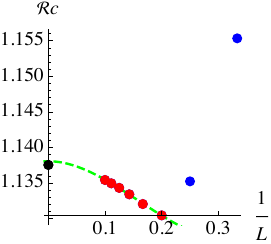}\fig{\locsize}{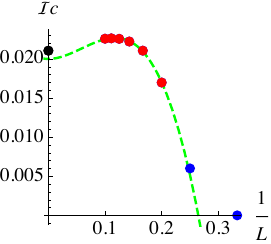}~\hfill~\fig{\locsize}{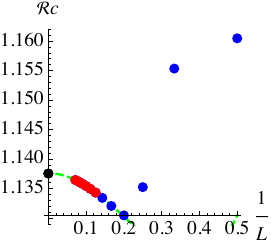}\fig{\locsize}{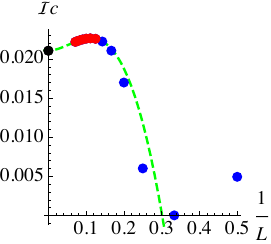}~\hfill~\fig{\locsize}{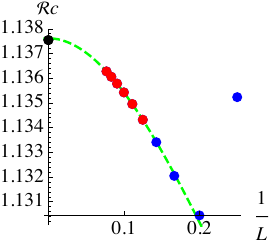}\fig{\locsize}{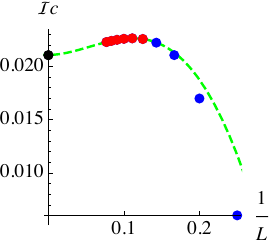}}
\centerline{\parbox{\locsize\textwidth}{\centerline{$c^{L\to \infty} = 1.13806 +0.020021 i$}}\hfill
\parbox{\locsize\textwidth}{\centerline{$c^{L\to \infty} = 1.13761 +0.0210886 i$}}\hfill
\parbox{\locsize\textwidth}{\centerline{$c^{L\to \infty }=1.13762 +0.0210937 i$}}}
\centerline{\fig{\locsize}{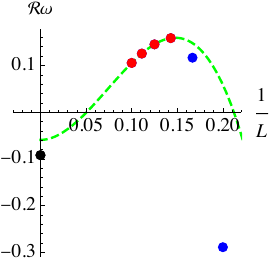}\fig{\locsize}{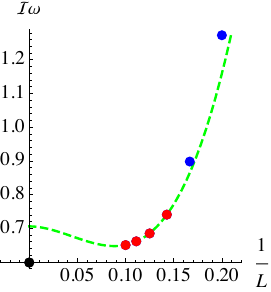}~\hfill~\fig{\locsize}{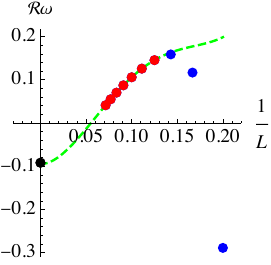}\fig{\locsize}{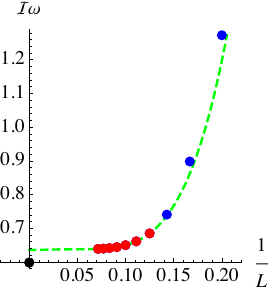}~\hfill~\fig{\locsize}{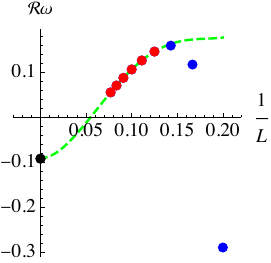}\fig{\locsize}{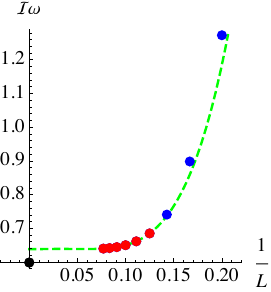}}
\centerline{\parbox{\locsize\textwidth}{\centerline{$\omega^{L\to \infty} = -0.0590247+0.705823  i$}}\hfill
\parbox{\locsize\textwidth}{\centerline{$\omega^{L\to \infty} = -0.0947766+0.635994 i$}}\hfill
\parbox{\locsize\textwidth}{\centerline{$\omega^{L\to \infty }=-0.0906943+0.638335 i$}}}
\centerline{\parbox{\locsize\textwidth}{\centerline{$\Delta=2+\omega^{L\to \infty} = 1.94097 +0.70582 i$}}\hfill
\parbox{\locsize\textwidth}{\centerline{$\Delta=2+\omega^{L\to \infty} = 1.90522 +0.635994  i$}}\hfill
\parbox{\locsize\textwidth}{\centerline{$\Delta=2+\omega^{L\to \infty }=1.90931 +0.638335 i$}}}
\caption{Values of $z_{\rm c}$ (top), $c$ (middle) and $\omega$ (bottom) for all available system sizes $L$. The left two columns are from the  grid data, the middle two from line data, and the right two from the minigrid; see Fig.~\ref{f:zerosQ=5} for the range covered by these data. The black dots represent the analytically predicted values for $c$ and $\omega$. The fitting polynomial is of the form $a_0 + a_2 L^{-2} + a_3L^{-3} $ for the grid, and $a_0 + a_2 L^{-2} + a_3L^{-3} + a_4 L^{-4}$ for the line and minigrid. Only red points are used for the fit, blue points are dropped.}
\label{f:extrapolationsQ=5}
\end{figure*}

\begin{figure*}
\begin{tabular}{|c|c|c|c|c|c|c|c|}
\hline  
$Q$, $b$ & $L_{\rm max}$ & $z_{\rm c}$& $c_{\rm ana}$ & $c_{L\to \infty}$ & $\omega_{L\to \infty}$ &  $2+\omega_{L\to \infty}$ & $2 h_{3,1}$ \\
\hline
\hline
$b=6$ & 10 & $0.1711 $ & 0.8 & $0.80001$ &  $0.774$ &    $2.774$ & 2.8   \\
\hline 
$b=7$  & 10 & 0.1426 & 0.85714 & 0.85715 & 0.604 &2.604 & 2.667 \\
\hline 
$b=-7$ & 10  & $-0.38$ & 0.89286 & $0.88812$ & -- & -- & 1.5\\
 \hline 
$b=8$  & 12 & $0.1299$ & $0.89286$ & $0.89287$ & $0.570$ & $2.570$ & 2.571 \\
\hline 
$b=-8$ & 12 & $-0.3404~$ &$0.91667$  &$0.91822 $  & $-0.264$ & $1.735$ & $1.556$\\
 \hline 
5 grid & 11 & $-0.0083{+}0.2879 i$ & ~~~  & $1.13806 {+}0.02002 i$ & $-0.059{+}0.706  i$ & $1.941 {+}0.706 i$ & \\
\cline{1-3} \cline{5-7} 
5 line & 15 & $ -0.0051{+}0.3008 i$ & $1.13755  {+} 0.02107 i$&  $1.13761 {+} 0.02109 i$  & $ -0.095{+}0.640 i$ & $1.905{+}0.636  i$& $1.908{+} 0.599 i$ \\
\cline{1-3} \cline{5-7} 
5 minigrid &14 & $-0.0049{+}0.3007i$ & ~~~ & $1.13762 {+}0.02109 i$ & $-0.091{+}0.638 i$ & $1.909{+}0.638 i$  &  ~~~ \\
\hline 
10 & 11 & $  0.1369 {+}0.6262 i$ & $ 1.58411 {+}0.19183 i$  & $ 1.58465 {+}0.19083 i $ & $ -0.481{+}1.999  i$ &$1.519 {+}1.999 i $ & $1.611 {+}1.186 i$\\
\hline 
20 & 9 & $0.2807{+}0.8234 i$ & $2.04608{+}0.48070 i $ & $2.04599{+}0.47921 i$ & $-0.482{+}2.930i$ & $1.518{+}2.930$  &  $1.303{+}1.518 i$\\
\hline 
40 & 8  & $0.3686 {+}0.9135i$ & $2.50576 {+}0.87158 i$ & $2.5122 {+}0.89312i$ & $-2.138{+}3.079 i$ & $-0.138{+}3.079i$& $0.996{+}1.734 i$\\
\hline 
$5{+}2i$ upper grid & 11 &$-0.1885{+}0.4978 i$  & $1.11224 {+}0.25058 i $  &$1.11029 {+}0.25394 i $ & $-0.470 {+}0.525$  & $1.530 {+}0.525 i$ & {$1.502 {+}0.635 i $}\\
\cline{1-3} \cline{5-7} 
$5{+}2i$ upper minigrid & 16 & $-0.1845{+}0.4831i$ & ~~~ & $1.11201 {+} 0.25059i$ & $-0.511 {+}0.617 i$ & $1.489 {+} 0.617 i$  & \\
\hline 
$5{+}2i$ lower grid &11 & $0.2298 {-}0.2968 i$ & $1.24645 {+}0.22938 i$ & $1.24678 {+}0.2295i$ & $0.648 {-}0.994 i$ & $2.648{-}0.994 i$ & 
$2.259 {-}0.955 i$\\
\cline{1-3} \cline{5-7} 
$5{+}2i$ lower minigrid & 16 & $0.2252 {-} 0.3330  i$ & ~~~ & $1.24647 {+}0.22935 i$ &  $0.603 {-}0.955  i$ & $2.603 {-}0.955 i$ & \\
\hline 
$8{+}i$ upper grid & 11 & $0.0546  {+}0.5700 i$ & $1.40726 {+}0.20408 i $ & $1.40773 {+}0.20324i$ & $0.534  {+}1.007 i$ & $2.534  {+}1.007 i$ &  $1.617  {+}0.996 i $ \\
\cline{1-3} \cline{5-7} 
$8{+}i$ upper minigrid & 16 & $0.0489 {+}0.5516 i$ & ~~~ & $1.40746 {+}0.20395 i$ & $-0.640 {+} 0.945 i$ &  $1.360{+}0.945 i$ & ~~~ \\
\hline
$8{+}i$ lower grid &11 & $0.1530 {-}0.5198 i $ & $1.47775 {-}0.04199  i $ & $1.47782 {-}0.04162i$ & $-0.321  {-}1.639 i$ & $ 1.679  {-}1.639 i$  &$1.793 {-}1.105 i$ \\
\cline{1-3} \cline{5-7} 
$8{+}i$ lower minigrid & 16 & $0.1273  {-}0.5027 i $ & ~~~& $1.47777 {-}0.04178  i$ & $-0.334 {-} 0.496 i$ & $1.666{-}0.496 i$  & ~~~\\
\hline
\end{tabular}
\caption{All our data. Data for the tricritical fixed point ${\cal B}_b$ with $b=6$ do not exist, and   are not shown.}
\label{f:All our data}
\end{figure*}
\begin{figure}[t]
\centerline{\fig{0.5}{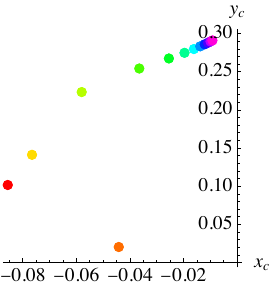}\hfill\fig{0.5}{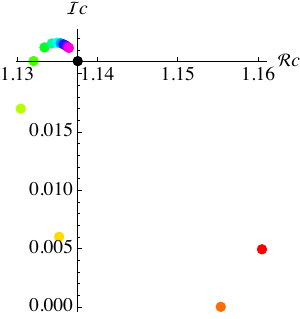}}
\caption{$z_{\rm c}$ (left) and $c$ (right), as a function of $L$, line data for $Q=5$,  sizes $L=4$ to $L=14$. A rather regular behavior starts at $L=7$.}
\label{zc and c for Q=5}
\centerline{\fig{0.5}{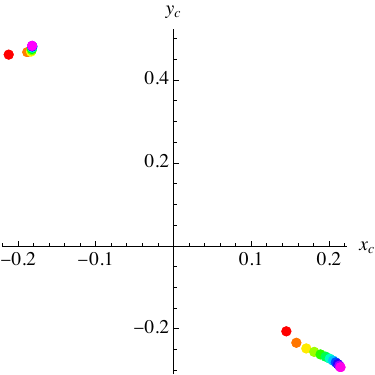}\hfill\fig{0.5}{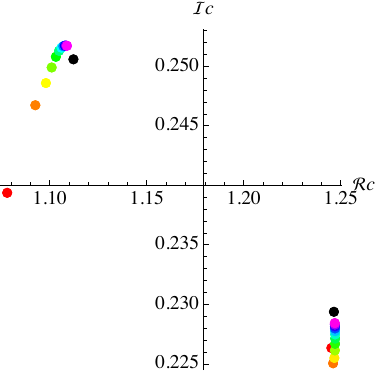}}
\caption{$z_{\rm c}$ (left) and $c$ (right), as a function of $L$ (minigrid data) for  $Q=5+2i$,  sizes $L=4$ to $L=14$.}
\label{Q=5+2i-minigrid-zeros-of-c-prime}
\end{figure}

\begin{figure}[t]
\Fig{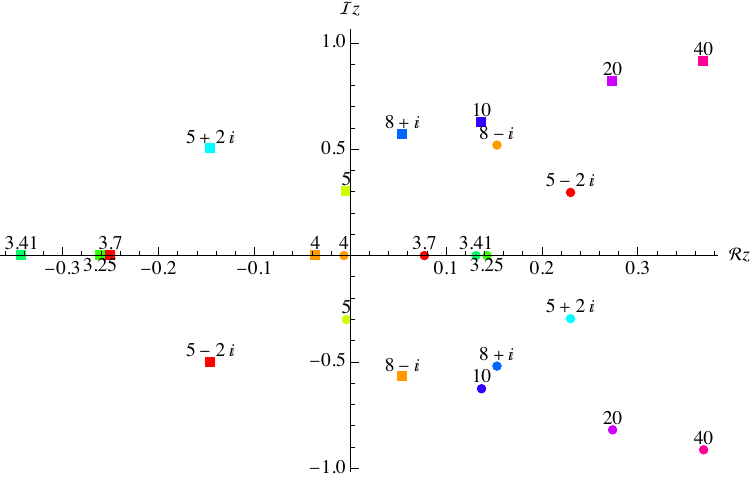}
\caption{$z_{\rm c}$ for all cases considered. Disks are the continuation of the critical FPs, squares the continuation of the tricritical ones. Labels are the values of $Q$.}
\label{location-of-all-FP}
\end{figure}

In Fig.~\ref{c-Q=5-top}, we show 
the effective $L$-dependent central charge as defined by \Eq{c-eff-L}, as a function of $z$. 
Since $Q$ is real, 
\be\label{c bar(z)=c(bar z)}
c_L(\bar z) = \overline {c_L(z)},
\ee
thus we know $c_L(z)$ for $-0.5\le x \le 0.5$, and $-1\le y \le 1$. 
In order to proceed, we wish to fit $c_L(z)$  with a high-order polynomial (order 130).  In order that this fit be precise, 
we exclude the ``spiky'' region in Fig.~\ref{c-Q=5-top}, and only fit to the data in Fig.~\ref{c-Q=5-middle}.
The resulting data minus this fit is shown in Fig.~\ref{c-Q=5-bottom}. Noting the multiplication by $10^5$, we conclude that the fit is excellent inside the domain chosen for it in Fig.~\ref{c-Q=5-middle}, with errors apparent on the boundary.

As the fit is a polynomial in $z$ with complex coefficients, we   conclude that to the best of our knowledge $c_L(z)$ is an analytic function in this domain. This is non-trivial, as $c_L(z)$ could have been a  function of both $z$ and $\bar z$.

To locate a critical point, analyticity strongly suggest to replace ``maximum'' and ``minimum'' used in section \ref{Extrapolation on the lattice for L to infty} by stationary point, 
\be\label{SP}
\partial_z c_{L}(z) \Big|_{z=z_{\rm c}}= 0.
\ee
Due to   \Eq{c bar(z)=c(bar z)}, all coefficients are real. 
Solving this polynomial equation (of degree 130) yields the zeros, represented as dots in the left of Fig.~\ref{f:zerosQ=5}.
The domain of data used are shown in yellow. 
We see zeros close to the boundary, as well as for $|z|\approx 1$; these spurious solutions are marked in black. 
The two physical solutions (in red) are 
close to the imaginary axis, for 
\be\label{86}
z_{\rm c}^{L=10} \approx -0.013767 \pm 0.283318 i.
\ee
The central charge at this point is
\bea
c_{L=10} &=& 1.13543 \pm  0.0225995 i, \\
c_{\rm ana} &=& 1.13755 \pm 0.0210687 i,
\eea
where on the second line we give the analytical result. 

It is instructive to study how the effective $c_L(z)$ behaves close to its saddle point. 
Recall that our data are well fit by a polynomial in $z$ with complex coefficients. 
As the leading Taylor coefficient vanishes, see \Eq{SP}, close to the saddle point the Taylor expansion has the form $c_{L}(z) = c_L(z_{\rm c}) + \frac12 c_L''(z_{\rm c})(z-z_{\rm c})^2 + ...$. 
This is shown in Fig.~\ref{|c_L(z)-c_L(zc)|^0.25}. What we also see are the strong singularities already observed
 in Fig.~\ref{c-Q=5-top}. We believe that as for real $z$ they can be interpreted as a level crossing between physically different eigenvalues. This is consistent with the observation that they seemingly  form curves in the complex plane.

 \subsection{Improving the precision}
 \label{Improving the precision}
 
 Knowing that $c_L(z)$ is an analytic function in $z$, and where the fixed point is approximately, there are several strategies to improve the measurements, both using much fewer points.  
The one we study first, and which we also use for the spectrum later, is to 
sample   on the imaginary axis (we call this numerical scheme ``line" for easy reference), with the same step size of $\delta z=0.01$. This strategy works since for $Q=5$, the fixed point is very close to the imaginary axis, see Figs.~\ref{|c_L(z)-c_L(zc)|^0.25} and \ref{f:zerosQ=5}.

The second strategy is to sample only the immediate vicinity of the fixed point, which was first determined coarsely by the ``grid'' scheme, followed by an $L \to \infty$ extrapolation. In this new scheme, nicknamed ``minigrid'', we use a $5\times 5$ mesh around the ``grid'' estimate of the fixed point, using a {\em larger} step size of $\delta z = 0.05$. Interpolations are then done via a  polynomial of maximal degree (with terms $z^0$ up to $z^{24}$).
This strategy can also be employed  for complex $Q$, namely $Q=5+2i$ (section \ref{s:Q=5+2i}), and $Q=8+i$ (section \ref{s:Q=8+i}).

The   advantage of the line and minigrid  strategies is that one needs far fewer points, allowing one to go to larger $L$: 
for the large grid we went up to $L=11$, while  on the line we could go up to $L=15$. This in turn leads to a much higher precision for $c$, and especially $\omega$, as is shown in Fig.~\ref{f:extrapolationsQ=5}.

 The data presented in Fig.~\ref{f:extrapolationsQ=5} should be compared to the analytical predictions
\bea
c(Q=5) &=& 1.13755  + 0.0210687 i\\
\Delta= 2 h_{3,1}(Q=5) &=& 1.9083 + 0.598652 i
\eea\begin{figure*}
{\setlength{\unitlength}{\columnwidth}\begin{picture}(2.07,0.85)
\put(0,0){\Fig{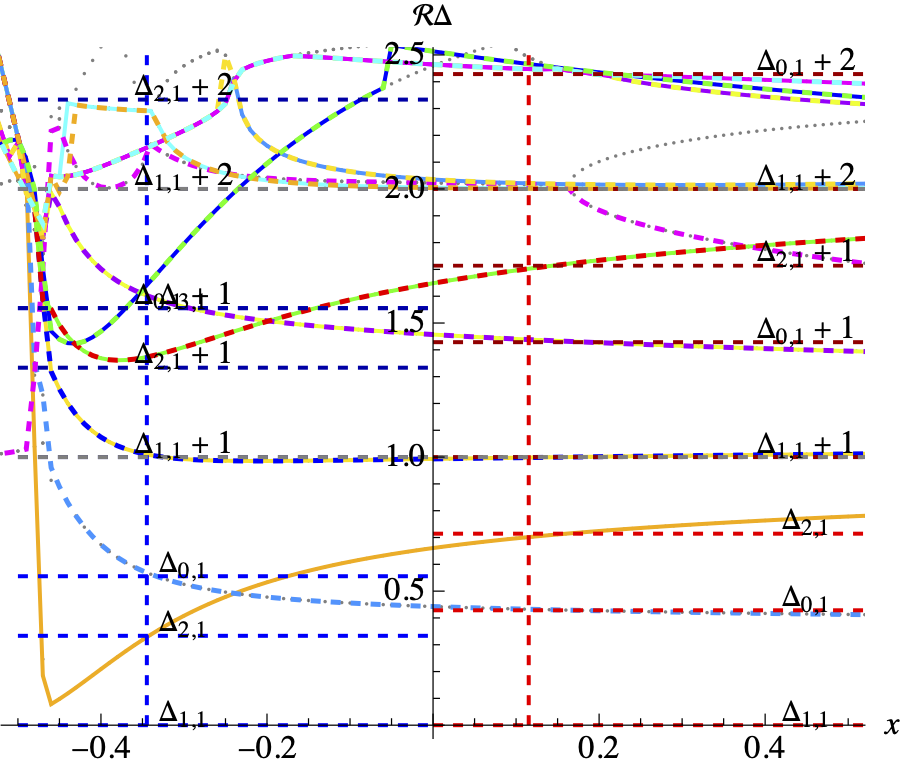}}
\put(1.07,0.027){\Fig{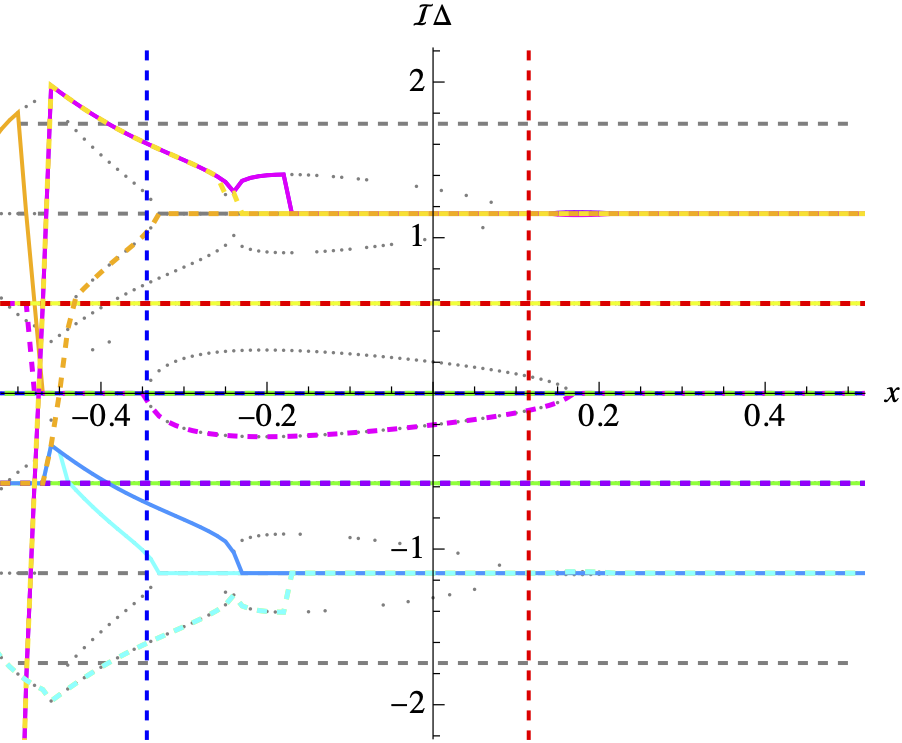}}
\put(0.57,0.01){$\ca B_8$}
\put(0.15,0.01){$\ca B_{-8}$}
\end{picture}}
\caption{The spectrum for $\ca B_8$ at $L=12$. Left real part, right imaginary part, in the same color code. Not  all levels could  be  reconstructed, due to a clipping of part of the spectrum for $\Re \Delta>2$. In the imaginary part, many curves lie indistinguishably on top of each other.}
\label{f:spectrum-B8}
\end{figure*}%
The agreement is good in all cases, especially for the line and minigrid data, which go to larger $L$. The central charge comes very close to its analytical prediction, with an error of about $10^{-4}$. The correction-to-scaling exponent $\omega$ takes longer to stabilize, and we see a clear advantage of having larger systems at our disposal.  The largest remaining discrepancy is in the imaginary part of $\Delta \equiv 2+\omega$,   which is of about $0.04$. Given that this is extracted from three consecutive sizes, this remains quite satisfactory. 

In Fig.~\ref{zc and c for Q=5},  we show the movement of the fixed point $z_{\rm c}$ and the central charge $c$ as a function of $L$. The data for $c$ follow a spiral, consistent with the complex exponent found for $\omega$. 
It is hard to make this quantitative, since both $z_{\rm c}^L$ and $c^L(z)$ change with $L$: a fit of the form $c(L) = c_{\rm ana} + a L^{-\omega}$ does not work. We did not pursue this further.

\section{Other values of  $Q$}
\label{Other values of  Q}
As discussed in section \ref{s:Theory}, our approach allows us to access any value of $Q$. To illustrate this, we first enlarge $Q$ up to $Q=40$. 
The results are collected in appendices \ref{s:Q=10} ($Q=10$),   \ref{s:Q=20} ($Q=20$) and   \ref{s:Q=40} ($Q=40$). 
Figs.~\ref{Q=10grid-c-cSP-contour-L=10} ($Q=10$, $L=10$), \ref{Q=20grid-c-cSP-contour-L=8} ($Q=20$, $L=8$), and \ref{Q=40grid-c-cSP-contour-L=7} ($Q=40$, $L=7$) show how the fixed point approaches the level-crossing lines.
While our analysis confirms that we can assess all practically relevant $Q$ values, we did not rule out  whether there is a critical $Q$, beyond which even for  larger $L$ we cannot locate the fixed point.
For the examples given, and within the limits of precision for these smaller sizes, we find  agreement between theory and numerics, both  for $c$ and $\omega$.   

We then proceed to complex values of $Q$, first $Q=5+2i$ (appendix \ref{s:Q=5+2i}) and $Q=8+i$ (appendix \ref{s:Q=8+i}).  
According to section \ref{Phase diagram as a function of Q}, see Fig.~\ref{Reh31=1-v2},  $Q=5+2i$  should have one stable and   one unstable fixed point, while $Q=8+i$  should  have two  unstable fixed points and no stable one. 
Since our ability to assess complex $Q$, and the agreement between theory and numerics are key claims of this work, we invested much effort in making our claims precise. Let us sketch our approach for $Q=5+2i$; the analysis for $Q=8+i$ proceeds in the same way. 

Our first step is to sweep the whole lattice in the ``grid'' scheme (appendix \ref{s:Q=5+2i:Grid}). Key results are the contour plot showing fixed points and level-crossing lines (Fig.~\ref{Q=5+2i-grid-c-cSP-contour-L=10.jpg}), and localizing the fixed points (Fig.~\ref{Q=5+2i-grid-zeros-of-c-prime-L=10}). On the next page, in Figs.~\ref{Q=5+2i-grid-Re-zc-upper} to \ref{Q=5+2i-grid-Re-omega-upper} we show the analysis for the fixed point in the upper half plane. Figs.~\ref{Q=5+2i-grid-Re-zc-lower} to \ref{Q=5+2i-grid-Re-omega-lower} repeats this analysis for the lower half plane. 

Knowing the location of both fixed points approximately, we repeat the analysis on a mini grid around our approximate fixed-point values for the coupling. Section \ref{Q=5+2i on minigrid, upper half plane} (Figs.~\ref{Q=5+2i-UHP-minigrid1-zeros-of-c-prime-L=15} to \ref{Q=5+2i-minigrid-zeros-of-c-prime-upper}) shows the analysis in the upper half plane, while 
section \ref{s:5+2i:Minigrid, lower half plane}  (Figs.~\ref{Q=5+2i-LHP-minigrid1-zeros-of-c-prime-L=15} to \ref{Q=5+2i-minigrid-zeros-of-c-prime-lower}) repeats the analysis for the lower half plane. 
This analysis  confirms  that theory and transfer matrix give the same value for $c$ with  4-digits accuracy at $L=15$, see \Eqs{D14} and  \eq{D19}.

Another key claim of ours is that in contrast to $Q=5$, the Potts model at complex $Q=5+2i$ has both a stable and an unstable fixed point. Via transfer matrix we find  an {\em unstable} fixed point in the upper half plane with $\omega = -0.51 +0.62 i $, see \Eq{omega-Q=5+2i-unstable} close to the theoretical prediction of $\omega = -0.49781 + 0.63539 i$ in \Eq{D17}. 
The fixed point in the lower half plane has $  \omega =  0.44  -1.48 i$,  see \Eq{omega-Q=5+2i-stable},  comparable to the analytic prediction of $\omega =  0.25868 -0.95536 i$, see \Eq{D22}; it is stable. We attribute the lack of precision to the closeness of further irrelevant couplings; the latter are  better separated at the unstable fixed point. 

We refer the reader to the results for $Q=8+i$ in appendix \ref{s:Q=8+i}, paralleling the analysis for $Q=5+2i$:
section \ref{s:Q=8+i:grid} with figures \ref{Q=8+i-grid-Re-c-L=10-raw} to \ref{Q=8+i-grid-zeros-of-c-prime-L=10} show the grid data. 
The fixed points are analyzed in Figs.~\ref{Q=8+i-grid-Re-zc-upper} to \ref{Q=8+i-grid-Re-omega-lower}. 
The refined analysis is performed separately for the fixed point in the upper and lower half planes. Results for the upper half plane are shown in appendix \ref{Q=8+i:minigid-upper}, Figs.~\ref{Q=8+i-UHP-minigrid-zeros-of-c-prime-L=15} to \ref{Q=8+i-minigrid-zeros-of-c-prime-upper}; appendix \ref{a:Q=8+1: Minigrid, lower half plane} with figures \ref{Q=8+i-LHP-minigrid-zeros-of-c-prime-L=14} to \ref{Q=8+i-minigrid-zeros-of-c-prime-lower} contains the results in the lower half plane. 

{\begin{figure*}[t]\newcommand{\here}{0.79} 
\leftline{{\parbox{\here\columnwidth}{\fig{\here}{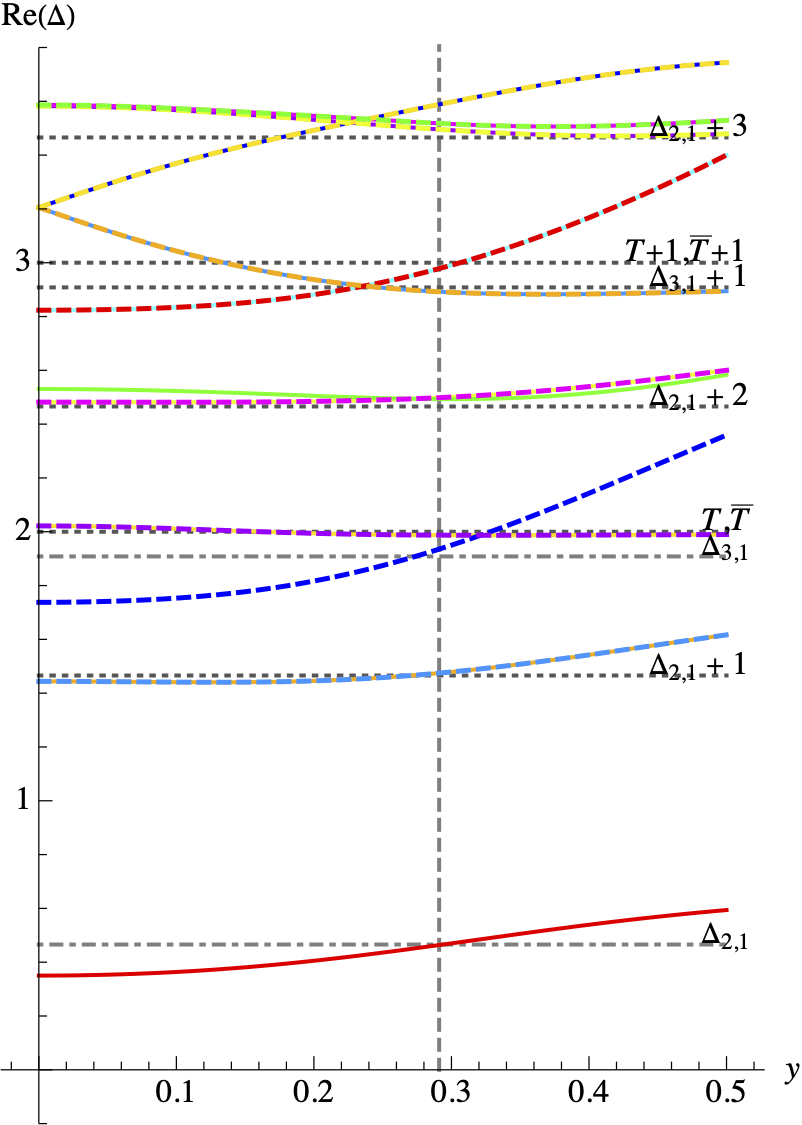}}}~~~{\parbox{\here\columnwidth}{\fig{\here}{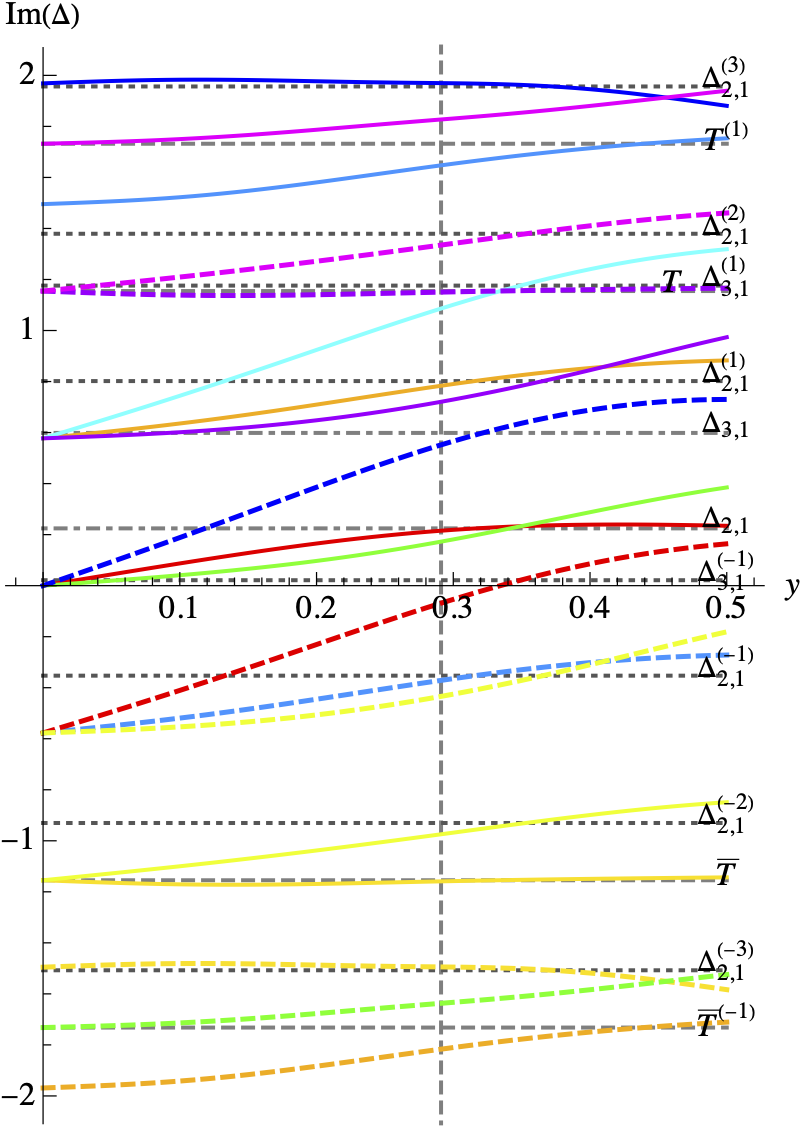}}}
~~~{\parbox{3.8cm}{\begin{tabular}{|c|c|r|}
\hline
state & operator & spin  \\
\hline
1 & $\mbox{GS}= V^d_{(1,1)}$ & 0  \\
2 & $\epsilon= V^d_{(2,1)}$ & 0  \\
3 & $L_{0,-1} V^d_{(2,1)}$ & $-1$  \\
4 & $ L_{-1,0} V^d_{(2,1)}$ & $1$   \\
5 & $\epsilon'= V^d_{(3,1)}$ & 0  \\
6 & $\overline T \equiv L_{0,-2} V^d_{(1,1)}$ & $-2$  \\
7 & $ T \equiv L_{-2,0} V^d_{(1,1)}$ & $2$  \\
8 & $L_{0,-2} V^d_{(2,1)}$ & $-2$  \\
9 & $L_{-2,0} V^d_{(2,1)}$ & $2$  \\
10 & $L_{-1,-1} V^d_{(2,1)}$ & $0$   \\
11 & $L_{0,-1} V^d_{(3,1)}$ & $-1$   \\
12 & $L_{-1,0} V^d_{(3,1)}$ & $1$ \\
13 & $L_{0,-1} \overline T$ & $-3$ \\
14 & $L_{-1} T$ & $3$  \\
15 & $L_{0,-3} V^d_{(2,1)}$ & $-3$  \\
16 & $L_{-3,0} V^d_{(2,1)}$ & $3$  \\
17 & $L_{-1,-2} V^d_{(2,1)}$ & $-1$ \\
18 & $L_{-2,-1} V^d_{(2,1)}$ & $1$  \\
19 & $L_{0,-3} V^d_{(2,1)}$ & $-3$  \\
20 & $L_{-3,0} V^d_{(2,1)}$ & $3$   \\
\hline
\end{tabular}}}}
\caption{Real and imaginary part $\Delta_{r,s}^{(S)}$ of spectrum in the ground-state sector for $Q=5$, $L=15$, taking only the first 20 EVs. 
Primaries in red, dashed, descendents in darker red, dotted. The red dashed vertical lines denote the position of the critical point.}
\label{f:SpectrumQ=5}
\end{figure*}}

\section{Spectrum}
\label{Spectrum}
\subsection{Outline}
Extracting the spectrum from the transfer matrix is an important check on our claim that the full conformal information can be extracted by analytic continuation of the $c \le 1$ loop-model CFTs. To sharpen our understanding we first consider $b=8$ in section \ref{s:B_8} for which the spectrum of a corresponding loop model is well established, both analytically and numerically \cite{JacobsenSaleur2019}. 
We then proceed to $Q=5$ in section \ref{s:Q=5 spectrum}. 

While we have not evaluated the spectrum  for  complex values of $Q$,  we are confident it   works there as well. This is confirmed by our ability to extract the first subleading scalar  $\Delta_{3,1}$ for $Q=5+2i$, see \Eqs{D16}-\eq{D17} for the fixed point in the  upper half plane, and   \Eqs{D21}-\eq{D22} for the fixed point in the  lower half plane. 
This analysis is repeated  for $Q=8+i$, see  \Eqs{E17}-\eq{E18} for the fixed point in the upper half plane, and \Eqs{E23}-\eq{E24} for the fixed point in the lower half plane.

\subsection{The case $b=8$ ($Q = 3.41421$)}
\label{s:B_8}

In Fig.~\ref{f:spectrum-B8} on page \pageref{f:spectrum-B8}, we show the spectrum 
for $b=8$. We have marked with a
red dashed vertical line the location of the minimum ($x_{\rm c} = 0.1152$) which we previously identified with the minimal model $\ca B_8$, and with a blue vertical line the location of the maximum ($x_{\rm c} = -0.3447$) which we previously identified with the minimal model $\ca B_{-8} \equiv \ca B_9$.
By horizontal lines, we show the expected spectrum  for this model.
This spectrum contains the spinless primary operators $V^d_{(r,1)}$ with exponents%
\footnote{We apologize to the reader for the requirement to distinguish between the notation $\Delta_{(r,s)} = h_{r,s}$ of section~\ref{sec:op-content-Potts} --- necessary to make contact with \cite{JacobsenSaleur2019,HeJacobsenSaleur2020,JacobsenRibaultSaleur2023,GransSamuelssonJacobsenNivesvivatRibaultSaleur2023,NivesvivatRibaultJacobsen2024,RouxJacobsenNivesvivatRibault2025,JacobsenNivesvivatRibaultRoux2025} --- and the notation $\Delta_{r,s} = 2 h_{r,s}$ used here and below, to remain consistent with our previous work \cite{WieseJacobsen2024}.}
\be
\label{91}
\Delta_{r,1} = 2 h_{r,1}({b= \mp 8}),  
\ee 
starting with $r=1$ for the vacuum (baseline). 
For the left FP $b=-8$, while for the right FP $b=8$.
We verify the appearance of $V_{(0,1)}$ and $V^d_{(2,1)}$,
with the proper dimensions both for the real and imaginary parts (it vanishes), as well as some of their descendents.  Descendents may have a non-vanishing spin $S$. The dimension of the primary increases by $1$ for each level of descendence, and by $i S/\sqrt3$ due to the spin. The latter imaginary contribution arises because our transfer matrix $T_L$ defined by \eqref{TM_R} is a mixture of a dilation and a rotation in radial quantization. Indeed, $T_L$ propagates by $\frac{\sqrt{3}}2$ lattice spacings upwards (the height of an equilateral triangle of side length 1), and by $\frac12$ lattice spacings rightwards, as can be seen from \eqref{labelled-triangle}. The former factor $\frac{\sqrt{3}}2$ is taken into account in the scaling analysis when normalizing the free energy per unit area, but the ratio $\sqrt{3}$ of the two factors remain. Using $L_{-N,-\bar{N}} V^d_{(r,1)}$ as a shorthand for any descendent on chiral level $N$ and antichiral
level $\bar{N}$, we get
\begin{figure}[t]\newcommand{\here}{0.5} 
\leftline{{\parbox{\here\columnwidth}{\fig{\here}{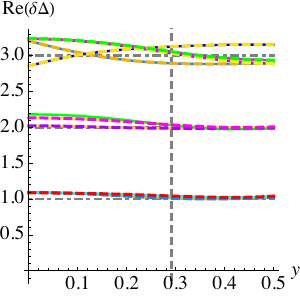}}}~{\parbox{\here\columnwidth}{\fig{\here}{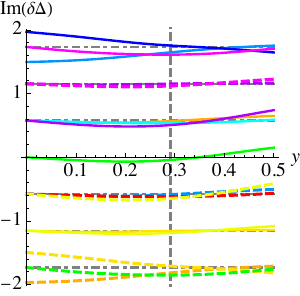}}}}
\caption{Difference in real and imaginary part for the descendents minus their primaries    in the ground-state sector for $Q=5$, $L=15$. Same color-code as in Fig.~\ref{f:SpectrumQ=5}.}
\label{f:SpectrumQ=5-descendents}
\end{figure}
\be\label{magic-spectrum}
\Delta (L_{-N,-\bar{N}} V^d_{(r,1)} )  = \Delta_{r,1} + (N+\bar{N}) + \tfrac{1}{\sqrt{3}} (N-\bar{N})  i .
\ee
The counting of descendents confirms that $V^d_{(r,1)}$ are indeed degenerate and form Kac modules with one null state  at level $r$. The content of primaries and the structure of the modules
were predicted from the torus partition function \cite{DiFrancescoSaleurZuber1987} and previously verified numerically for real $Q < 4$ in a square-lattice loop model with only nearest neighbor interactions
\cite{JacobsenSaleur2019} (see also section~\ref{sec:op-content-Potts}).

{\begin{figure}[t]\newcommand{\here}{0.5} 
\leftline{{\parbox{\here\columnwidth}{\fig{\here}{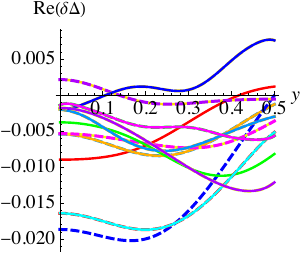}}}~{\parbox{\here\columnwidth}{\fig{\here}{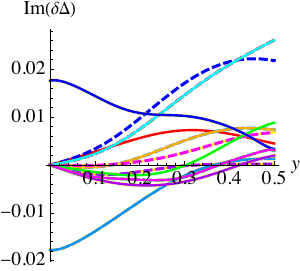}}}}
\caption{Difference in real and imaginary part for the   spectrum in the ground-state sector for $Q=5$, $L=15$, when moving the path from $i \mathbb R$  to $i \mathbb R -0.01$.  Same color-code as in Fig.~\ref{f:SpectrumQ=5}.}
\label{f:SpectrumQ=5-displaced-path}
\end{figure}}

\subsection{$Q=5$}
\label{s:Q=5 spectrum}
\subsubsection{Degenerate operators $ V^d_{(r,1)}$}
\label{quotient_rep}

We now study  the spectrum at $Q=5$ in the ground-state sector,
using the quotient representation $\overline{\mathcal W}_{0,\mathfrak{q}^2}$ of defect-free link patterns of dimension
$\tfrac{1}{L+1} {2L \choose L}$, where $ \sqrt{Q} = \mathfrak{q} + \mathfrak{q}^{-1}$; see sections \ref{sec:op-content-Potts} and \ref{sec:global-sym} as well as \cite{JacobsenRibaultSaleur2023} for details. 
Fig.~\ref{f:SpectrumQ=5} shows the exponents corresponding to the first 20 eigenvalues (EV) in this sector.
As the critical points have $x \approx 0$ (see Fig.~\ref{f:zerosQ=5}), we take $x=0$ and plot against $y \ge 0$.
The vertical red dashed line indicates the fixed point.

This spectrum contains the spinless primary operators $V^d_{(r,1)}$ with exponents $\Delta_{r,1}= 2 h_{r,1}^{-b}$, 
starting with $r=1$ for the vacuum (baseline). The general formulas are given in \Eqs{91}-\eq{magic-spectrum} and the paragraph in between. We verify the appearance of $\epsilon = V^d_{(2,1)}$ and $\epsilon' = V^d_{(3,1)}$, with the proper dimensions both for the real and imaginary parts.  
 We   see in Fig.~\ref{f:SpectrumQ=5} that 
all of this is   perfectly respected  at $Q=5$ for the 20 lowest-lying states: 
 we confimr that the results for the spectrum carry over to the complex CFT by analytic continuation.

\smallskip

\noindent
{\bf Descendents}:
As discussed above, 
descendents may have a non-vanishing spin $S$: as a result,  the dimension of the  primary increases by $1$ for each level, and by $i S/\sqrt3$
due to the spin. This is verified on Fig.~\ref{f:SpectrumQ=5-descendents}. Apart from errors which we associate to system size $L=15$ being too small, the spacing of the real part is indeed integer, and for the imaginary part   multiples of $1/\sqrt3$.

\smallskip

\noindent
{\bf Double-layer transfer matrix}:
It is of course possible to study the spectrum using a pure dilation operator.
{\begin{figure}[t]\newcommand{\here}{0.5} 
\leftline{{\parbox{\here\columnwidth}{\fig{\here}{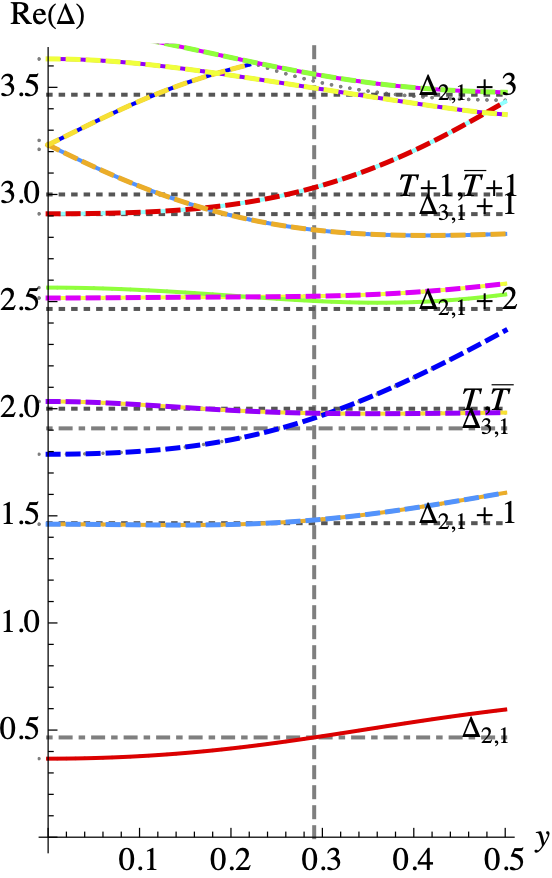}}}~{\parbox{\here\columnwidth}{\fig{\here}{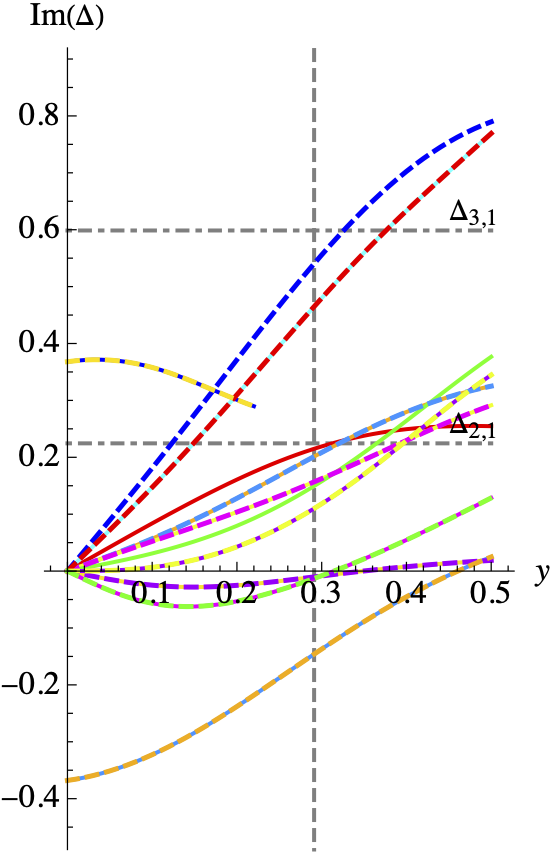}}}}
\caption{Real and imaginary part $\Delta_{r,s}^{(S)}$ of spectrum in the ground-state sector for $Q=5$, $L=12$, maximally 20 EVs, two consecutive layers. 
Primaries in dot-dashed, descendents  dotted. The   dashed vertical lines denote the position of the critical point.}
\label{f:SpectrumQ=-5-2layers}
\end{figure}}%
To this end,  we performed transfer-matrix calculations on two consecutive layers, such that the first step is up right, and the second up left, with no resulting translation. (In the notation of section~\ref{sec:TM} this means diagonalizing the operator $u^{-1} T_L^2$.)
We then expect the last term $(N-\bar N)/\sqrt{3}$ in \Eq{magic-spectrum} to be absent. This can be checked in Fig.~\ref{f:SpectrumQ=-5-2layers}. In particular, for real $z$, i.e.\ $y=0$, the imaginary part of the spectrum vanishes. This is indeed the case for all levels, except two. The latter belong to levels degenerate at $z=0$, with $\Delta=3.231 \pm 0.368 i$. We believe that 
this feature is a finite-size effect. From a CFT point of view, values outside the fixed point are not universal, so we should focus on what happens at the fixed point: there  seem to be both level-3 descendents of the identity, or level-1 descendents of the stress tensor.

\smallskip

\noindent {\bf Insensitivity to the chosen path}:
In Fig.~\ref{f:SpectrumQ=5-displaced-path} we show the difference in real and imaginary part for the   spectrum in the ground-state sector for $Q=5$,   when moving the path from $i \mathbb R$  to $i \mathbb R -0.01$. The latter is   the approximate uncertainty     in the determination of the critical point $z_{\rm c}$. We conclude that an uncertainty of about $0.01$ in the location of $z_{\rm c}$ translate into a similar uncertainty in the spectrum.

\smallskip 

\subsubsection{Non-diagonal operators $V_{(r,s)}$}

We now study the non-diagonal operators $V_{(r,s)}$ by diagonalisation in the standard modules ${\cal W}_{j,\tilde{z}}$ with $j=2s$ defects. For simplicity we choose the trivial phase factor $\tilde{z}=1$ corresponding to the spinless non-diagonal operators $V_{(0,s)}$. We shall use the notation $X_j(L)$ for the effective exponent obtained from the finite-size scaling analysis, to be compared to $2h$. Notice that although the operator $V_{(0,1)}$ is absent from the torus partition function, as explained in section~\ref{sec:op-content-Potts}, it can still be realized in the transfer-matrix formalism within the corresponding standard module.

\begin{figure}[t]
\fig{0.54}{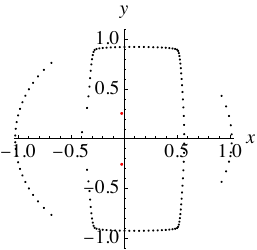}\fig{0.46}{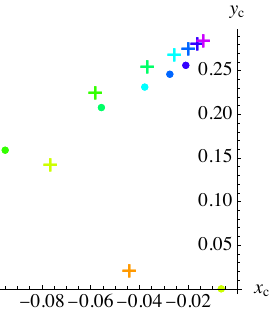}
\caption{Left: Loci of $X_2(L)-X_2(L+1)=0$, at size $L=9$. Red dots show the estimates of $z_{\rm c}$. Right: Movement of $z_{\rm c}$ as a function of $L$, from red ($L=4$) to dark blue ($L=9$) to magenta ($L=10$). Crosses come from the analysis of $c$, dots from $X_2$.}
\label{f:X2}
\end{figure}

We were able to use 
\be
X_2(L) = X_2(L+1)
\ee
to locate the FP. An example is shown in   Fig.~\ref{f:X2} on the left, with the movement as a function of $L$ on the right. 
Extrapolation yields
\be
z_{\rm c} \simeq 0.005 + 0.294 i .
\ee
This is comparable to what we got from the central charge, see table \ref{Q=5+2i-minigrid-zeros-of-c-prime}, even though  our largest system size   was only $L=10$.

Fitting the exponents for $j=2$ and $j=4$ defects at finite sizes (not shown), and extrapolating to $L\to \infty$ yields  
\bea
\Delta_{X_2}&=&0.512959 -0.0772247 i, \\
2h_ {0,1}&=& 0.511462 -0.0748315 i, \\
\Delta_{X_4}&=&2.01179 -0.317301 i, \\
2h_ {0,2}&=&  2.01146 -0.304593 i.
\eea
This again provides an excellent agreement between the transfer matrix results and the analytic continuation of the CFT.

\subsubsection{Additional states in the non-quotient spectrum}
\begin{figure}
\fig{0.5}{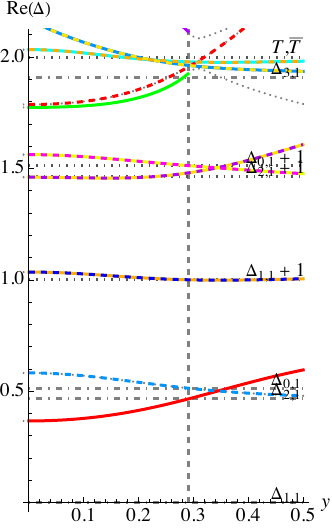}\fig{0.5}{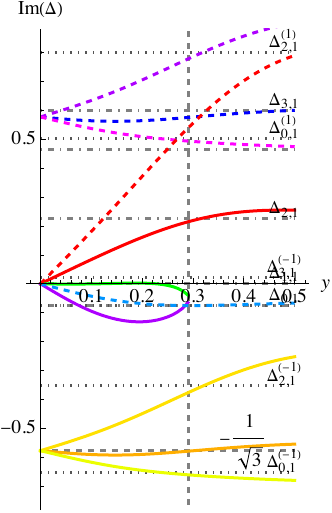}
\caption{Full spectrum. Note the additional states as compared to Fig.~\ref{f:SpectrumQ=5}.}
\label{spectrum-full-Re-Q=5-L=12-maxlevel=20-right}
\end{figure}
Rather than diagonalizing $T_L$ in the quotient representation $\overline{\mathcal W}_{0,\mathfrak{q}^2}$, as in section~\ref{quotient_rep}, we also examined the non-quotient representation ${\mathcal W}_{0,\mathfrak{q}^2}$.
From the point of link patterns this means that we now work in a larger space that distinguishes whether the link
between two sites straddles the periodic boundary condition. In particular, if we changed the twist parameter
in this representation, using ${\mathcal W}_{0,\tilde{\mathfrak{q}}^2}$, we would be able to give a different weight
$\tilde{n} = \tilde{\mathfrak{q}} + \tilde{\mathfrak{q}}^{-1}$ to non-contractible loops, while still giving the weight
$n = \mathfrak{q} + \mathfrak{q}^{-1}$ to  contractible loops. But even taking $\tilde{\mathfrak{q}} = \mathfrak{q}$,
as we did, the algebraic structure changes: the non-quotient representation ${\mathcal W}_{0,\mathfrak{q}^2}$
decomposes as a direct sum of $\overline{\mathcal W}_{0,\mathfrak{q}^2}$ and the 2-defect standard module
${\mathcal W}_{1,\pm 1}$.

As a result, we obtain more states than in Fig.~\ref{f:SpectrumQ=5}. In
Fig.~\ref{spectrum-full-Re-Q=5-L=12-maxlevel=20-right} we show the spectrum without taking the quotient,
restricted to the lowest 20 eigenvalues. We notice in particular the appearance of the   non-diagonal  primaries
$V_{(0,1)}$ and $V_{(1,1)}$, in   agreement with section~\ref{sec:op-content-Potts} and Conjecture~2.

\begin{figure}
\fig{0.5}{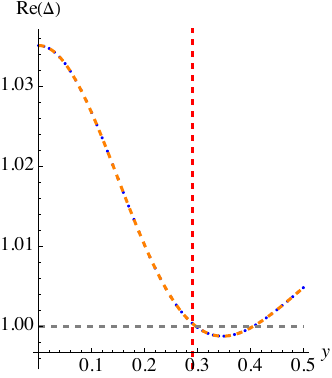}\fig{0.5}{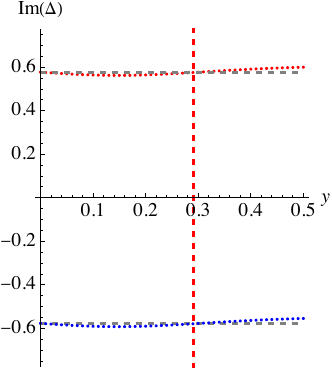}
\caption{$\Delta$ for EVs 4 and 5 at $L=12$. At the fixed point this becomes $\Delta = 1\pm i/\sqrt{3}$.}
\label{level4and5-Re-L=12-v2}
\end{figure}
\begin{figure}
\Fig{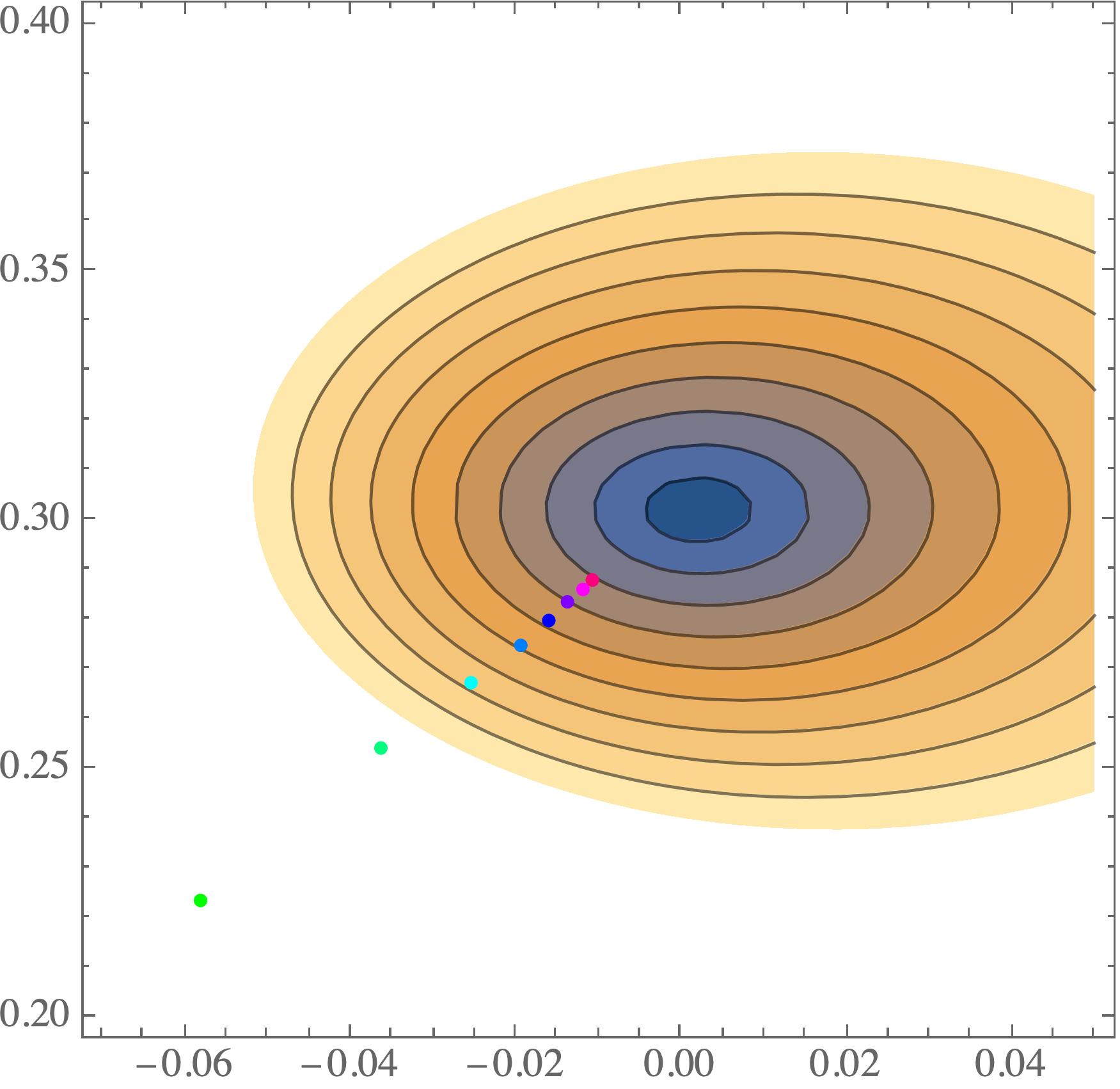}
\caption{The location $z_c = x+iy$ where  $\Delta =1+i/\sqrt{3} $. The contour plot shows $|\Delta -1-i/\sqrt{3}|^2 $. The dots show the the successive approximations from $c'(z)=0$.  }
\label{Jlocation-Q=5-v2-L=12}
\end{figure}

The field $\bar J = V_{(1,1)}$ with conformal weights
\begin{equation}
 (\Delta_{(1,1)},\Delta_{(1,-1)}) = (0,1)
\end{equation}
is of particular interest.
It was shown in \cite{JacobsenNivesvivatSaleur2024}
to be one of a pair of  non-conserved  currents, the other one being $J = V_{(1,-1)}$ with conformal weights $(1,0)$, that satisfy
\begin{equation}
 \bar \partial J = \partial \bar J \neq 0 \,.
\end{equation}
and which do {\em not} giving rise to a Kac-Moody algebra.

Using here our numerical approach at $Q=5$, we find
at the critical point, indicated with a red dashed line for $L=12$ in Fig.~\ref{level4and5-Re-L=12-v2} 
\be
\Delta =1.00047 +0.575189 i \approx 1 + \frac i{\sqrt{3}} \,,
\ee
in fine agreement with $(h,\bar{h}) = (1,0)$ for the field $J$ [cf.~\Eq{magic-spectrum}].

As before, we fit the spectrum as a function of $y$ by a polynomial. 
We can then use this high-quality fit to get the location where $J =1 +  i/{\sqrt{3}}$, as shown in the contour plot of Fig.~\ref{Jlocation-Q=5-v2-L=12}. We also show the convergence of $z_c$ obtained via $c'(z)=0$. 
The location is seemingly right on the imaginary axes. We can do this more precisely, by fitting a polynomial, and searching for $\Delta=1+i/\sqrt3$. The result is
\be
z_c = 0.00182 \pm  0.30206 i.
\ee
This should be compared to our best estimate from $c'(z)=0$, namely $z_{\rm c} = 0.0049+0.3007i$, see Fig.~\ref{f:All our data}.

We finally note that we found another element in the spectrum, with $\Delta \approx 2\times (1+i/\sqrt 3)$.
It can be identified with the stress tensor $T$ with conformal weights $(h,\bar h) = (2,0)$.

\subsubsection{OPE coefficients: Quotient spectrum corrected by $\epsilon'$}
\begin{figure}[t]
\centerline{\fig{0.5}{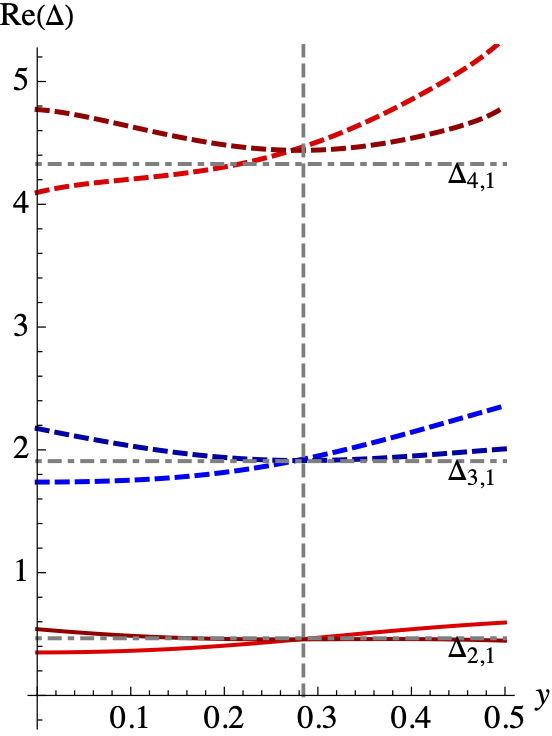}\hfill\fig{0.5}{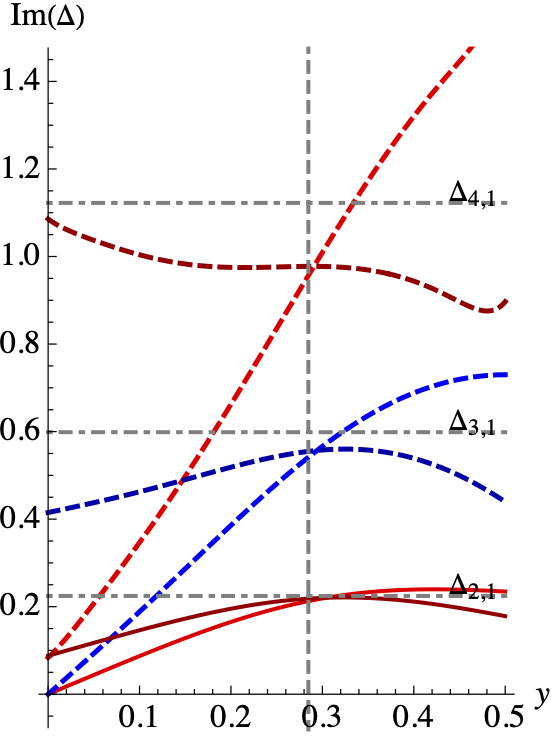}}
\caption{Left: real part of the spectrum for $L=15$ with the first three primaries as a function of $y$: $\Delta_{2,1}$ (red, solid, bottom), $\Delta_{3,1}$ (blue dashed, middle) and $\Delta_{4,1}$ (top, red dashed). We use the  same colors as in Fig.~\ref {f:SpectrumQ=5}. Note that $\Delta_{4,1}$ is not shown there, and we changed the  auxiliary lines to grey to improve readability.  The same levels  corrected around $z_{\rm c}$ for the influence of $V^d_{(3,1)}$ are shown in the same darkened color. Right: {\em idem} for the imaginary part. The darker lines are the improvements from \Eq{101} with parameters as given in \Eqs{102}-\eq{103}.}
\label{primaries-corrected-v2}
\end{figure}
Up to now  we read off exponents as $\Delta = \Delta(z_{\rm c})$. When one is not exactly at the fixed point, there are corrections due to subdominant operators. The leading one is $\epsilon' = V^d_{(3,1)}$, integrated over all of space.  Denoting by $g$ its effective coupling constant, we expect to see a corrected value $\Delta_{i,1}^{\rm corr}(z)$ \cite{LaoRychkov2023}
\be\label{101}
\Delta_{i,1}^{\rm corr}(z) =   \Delta_{i,1} (z) -g  (z-z_{\rm c}) C_{(i,1),(i,1),(3,1)}.
\ee
Here $C_{(i,1),(i,1),(3,1)}$ denotes the OPE coefficient (3-point structure constant) of twice the primary $V^d_{(i,1)}$ and once $V^d_{(3,1)}$, which is evaluated in \Eq{wott} and  in appendix \ref{OPE-coefficients following Dotsenko-Fateev}. 
In  Fig.~\ref{primaries-corrected-v2} we used
\bea\label{102}
z_{\rm c} &=&-0.00852 + 0.28466  i, \\
 g &=& 0.2 - 0.8  i .
 \label{103}
\eea
In principle the effective coupling  can be gotten as 
\bea
g &\approx&g_2:= \frac{\partial_z \Delta_{2,1}(z)}{C_{(2,1),(2,1),(3,1)} } =0.349 - 0.885 i\\
g &\approx &g_3:= \frac{\partial_z \Delta_{3,1}(z)}{C_{(3,1),(3,1),(3,1)} }= 0.297 - 0.874 i \\
g &\approx& g_4: = \frac{\partial_z \Delta_{4,1}(z)}{C_{(4,1),(4,1),(3,1)} }= 0.210 - 0.829 i
\eea
The value used is a compromise, to get  the spectrum
(dark lines in Fig.~\ref{primaries-corrected-v2}) as flat as possible around $y_{\rm c}$. Since $\Delta_{4,1}$ has uncorrected the largest slope, this value is close to the coupling constant extracted from the measurement of this OPE coefficient. 
This procedure can be improved by adding more operators. Another possibility is to directly measure a 3-point function. This was persued in \cite{TangMaTangHeZhu2024,TangMaTangHeZhuUnpublished}, and yields much more precise results.

\enlargethispage{1.3cm}

\section{Conclusions}
\label{Conclusions}
In this paper we have significantly substantiated our claims, first summarized in \cite{JacobsenWiese2024},
that complex CFTs can be realized as the continuum limit of lattice models, both for $Q > 4$ and
complex $Q$. The corresponding lattice realization \eqref{Zloop} is a loop model that stems from a triangular-lattice Potts model
with competing two- and three-spin interactions. Along its self-dual manifold, parametrized by $z$ for $w=1$
[see \Eq{w=1}], it possesses a pair of complex fixed points $z_{\rm c}$ for well-chosen values of $z \in \mathbb{C}$.
Their respective continuum limits are complex CFTs (see Conjecture 1).

Our numerical study is based on the diagonalization of the transfer matrix $T_L$ in the basis of standard modules of the
affine Temperley-Lieb algebra. We have first established a robust and versatile methodology for determining the
critical points $z_{\rm c}$ for $Q \in \mathbb{C}$. Using this, we have determined their conformal data:
central charge, critical exponents and three-point structure constants (in the latter case comparing to   published and unpublished work by
\cite{TangMaTangHeZhu2024}). We have established that all of this is in agreement with Conjecture 2,
which identifies the complex CFT with the analytic continuation of the non-unitary CFT studied in a body of recent papers
\cite{JacobsenSaleur2019,HeJacobsenSaleur2020,JacobsenRibaultSaleur2023,GransSamuelssonJacobsenNivesvivatRibaultSaleur2023,NivesvivatRibaultJacobsen2024,RouxJacobsenNivesvivatRibault2025,JacobsenNivesvivatRibaultRoux2025}. This result is accompanied   by a careful discussion of how the analytical continuation should be taken.

Concerning the operator content of the complex CFT, we have checked in   detail the presence of degenerate and
non-diagonal operators predicted in the table of section~\ref{sec:op-content-Potts}. The diagonal operators, including
the special case of order-parameter operators, could be similarly identified by diagonalizing $T_L$ in the representation
${\cal W}_{0,\tilde{\mathfrak{q}}^2}$, for appropriate values of the twist parameter $\tilde{\mathfrak{q}}$.

The ongoing work on the non-unitary CFT of loop models already contains, or is about to soon produce, more
detailed results, including higher-order correlation functions, families of boundary conditions parametrized by
discrete or continuous parameters, and results in geometries with non-trivial topology. We would expect this
to carry over to the complex CFT setting as well. A few examples for boundary data have already appeared in
\cite{TangLiuTangZhu2025}.

We finally point out that while we have framed this work in the context of the $Q$-state Potts model, a very
similar scenario plays out in the complex $O(n)$ model. The relevant predictions  gathered in section~\ref{sec:op-content-On}, can again be obtained by analytic continuation.

The present work raises the question of its possible experimental realization. 
Proposals for  non-Hermitian quantum mechanics  are widely discussed  in the   literature, see  \cite{WisemanMilburn2009,AshidaGongUeda2020} for a review. 
The most direct way seems to be via post-selection in monitored quantum systems \cite{LiChenFisher2018,SkinnerRuhmanNahum2019}, which has successfully been implemented in   experiments  
on   trapped ions   \cite{NoelNiroulaZhuRisingerEganBiswasCetinaGorshkovGullans2022,HokeIppolitiRosenbergAbaninAcharyaAndersenAnsmannAruteArya2023} or superconducting Q-bits \cite{KohSunMottaMinnich2023}.
In contrast, realizing a complex CFT has to our knowledge not been achieved, even though walking behavior, i.e.\ the presence of a close-by complex CFT is widely discussed, see e.g.\ \cite{KumarPujariDEmidio2026}.
There are   attempts to get closer to a complex CFT in quantum Monte Carlo simulations \cite{ZouYinLiYao2025}.  A  recent proposal to realize the $O(3)$  complex CFT is given in  \cite{YangScaffidi2026}, using a no-clicks measurement protocol which can be mapped to a Lindblad master equation. 

To realize a complex CFT in an experiment,  one can use the equivalence of a $d+1$ dimensional classical system to a $d$-dimensional quantum system. There the spectrum of a CFT maps onto energy eigenvalues of a quantum Hamiltonian via $\Delta = v_{\rm F} (E + i /\tau)$, where $v_{\rm F}$ is the Fermi velocity, and  $E$ the energy of the state. The wave function  in time of a mode $\Delta$ evolves as  $|\psi_\Delta(t)\rangle \sim \rme^{i v_{\rm F} t (E+ i/\tau)} $. 
This is  a   {\em quasiparticle} interpretation, with a positive imaginary part of $\Delta$ indicating decay, and 
a negative imaginary part  growth. The most easily   realizable CFT arises  when the imaginary parts all have  the same sign. As complex CFTs come in pairs, let us choose the branch where the operators of interest have $\Delta$ with positive imaginary part, and let us further assume that the imaginary part grows with the energy. 
This   means that  higher excitations decay  faster, which    naturally select the lowest-lying states over time, ensuring converges to the complex CFT.  
We have verified that both the family of degenerate operators $V_{(i,1)}$ which contains $\epsilon$, and the family of   order-parameter operators, $V_{(r,0)}$ with $r\in \mathbb{N}+\tfrac12$, which contain $\sigma$, in our choice of branch have $\Im \Delta > 0$. 
Moreover, $\Im \Delta \ge \frac{1}{|b|} \Re \Delta$ for any $Q > 4$ (e.g., $\Im \Delta \ge 0.153\, \Re \Delta$ for $Q=5$). Thus higher excited states would indeed be decaying faster. 
On the other hand, defect operators have $\Im \Delta < 0$. While it is not a contradiction to have both types of states, the latter ones would grow exponentially. As they do not result from the fusion of spin  or energy operators, they may simply be absent from an experiment.

We hope that  efforts to realize the complex CFT within an experiment will succeed, and make the present study more broadly  relevant.

\vspace*{-2mm}

\acknowledgments

\vspace*{-2mm}

We acknowledge stimulating discussions with C.~Bachas, Y.-C.~He, A.W.W.~Ludwig, A.~Nahum, S.~Ribault, S.~Rychkov, M.~Salmhofer, H.~Saleur, Y.~Tang and W.~Zhu. We  thank   Y.~Tang and  W.~Zhu for providing us with  their unpublished OPE coefficients, and P.~Roux for his julia code to evaluate OPE coefficients. 
This work was supported by the French Agence Nationale de la Recherche (ANR) under grant ANR-21-CE40-0003 (project CONFICA), and in part by grant NSF PHY-2309135 to the KITP.

\appendix

\clearpage

\section{$Q=10$}
\label{s:Q=10}
\begin{figure}[t!]
\fig{0.5}{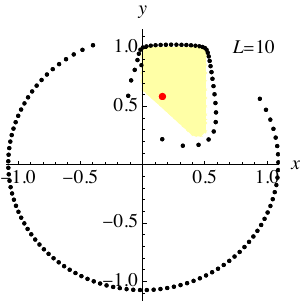}
\caption{Zeros of $c'(z)$ at $L=10$.}
\label{Q=10grid-zeros-of-c-prime-L=10}
\centerline{\fig{0.49}{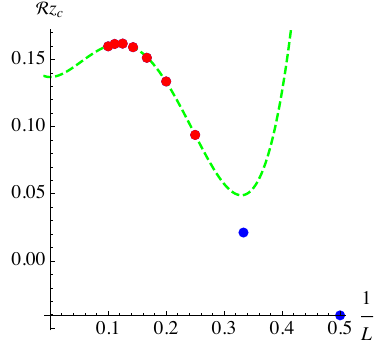}~\fig{0.49}{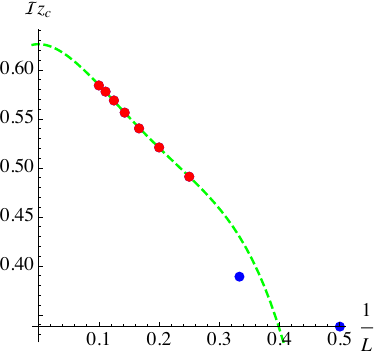}}
\caption{Extrapolation of  $z_{\rm c}$ for $Q=10$.
This gives $ z_{{\rm c},L\to \infty}^{\rm upper} =  0.136932 +0.62623 i. $}
\label{Q=10-zc}
\centerline{\fig{0.49}{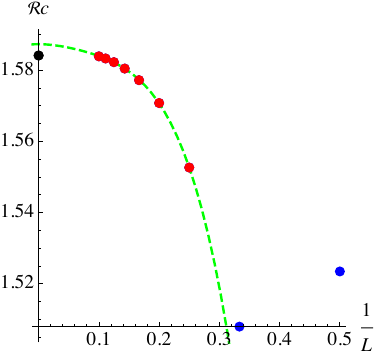}~\fig{0.49}{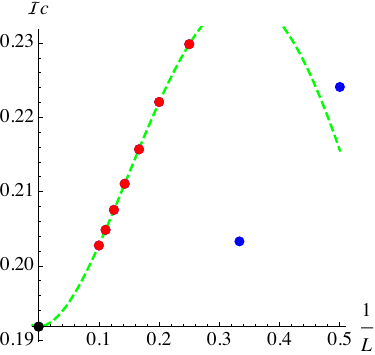}}
\caption{Extrapolation of  $c$ for $Q=10$ yields  $c^{\rm upper}_{L\to \infty} =  1.58465 +0.190826 i$. Compare to $c^{\rm upper}_{\rm ana} = 1.58411 +0.191825 i $.}
\label{Q=10grid-Re-c-upper}
\centerline{\fig{0.49}{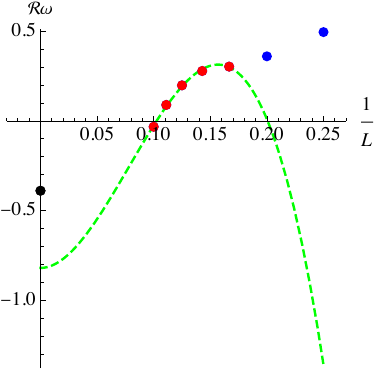}~\fig{0.5}{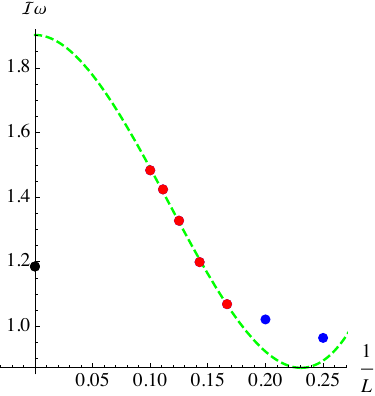}}
\caption{Extrapolation for $\omega$.  $\omega^{\rm upper}({L\to \infty}) =  -0.480646+1.99852  i$, 
$\Delta^{\rm upper}({L\to \infty}) =2+ \omega^{\rm upper}({L\to \infty})= 1.51935 +1.99852 i $, 
$ 2 h_{3,1}= 1.6106 +1.18574 i.$}
\label{Q=10grid-Re-omega-upper}
\vspace*{-.4cm}
\end{figure}

\enlargethispage{1cm}
\begin{figure}[h!]
\centerline{\fig{0.5}{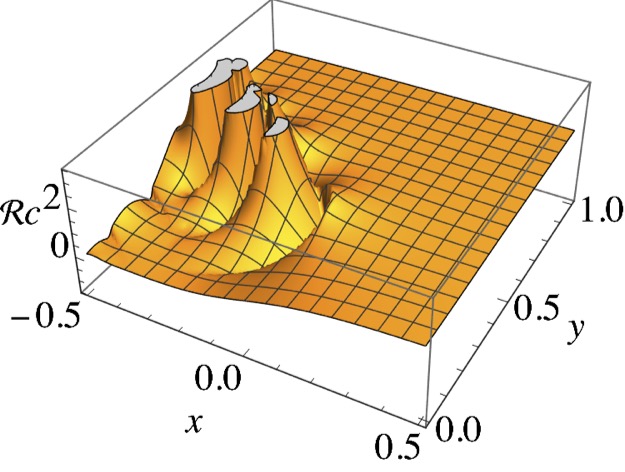}~\fig{0.5}{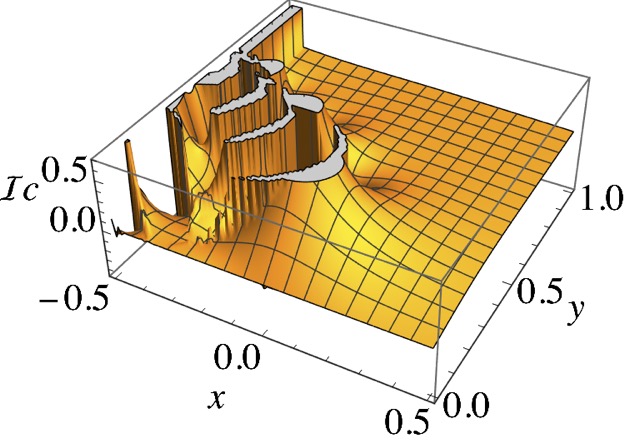}}
\centerline{\fig{0.5}{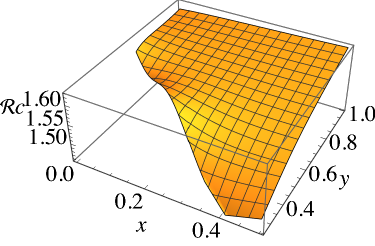}~\fig{0.5}{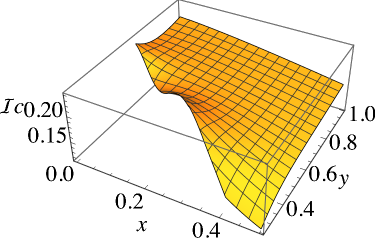}}
\centerline{\fig{0.5}{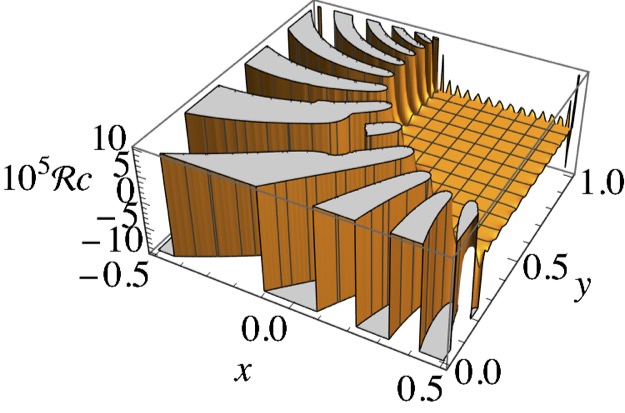}~\fig{0.5}{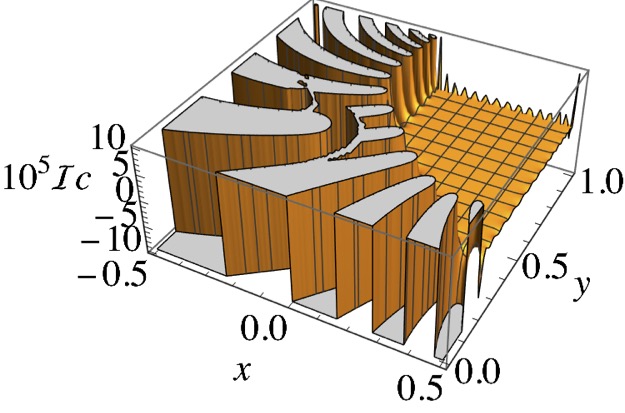}}
\caption{$c(z)$ as in Figs.~\ref{c-Q=5-top}-\ref{c-Q=5-bottom}. Top: full data. Middle: selected data. Bottom: fitting error.}
\label{c-Q=10-bottom}
\Fig{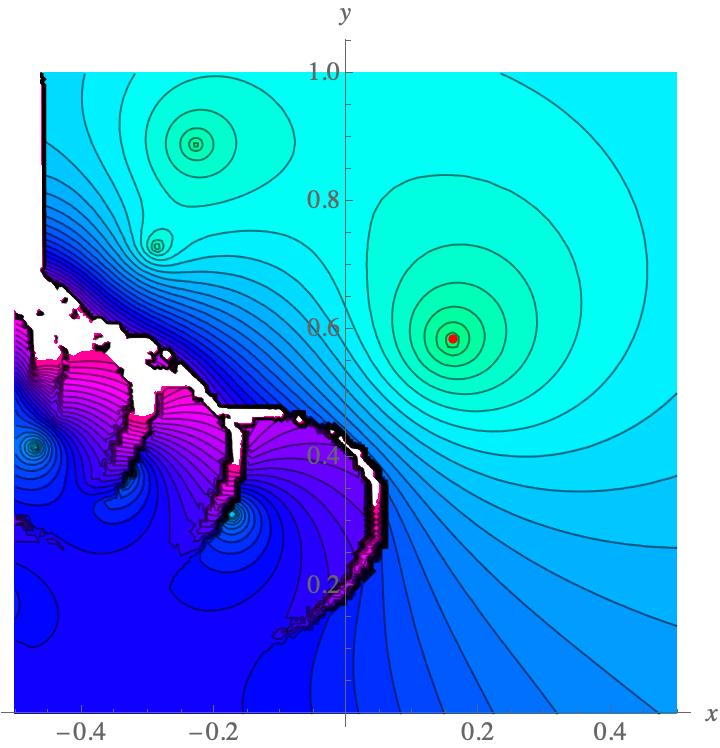}
\caption{Contour plot for $|c(z)-c_{\rm theory}|^{0.3}$ for $Q=10$, $L=10$. Saddle point marked by red dot.}
\label{Q=10grid-c-cSP-contour-L=10}
\end{figure}
Here we repeat for $Q=10$ the 
data analysis performed  in section \ref{s:Q=5} for $Q=5$.  
Fig.~\ref{c-Q=10-bottom} shows our data on the grid; Fig.~\ref{Q=10grid-c-cSP-contour-L=10} the effective central charge $c_L(z)$.
Fig.~\ref{Q=10grid-zeros-of-c-prime-L=10} 
show the location of $c'(z)=0$, Fig.~\ref{Q=10-zc} extrapolations for $z_{\rm c}$ to $L=\infty$, Figs.~\ref{Q=10grid-Re-c-upper} and \ref{Q=10grid-Re-omega-upper} extrapolations for $c$ and $\omega$. 
 The fitting polynomial   consists of  $\{1,x^2,x^3\}$, with $x:=1/L$.

\newpage
\enlargethispage{3cm}
\noindent
Numerical values for $c$ and $\omega$ are compared to theoretical predictions in Figs.~\ref{Q=10grid-Re-c-upper} and \ref{Q=10grid-Re-omega-upper}. The agreement for $c$ is good, while larger sizes would be needed for a better agreement of $\omega$.

\clearpage

\section{$Q=20$}
\label{s:Q=20}
\begin{figure}[h]
\centerline{\fig{0.5}{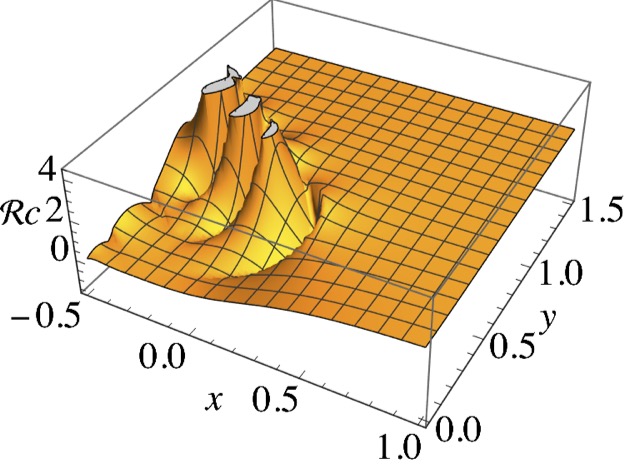}~\fig{0.5}{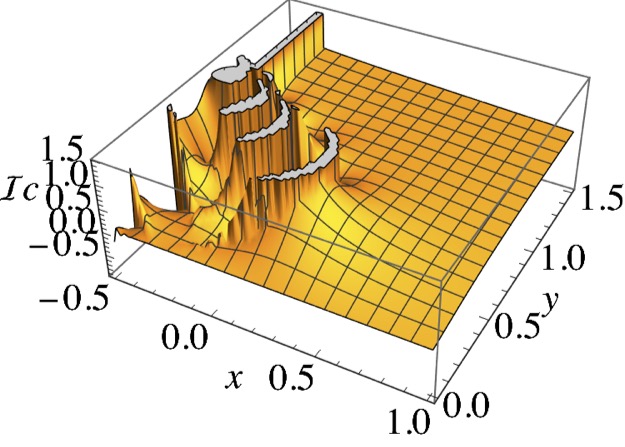}}
\centerline{\fig{0.5}{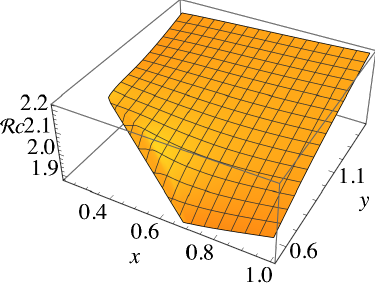}~\fig{0.5}{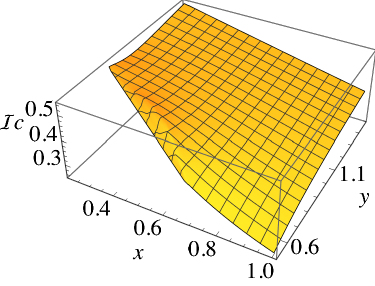}}
\centerline{\fig{0.5}{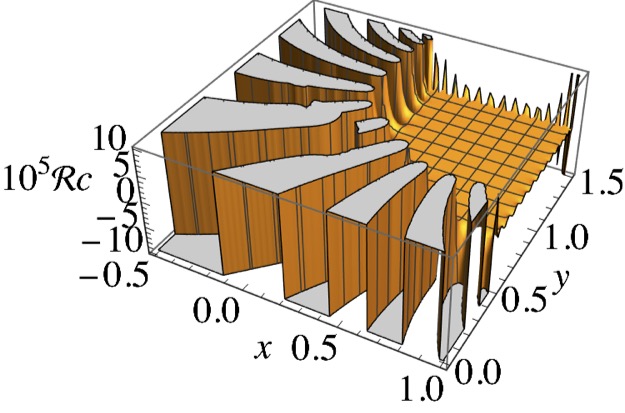}~\fig{0.5}{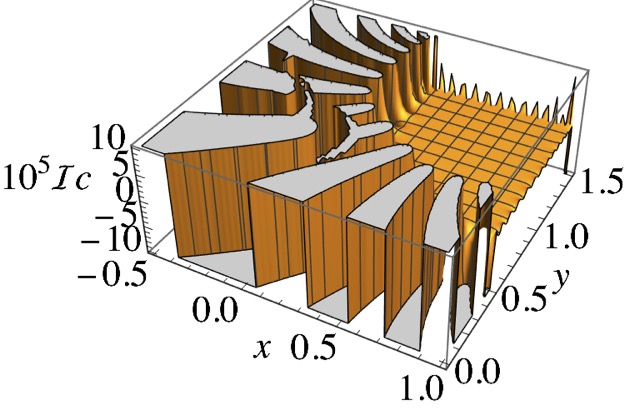}}
\caption{$c(z)$ as in Figs.~\ref{c-Q=5-top}-\ref{c-Q=5-bottom}. Top: full data. Middle: selected data. Bottom: fitting error.}
\label{c-Q=20-bottom}
\Fig{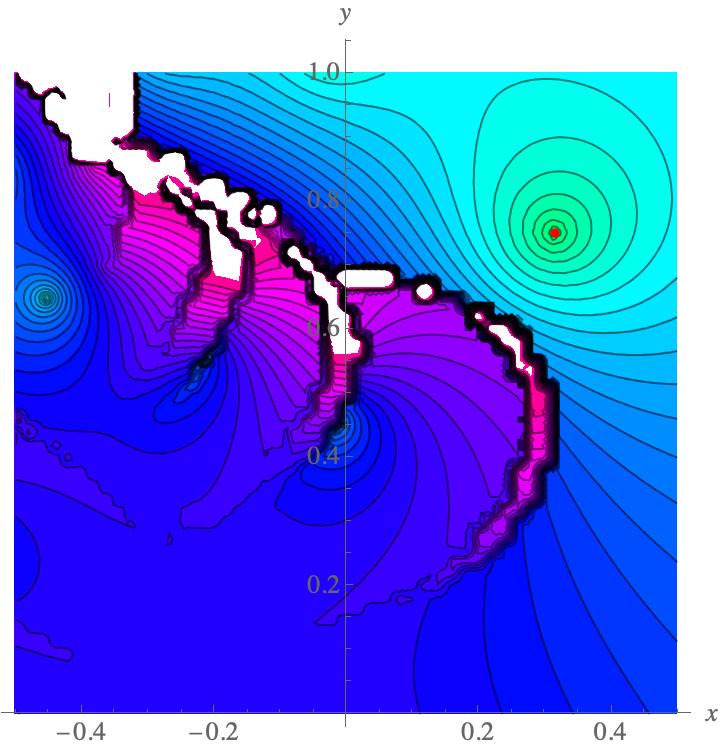}
\caption{Contour plot for $|c(z)-c_{\rm theory}|^{0.3}$ for $Q=20$, $L=8$. Saddle point marked by red dot.}
\label{Q=20grid-c-cSP-contour-L=8}
\end{figure}
\begin{figure}
\fig{0.5}{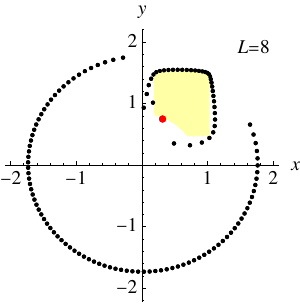}
\caption{Zeros of $c'(z)$ at $L=8$.}\centerline{\fig{0.49}{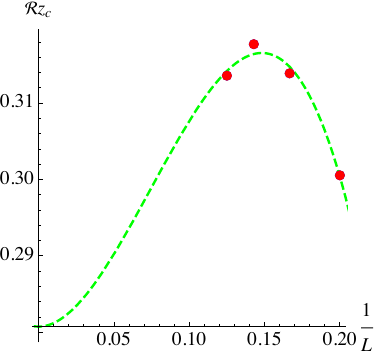}~\fig{0.49}{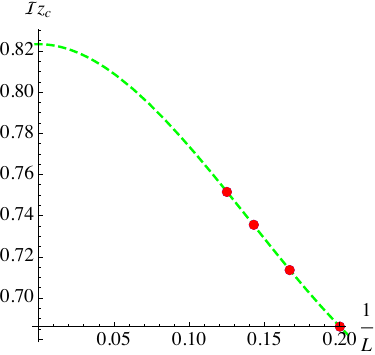}}
\caption{Extrapolation of  $z_{\rm c}$ for $Q=20$.
This gives $ z_{{\rm c},L\to \infty}^{\rm upper} =  0.136932 +0.62623 i. $}
\label{Q=20-zc}
\centerline{\fig{0.49}{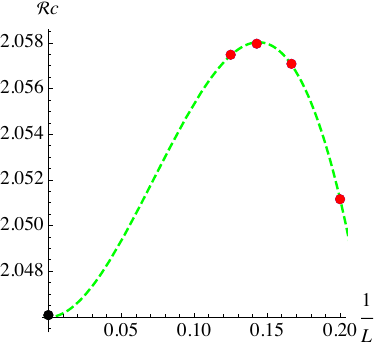}~\fig{0.49}{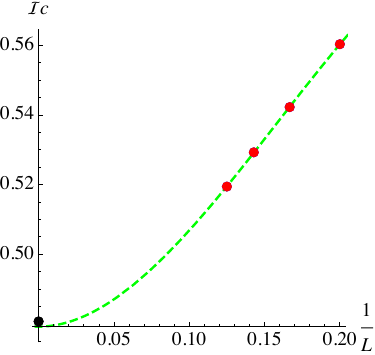}}
\caption{Extrapolation of  $c$ for $Q=20$ yields  $c^{\rm upper}_{L\to \infty} =  1.58465 +0.190826 i$. Compare to $c^{\rm upper}_{\rm ana} = 1.58411 +0.191825 i $.}
\centerline{\fig{0.49}{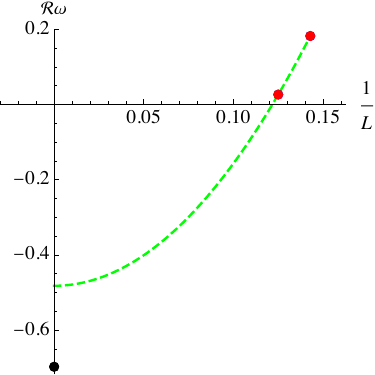}~\fig{0.5}{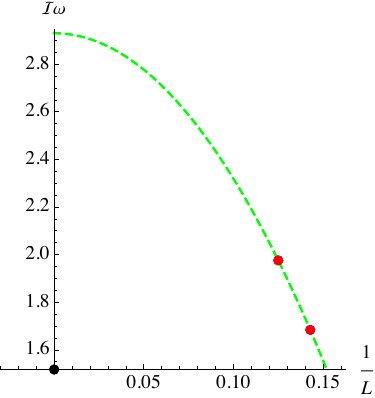}}
\caption{Extrapolation for $\omega$: $\omega_{L\to \infty} = -0.482+2.930i$, $2+\omega =   1.518 +2.930i $, $2 h_{3,1} = 1.303 +1.518 i$.}
\label{Q=20-omega}
\end{figure}

\setlength{\thislength}{6.5mm}
\enlargethispage{\thislength}
Here we repeat for $Q=20$ the 
data analysis performed  in section \ref{s:Q=5} for $Q=5$.  
Fig.~\ref{c-Q=20-bottom} shows the data on the grid, Fig.~\ref{Q=20grid-c-cSP-contour-L=8} the effective central charge $c_L(z)$, and 
Figs.~\ref{Q=20-zc}--\ref{Q=20-omega}  the extrapolations for $z_{\rm c}$, $c$, and $\omega$.
\noindent   
The fitting polynomial in $x:=1/L$  consists of $\{1,x^2,x^3\}$. Extrapolations for $\omega$ are shaky 
\enlargethispage{\thislength} due to the small system size, and the closeness to the singularity visible in Fig.~\ref{Q=20grid-c-cSP-contour-L=8}. \nopagebreak

\clearpage

\section{$Q=40$}
\label{s:Q=40}
\begin{figure}[h]
\centerline{\fig{0.5}{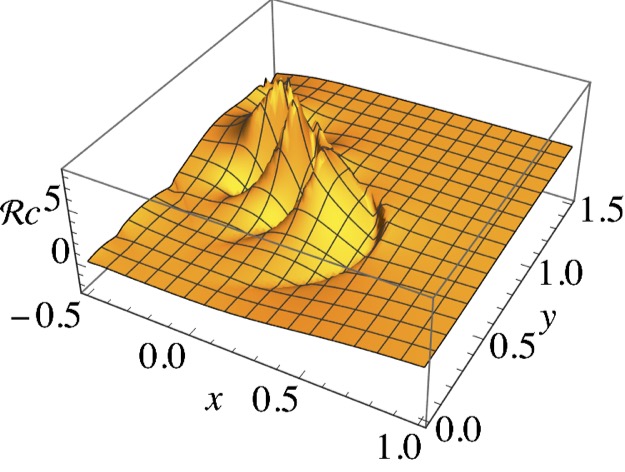}~\fig{0.5}{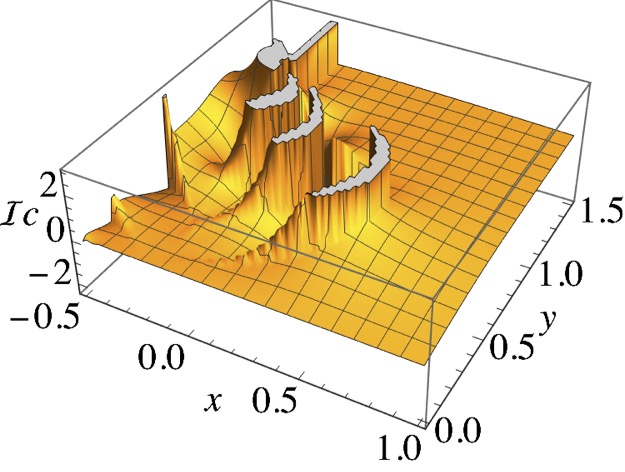}}
\centerline{\fig{0.5}{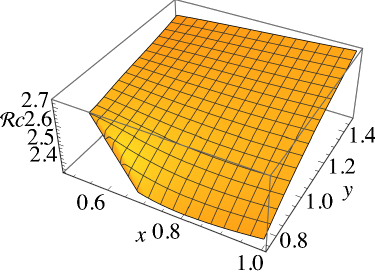}~\fig{0.5}{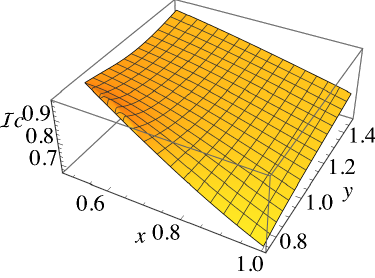}}
\centerline{\fig{0.5}{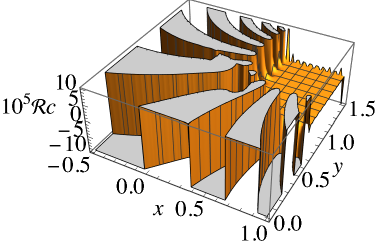}~\fig{0.5}{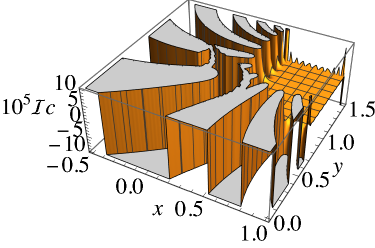}}
\caption{$c(z)$ as in Figs.~\ref{c-Q=5-top}-\ref{c-Q=5-bottom}. Top: full data. Middle: selected data. Bottom: error.}
\label{c-Q=40-bottom}
\Fig{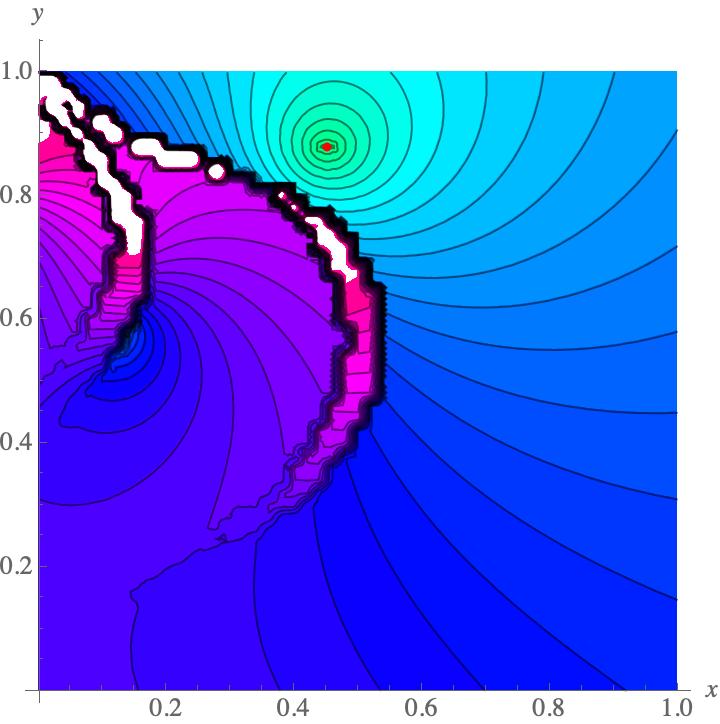}
\caption{Contour plot for $|c(z)-c_{\rm theory}|^{0.3}$ for $Q=40$, $L=7$.  Saddle point marked by red dot.}
\label{Q=40grid-c-cSP-contour-L=7}
\end{figure}
\begin{figure}
\fig{0.5}{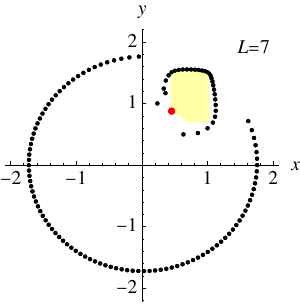}
\caption{Zeros of $c'(z)$ at $L=7$.}
\centerline{\fig{0.49}{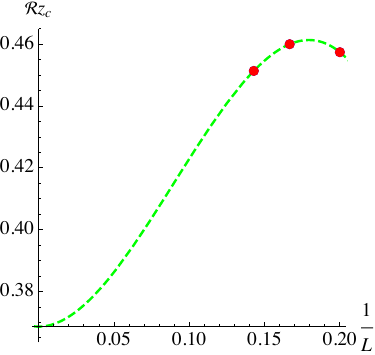}~\fig{0.49}{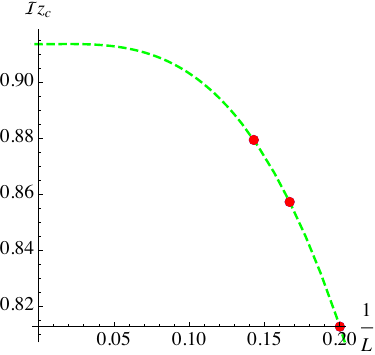}}
\caption{Extrapolation of  $z_{\rm c}$ for $Q=40$.
This gives $ z_{{\rm c},L\to \infty}^{\rm upper} =  0.136932 +0.62623 i. $}
\label{Q=40-zc}
\centerline{\fig{0.49}{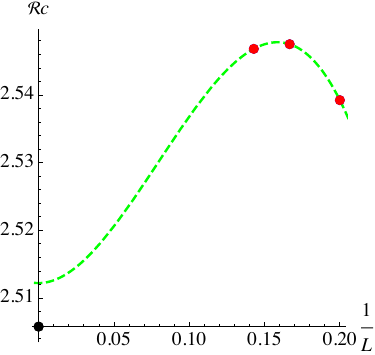}~\fig{0.49}{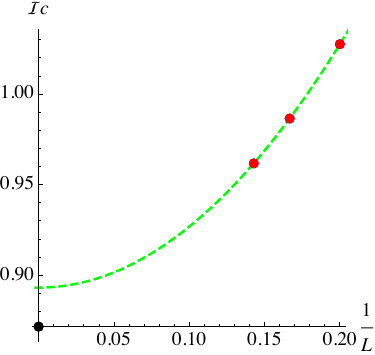}}
\caption{Extrapolation of  $c$ for $Q=40$ yields  $c^{\rm upper}_{L\to \infty} =  2.5122 +0.89312i$. Compare to $c^{\rm upper}_{\rm ana} = 2.50576 +0.871581 i $.}
\centerline{\fig{0.49}{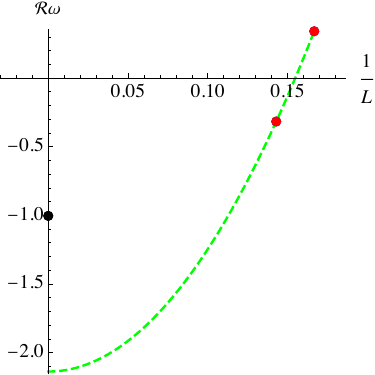}~\fig{0.5}{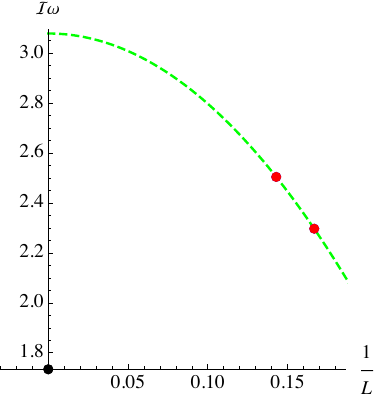}}
\caption{Extrapolation for $\omega$: $\omega_{L\to \infty} = -2.138+3.079 i$,  $2+\omega_{L\to \infty}=-0.138{+}3.079i$, $\Delta_{3,1}=0.996{+}1.734 i$}
\label{Q=40-omega}
\end{figure}

\setlength{\thislength}{6.5mm}
\enlargethispage{\thislength}
Here we repeat for $Q=40$ the 
data analysis performed  in section \ref{s:Q=5} for $Q=5$.  
Fig.~\ref{c-Q=40-bottom} shows the data on the grid, Fig.~\ref{Q=40grid-c-cSP-contour-L=7} the effective central charge $c_L(z)$, and 
Figs.~\ref{Q=40-zc}--\ref{Q=40-omega}  the extrapolations for $z_{\rm c}$, $c$, and $\omega$.
\noindent 
The fitting polynomial   contains $\{1,x^2,x^3\}$, with $x:=1/L$. Extrapolations for $\omega$ are shaky 
\enlargethispage{\thislength} due to the small system size, and the closeness to the singularity visible in Fig.~\ref{Q=40grid-c-cSP-contour-L=7}. \nopagebreak
\clearpage

\clearpage


\section{$Q=5+2i$}
\label{s:Q=5+2i}

\subsection{Grid}
\label{s:Q=5+2i:Grid}
\begin{figure}[h!]
\centerline{\fig{0.5}{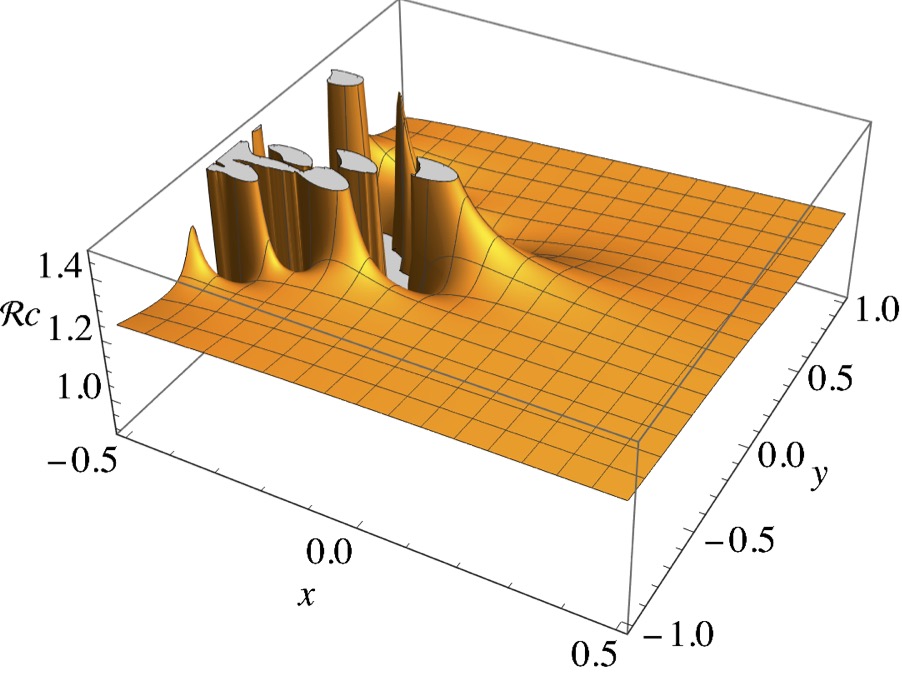}\hfill\fig{0.5}{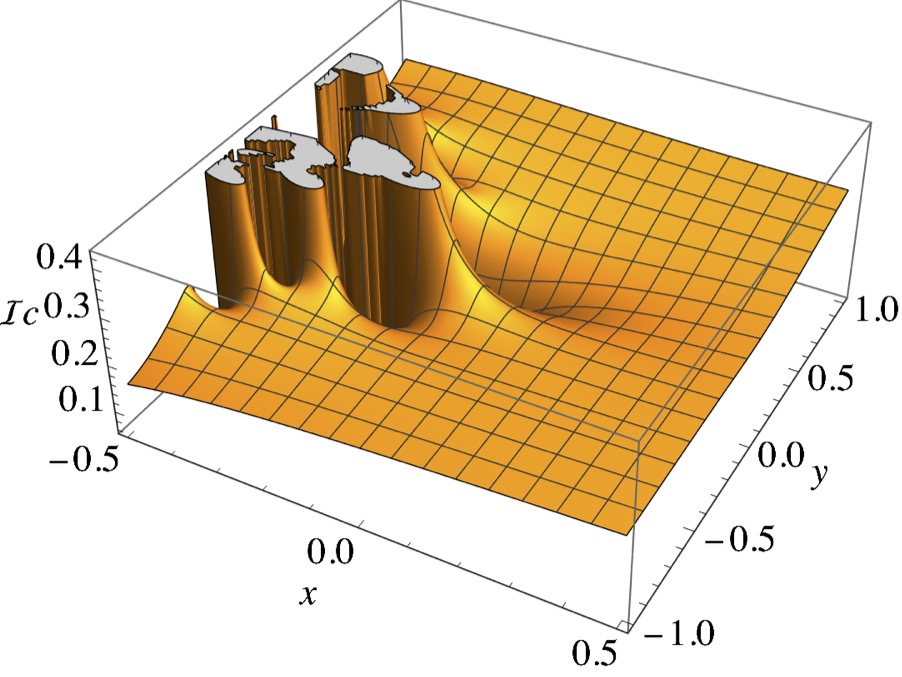}}
\caption{$c$, $L=10$,  raw data.}
\end{figure}
\begin{figure}[h!]
\centerline{\fig{0.5}{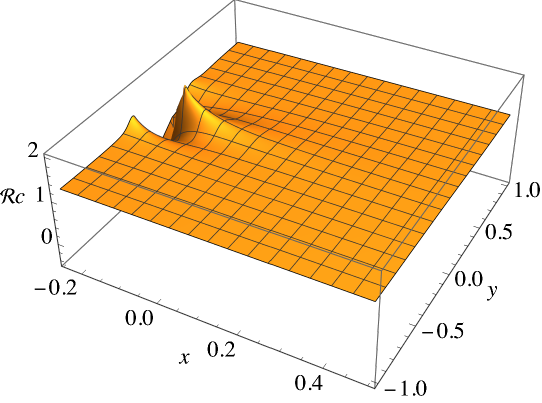}\hfill\fig{0.5}{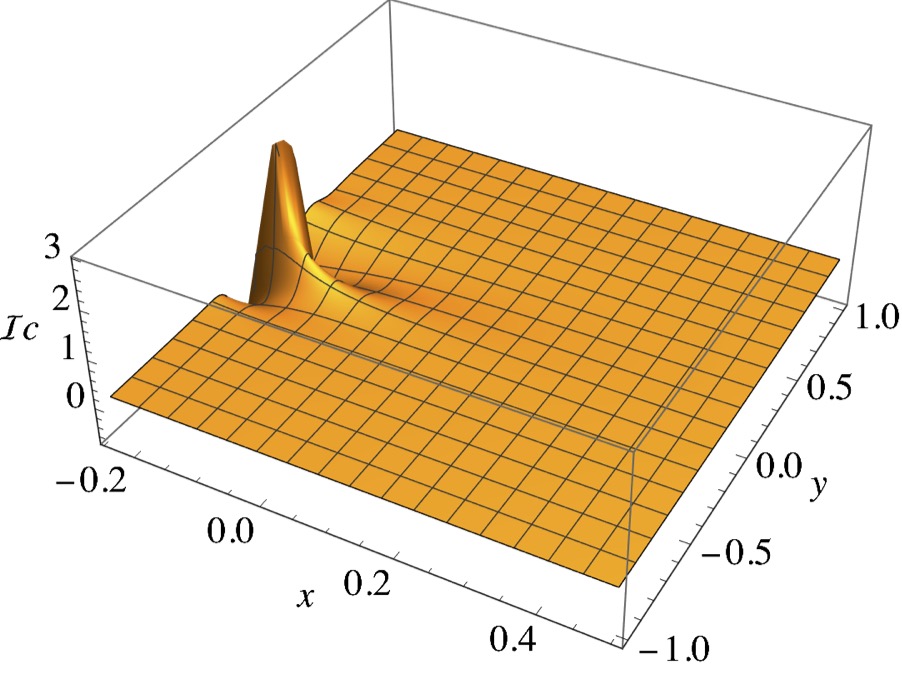}}
\caption{$c$,  $L=10$, selected data.}
\end{figure}
\begin{figure}[h!]
\centerline{\fig{0.5}{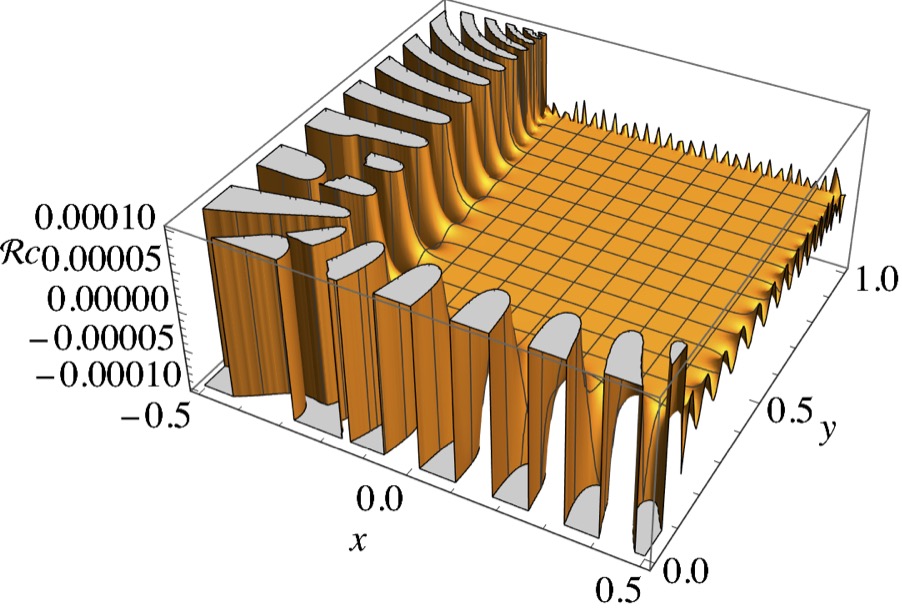}\hfill\fig{0.5}{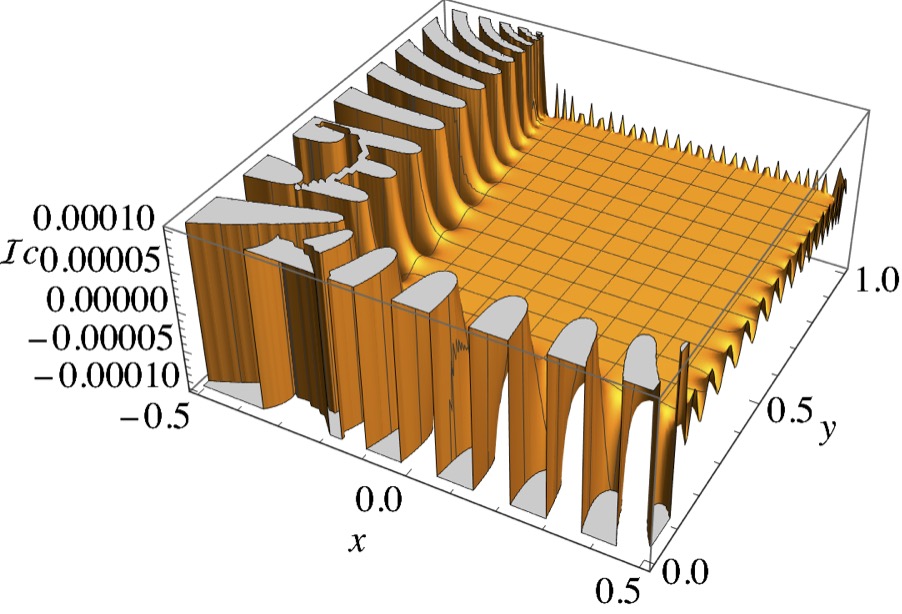}}
\caption{$c$, $L=10$, subtracted data, upper half plane.}
\end{figure}
\begin{figure}[h!]
\centerline{\fig{0.5}{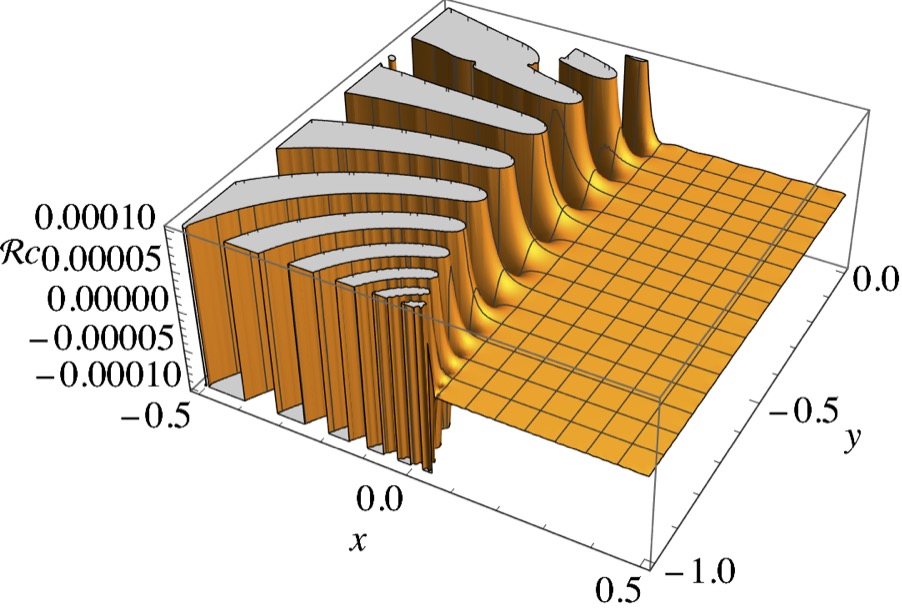}\fig{0.5}{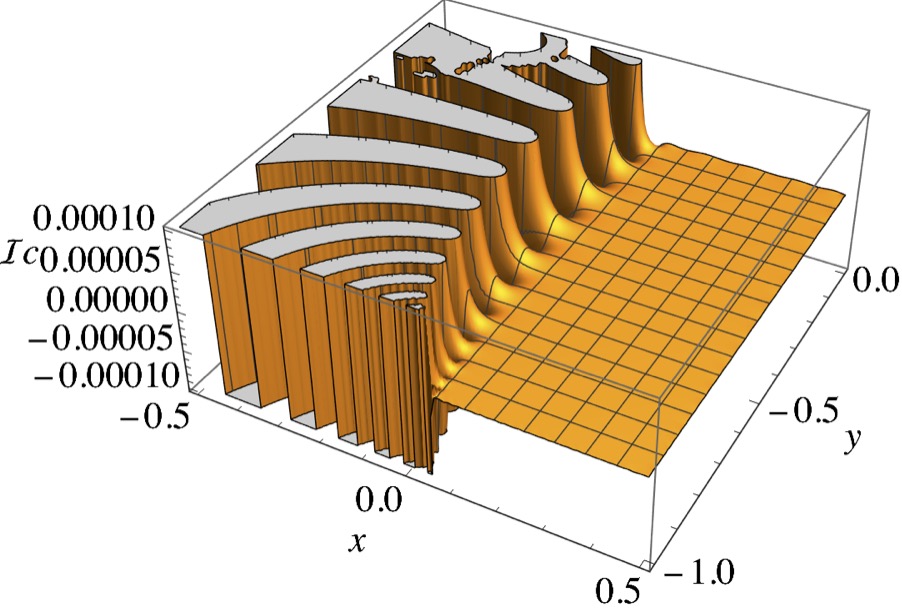}}
\caption{$c$,  $L=10$,  subtracted data, lower half plane.}
\end{figure}

\begin{figure}[h!]
\centerline{\fig{0.72}{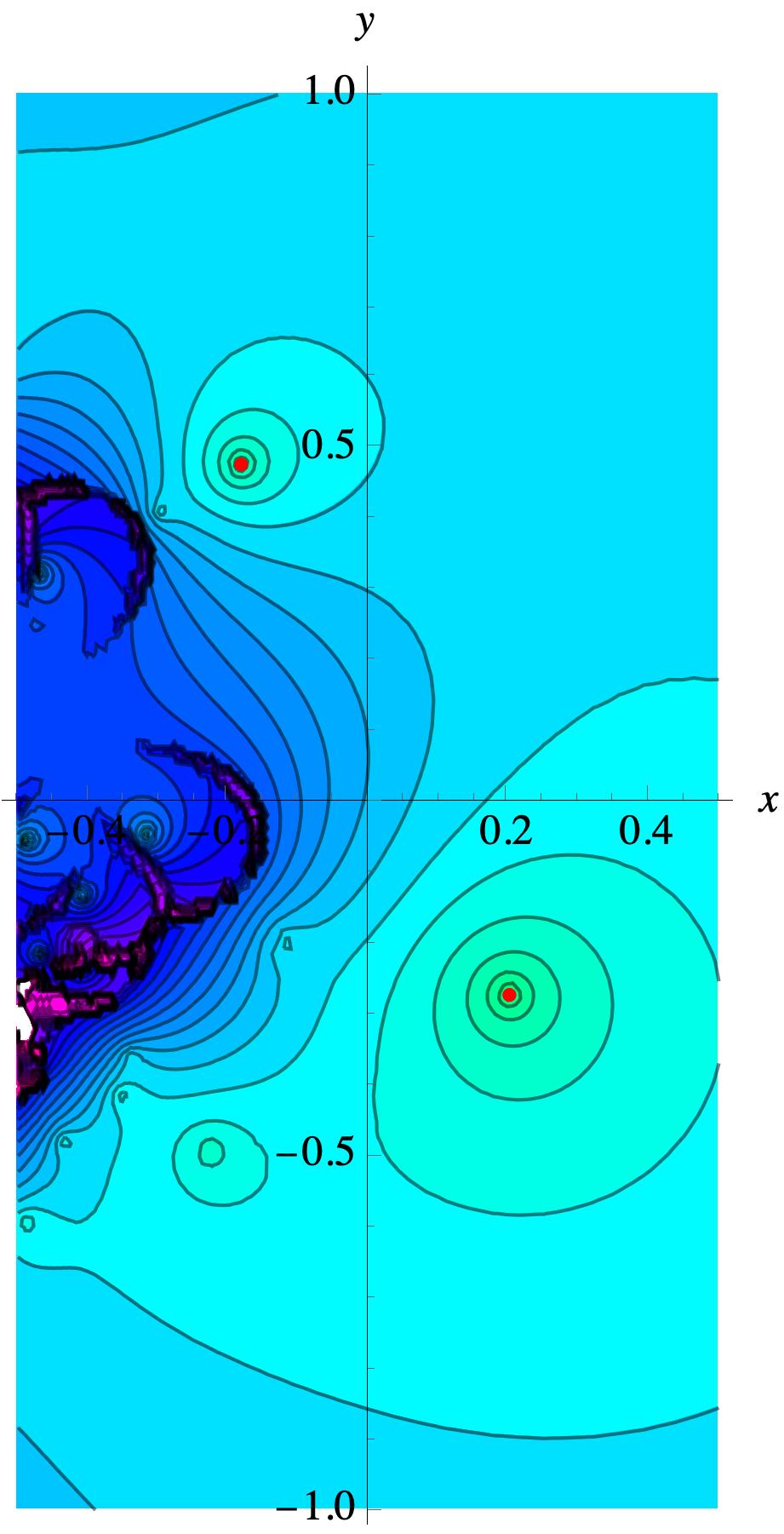}}
\caption{Contour plot for $|c(z)-c_{\rm theory}|^{0.3}$ for $Q=5+2i$, $L=10$.  Saddle points $c'(z)=0$ are marked by a red dot. The level-crossing lines are clearly visible.}
\label{Q=5+2i-grid-c-cSP-contour-L=10.jpg}
\end{figure}

\begin{figure}[h!]
\centerline{\fig{0.63}{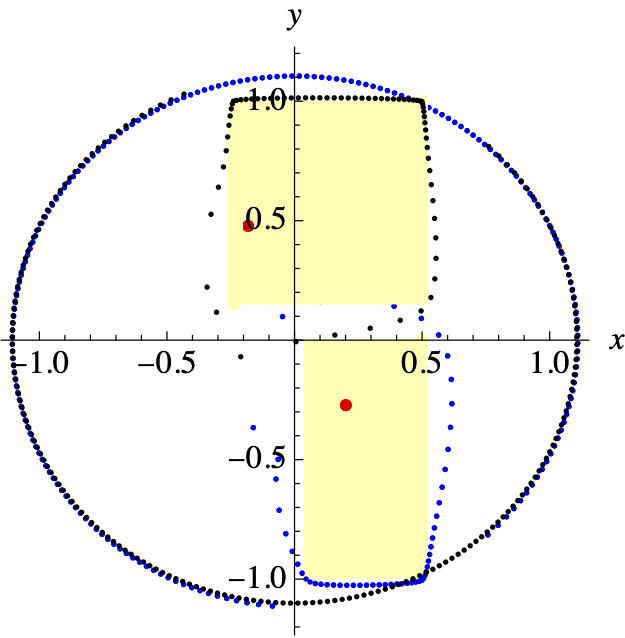}}
\caption{Zeros of $c'(z)$ at $L=10$. The yellow areas denote the points used to generate the fitting polynomials for upper and lower half plane.}
\label{Q=5+2i-grid-zeros-of-c-prime-L=10}
\end{figure}

The contour plot \ref{Q=5+2i-grid-c-cSP-contour-L=10.jpg} gives a good idea about the location of the singularities. 
Note that the  two non-trivial fixed points marked by the red dots in Fig.~\ref{Q=5+2i-grid-c-cSP-contour-L=10.jpg} are not complex conjugate of each other. The precise locations of $c'(z)=0$ are shown in Fig.~\ref{Q=5+2i-grid-zeros-of-c-prime-L=10}; again the non-trivial ones are   in red. 

\newpage

\noindent

\noindent 
Extrapolations for $z_c$, $c$ and $\omega$ are given on the next page, separately for the two fixed points. 

\newpage
\enlargethispage{2cm}
\leftline{\bf Fixpoint $B_{-b}$ (upper half plane)}
\noindent Fits for upper half plane. The fitting polynomial always contains $\{1,x^2,x^3\}$, with $x:=1/L$.
\begin{figure}[h!]
\centerline{\fig{.5}{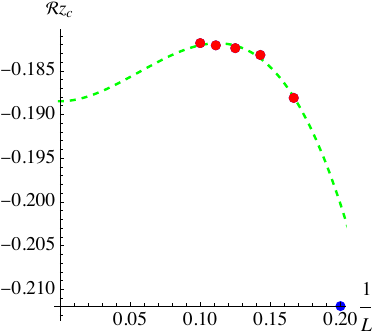}\hfill\fig{0.48}{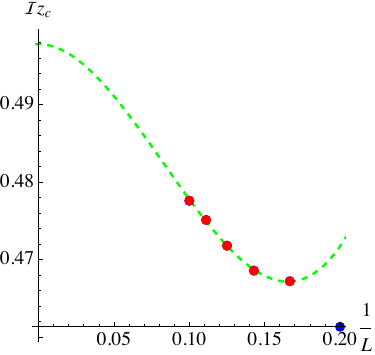}}
\caption{Extrapolation of $z_c$.}
\label{Q=5+2i-grid-Re-zc-upper}
\end{figure}
\be
  z_c^{\rm upper}({L\to \infty})=  -0.188476+0.497827  i. 
\ee
\begin{figure}[h!]
\centerline{\fig{0.5}{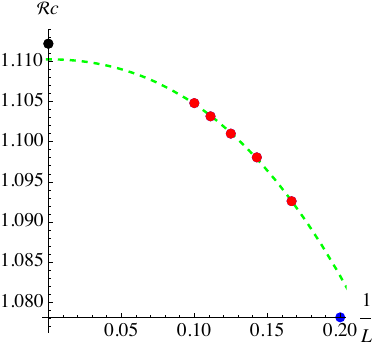}\hfill\fig{0.5}{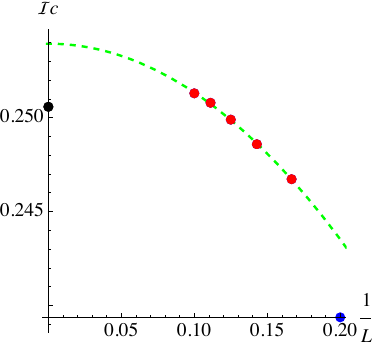}}
\caption{Extrapolation of $c$.}
\end{figure}
\bea
  c^{\rm upper}_{L\to \infty} &=& 1.11029 +0.253942 i \\
  c^{\rm upper}_{\rm ana} &=&1.11224 +0.250583 . 
\eea
\begin{figure}[h!]
\centerline{\fig{.5}{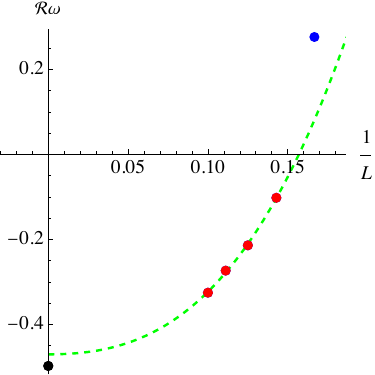}\hfill\fig{0.5}{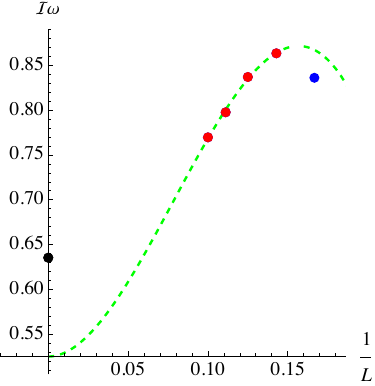}}
\caption{Extrapolation of $\omega$.}
\label{Q=5+2i-grid-Re-omega-upper}
\end{figure}
\bea
  \omega^{\rm upper}_{L\to \infty} &=&  -0.470446+0.525107   i, \\
 \Delta^{\rm upper}_{L\to \infty} &=&2+ \omega^{\rm upper}_{L\to \infty} \nn\\
 &=& 1.52955 +0.525107 i \\
 {2 h_{3,1}}&=& 1.50219 +0.635388 i.
\eea

\newpage 

\enlargethispage{2cm}
\leftline{\bf Fixpoint $B_{b}$ (lower half plane)}
\noindent Fits for lower half plane as for upper half plane (left). 
\begin{figure}[h!]
\centerline{\fig{.48}{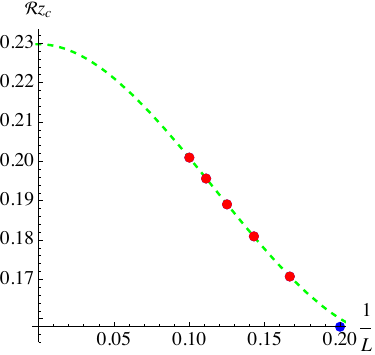}\hfill\fig{.5}{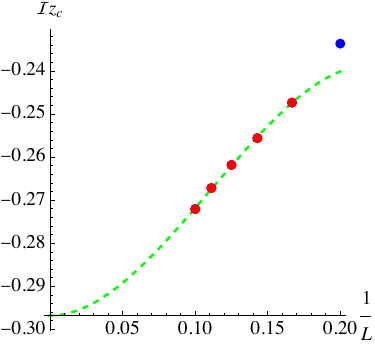}}
\caption{Extrapolation of $z_c$.}
\label{Q=5+2i-grid-Re-zc-lower}
\end{figure}
\be
  z_c^{\rm lower}({L\to \infty})= 0.229789 -0.296768 i. 
\ee
\begin{figure}[h!]
\centerline{\fig{.5}{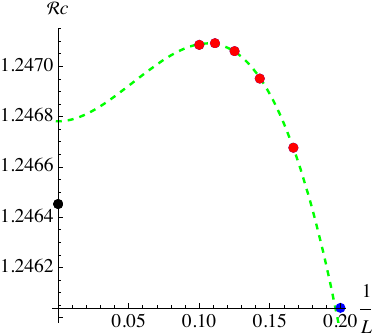}\hfill\fig{.5}{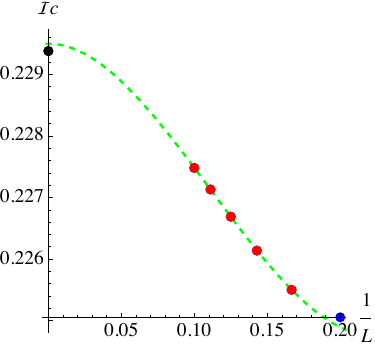}}
\caption{Extrapolation of $c$.}
\end{figure}
\bea
  c^{\rm lower}_{L\to \infty}&=&  1.24678 +0.2295 i . \\
  c^{\rm lower}_{\rm ana} &=& 1.24645 +0.22938 i
\eea
\begin{figure}[h!]
\centerline{\fig{.5}{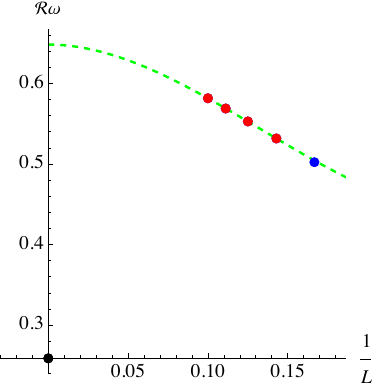}\hfill\fig{.5}{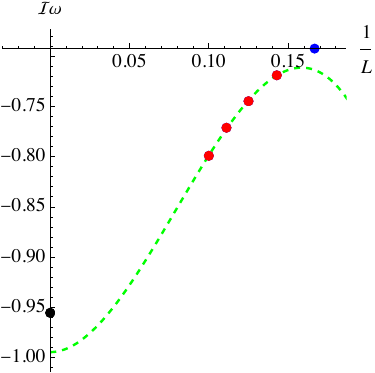}}
\caption{Extrapolation of $\omega$.}
\label{Q=5+2i-grid-Re-omega-lower}
\end{figure}
\bea
  \omega^{\rm lower}_{L\to \infty}&=&  0.648385 -0.994432  i\\
 \Delta^{\rm lower}_{L\to \infty}&=& 2+\omega^{\rm lower}_{L\to \infty} \nn\\
 &=&  2.64838 -0.994432 i \\
{ 2 h_{3,1}} &=& 2.25868 -0.955363  i.
\eea

\clearpage

\subsection{Minigrid, upper half plane}
\label{Q=5+2i on minigrid, upper half plane}
\begin{figure}[h!]
\centerline{\fig{0.6}{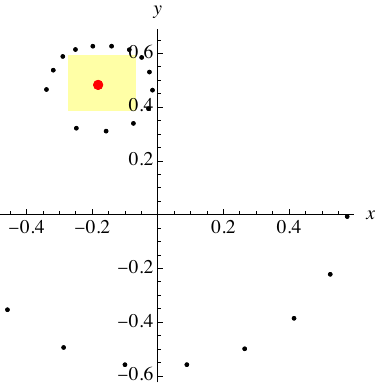}}
\caption{Zeros of $c'(z)$, $L=15$.}
\label{Q=5+2i-UHP-minigrid1-zeros-of-c-prime-L=15}
\end{figure}

Here we use a minigrid of $5 \times 5$ points, centered around the non-trivial fixed point identified in the preceding section. In Fig.~\ref{Q=5+2i-UHP-minigrid1-zeros-of-c-prime-L=15} 
we show the area covered by it, as well as the zeroes of the fitting polynomial, for our largest system sizes, $L=15$. The non-trivial solution marked in red is clearly visible. 

\begin{figure}[h!]
\centerline{\fig{.5}{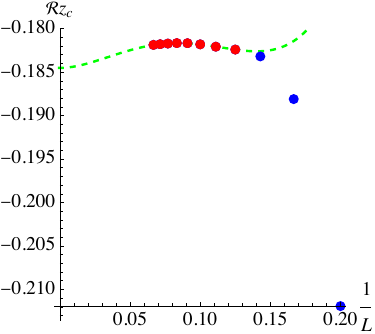}\hfill\fig{.5}{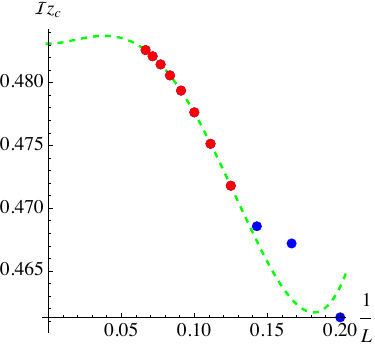}}
\caption{Extrapolation of $z_c$.}
\label{Q=5+2i-minigrid-Re-zc-upper}
\end{figure}

\begin{figure}[h!]
\centerline{\fig{.49}{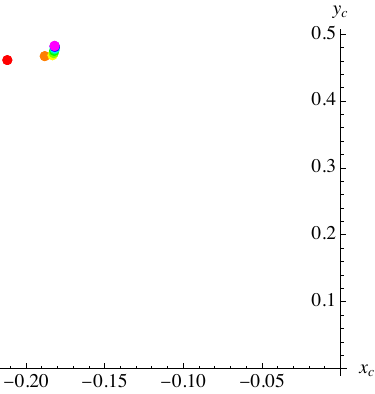}%
\hfill\fig{.51}{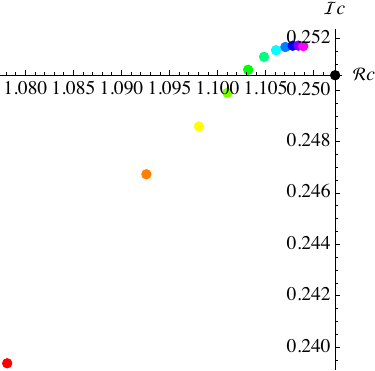}%
}
\caption{Movement of $z_c$ (left) and $c$ (right) with $L$ from $L=4/5$ (red)
to $L=15/16$ (magenta). The black dot is the analytic result.}
\label{Q=5+2i-minigrid-zeros-of-c-prime-upper}
\end{figure}

\noindent
In Figs.~\ref{Q=5+2i-minigrid-Re-zc-upper} to \ref{Q=5+2i-minigrid-zeros-of-c-prime-upper} we show fits using a  polynomial   with $\{1,x^2,x^3\}$,  $x:=1/L$.

\noindent
Extrapolation yields
\be
  z_c^{\rm upper}({L\to \infty})=  -0.184504+0.483112 i  . 
\ee

\begin{figure}[h!]
\centerline{\fig{.5}{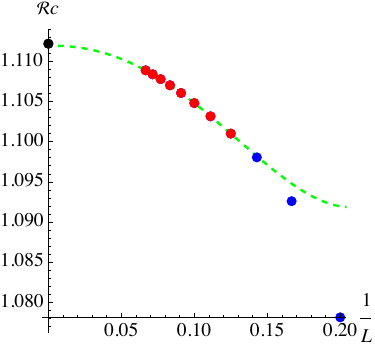}\hfill\fig{.5}{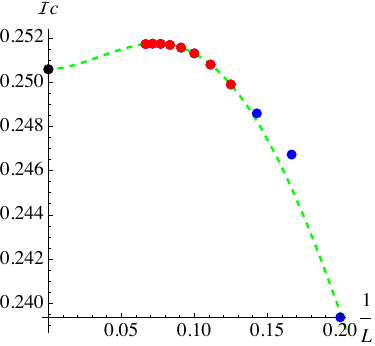}}
\caption{Extrapolation of $ c$.}
\end{figure}

\noindent
For the central charge we find from extrapolation and analytics
\bea \nn
  c^{\rm upper}_{L\to \infty} &=& 1.11201 +0.250589 i\\
  c^{\rm upper}_{\rm ana} &=&1.11224 +0.250583 i. 
\label{D14}
\eea
The agreement is excellent.

\begin{figure}[h!]
\centerline{\fig{.5}{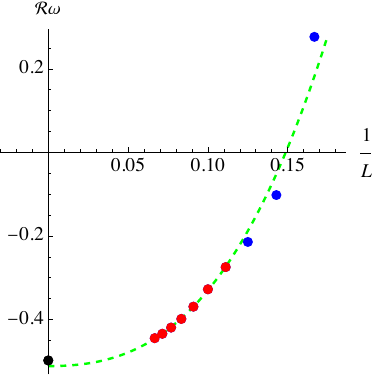}\hfill\fig{.5}{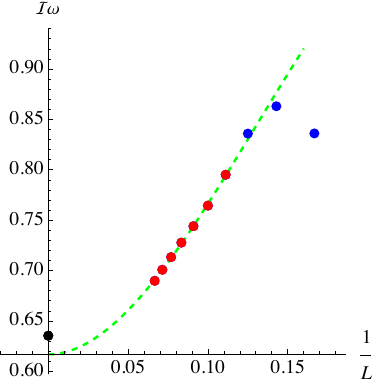}}
\caption{Extrapolation of $\omega$.}
\end{figure}
\bea
\label{omega-Q=5+2i-unstable}
  \omega^{\rm upper}_{L\to \infty} &=& -0.511328+0.616934 i , \\
 \Delta^{\rm upper}_{L\to \infty} &=&2+ \omega^{\rm upper}_{L\to \infty} \nn\\
 &=& 1.48867 +0.61693 i \label{D16} \\
 {2 h_{3,1}}&=& 1.50219 +0.63539 i.
 \label{D17}
\eea

\vspace*{3cm}

\newpage

\subsection{Minigrid, lower half plane}
\label{s:5+2i:Minigrid, lower half plane}
\begin{figure}[h!]
\centerline{\fig{.6}{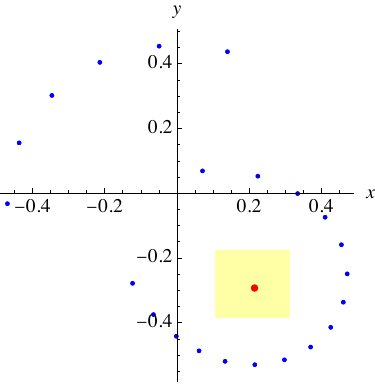}}
\caption{Zeros of $c'(z)$, $L=15$.}
\label{Q=5+2i-LHP-minigrid1-zeros-of-c-prime-L=15}
\end{figure}

As in subsection \ref{Q=5+2i on minigrid, upper half plane}, 
 we use a minigrid of $5 \times 5$ points, centered around the non-trivial fixed point identified in the preceding section. In Fig.~\ref{Q=5+2i-LHP-minigrid1-zeros-of-c-prime-L=15} 
we show the area covered by it, as well as the zeroes of the fitting polynomial, for our largest system sizes $L=15/16$. The non-trivial solution marked in red is clearly visible. 

\enlargethispage{1cm}

\begin{figure}[h!]
\centerline{\fig{.5}{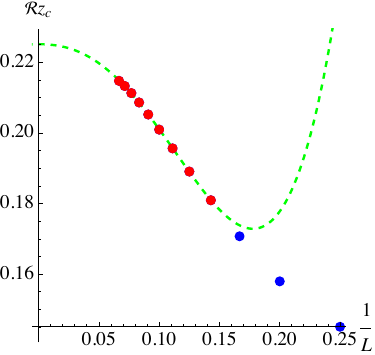}\hfill\fig{.5}{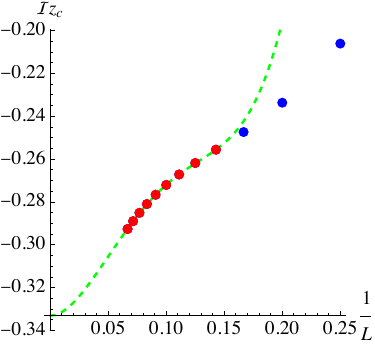}}
\caption{Extrapolation of $z_c$.}
\label{Q=5+2i-minigrid-Re-zc-lower}
\end{figure}

\begin{figure}[h!]
\centerline{\fig{.5}{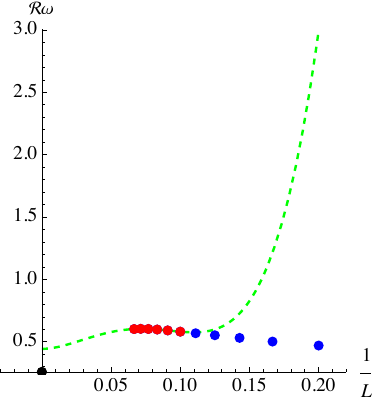}\hfill\fig{.5}{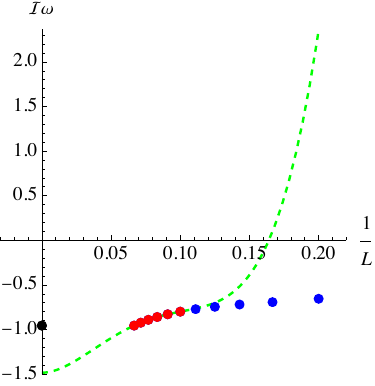}}
\caption{Extrapolation of $\omega$.}
\label{omega-5+2i-lower}
\end{figure}

\begin{figure}[h!]
\centerline{\fig{.52}{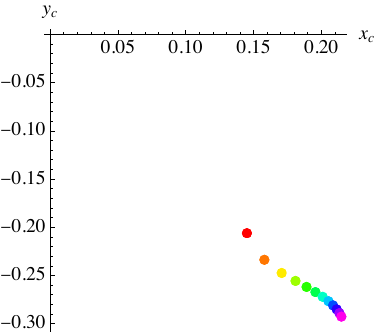}%
\hfill\fig{.475}{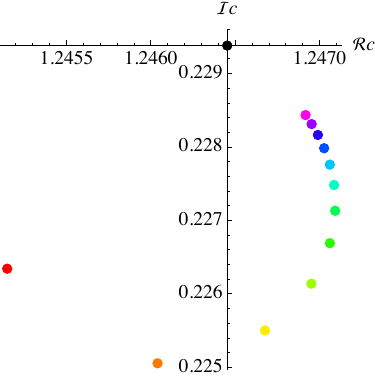}}%
\caption{Movement of $z_c$ (left) and $c$ (right) with $L$ from $L=4$ (red)
to $L=15$ (magenta). The black dot is the analytic result.}
\label{Q=5+2i-minigrid-zeros-of-c-prime-lower}
\end{figure}

In Figs.~\ref{Q=5+2i-minigrid-Re-zc-lower} to \ref{Q=5+2i-minigrid-zeros-of-c-prime-lower} we show fits using a  polynomial in $x:=1/L$   with monomials $\{1,x^2,x^3\}$.

\be
  z_c^{\rm lower}({L\to \infty})=  0.225171 -0.333031 i. 
\ee
\begin{figure}[h!]
\centerline{\fig{.5}{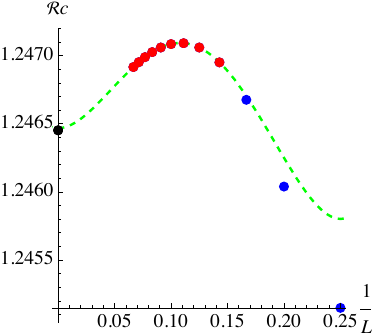}\hfill\fig{.5}{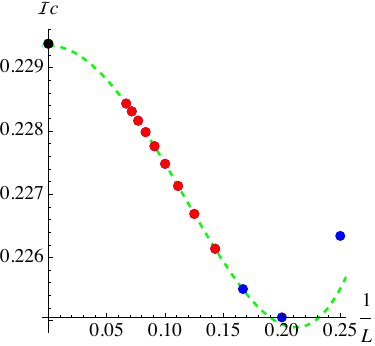}}
\caption{Extrapolation of $ c$.}
\end{figure}
\newpage

\noindent
For the central charge we find from extrapolation and analytics
\bea\nn
  c^{\rm lower}_{L\to \infty} &=& 1.24647 +0.22935 i \\
  c^{\rm lower}_{\rm ana} &=&1.24645 +0.22938 i . 
  \label{D19}
\eea
The agreement is excellent.

\bea
\label{omega-Q=5+2i-stable}
  \omega^{\rm lower}_{L\to \infty} &=&   0.443876 -1.48309 i, \\
 \Delta^{\rm lower}_{L\to \infty} &=&2+ \omega^{\rm lower}_{L\to \infty} \nn\\
 &=& 2.44388 -1.48309 i   
 \label{D21}
 \\
 {2 h_{3,1}}&=& 2.25868 -0.955363 i.
 \label{D22}
\eea

\clearpage

\section{$Q=8+i$}
\label{s:Q=8+i}
\subsection{Grid}
\label{s:Q=8+i:grid}
\begin{figure}[h!]
\centerline{\fig{.5}{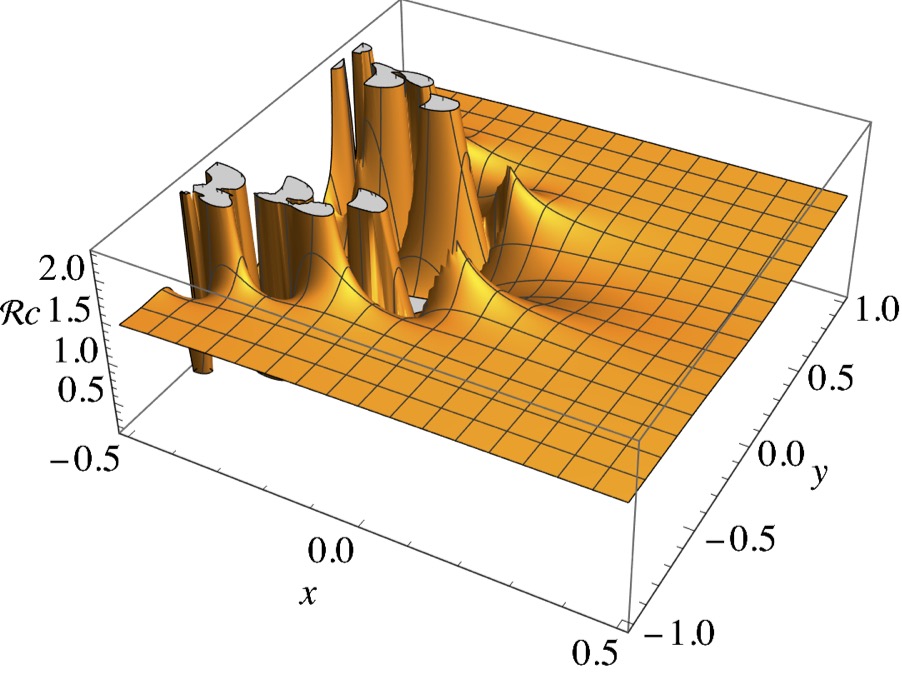}\fig{.5}{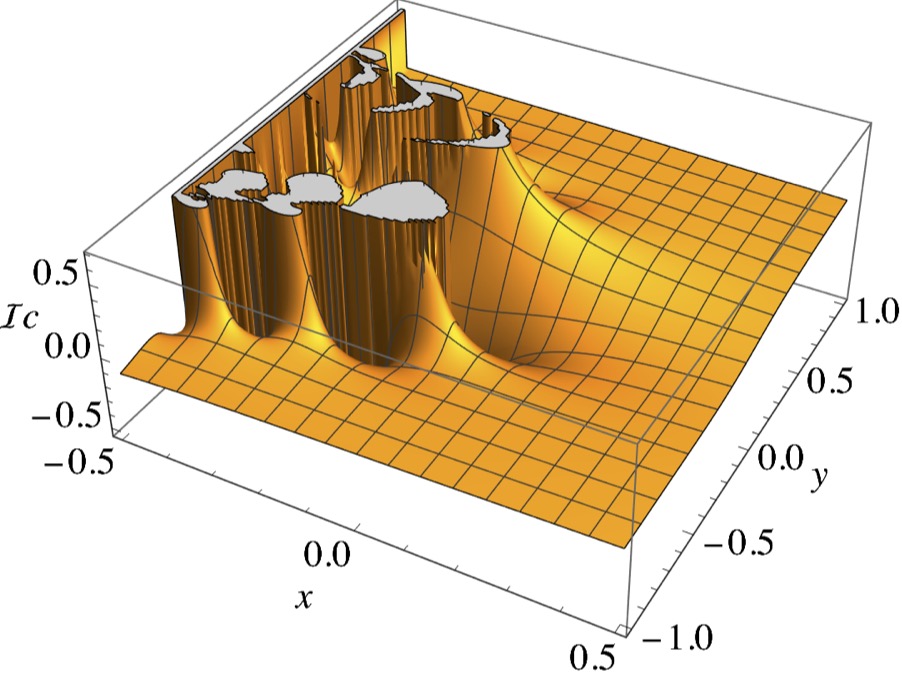}}
\caption{$c$ at $L=10$, raw data.}
\label{Q=8+i-grid-Re-c-L=10-raw}
\end{figure}

\begin{figure}[h!]
\centerline{\fig{0.5}{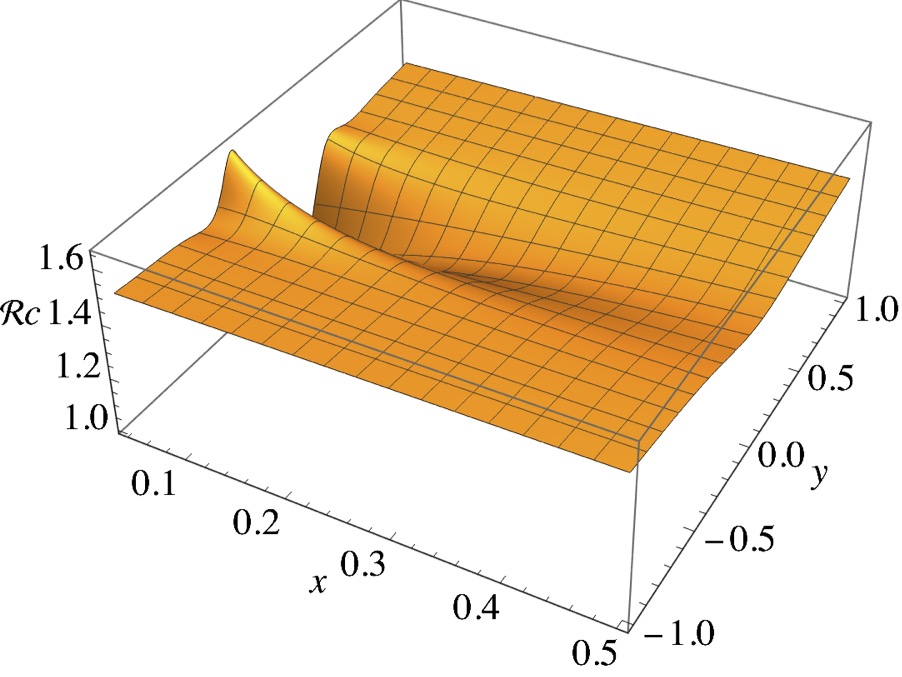}\hfill\fig{0.5}{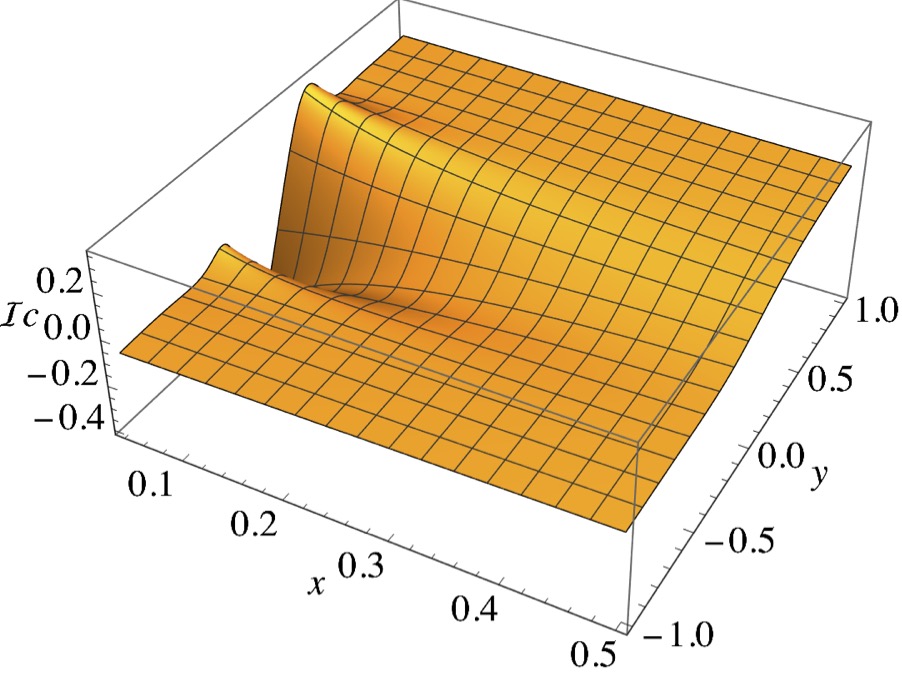}}
\caption{$c$ at $L=10$, selected data.}
\end{figure}

\begin{figure}[h!]
\centerline{\fig{0.5}{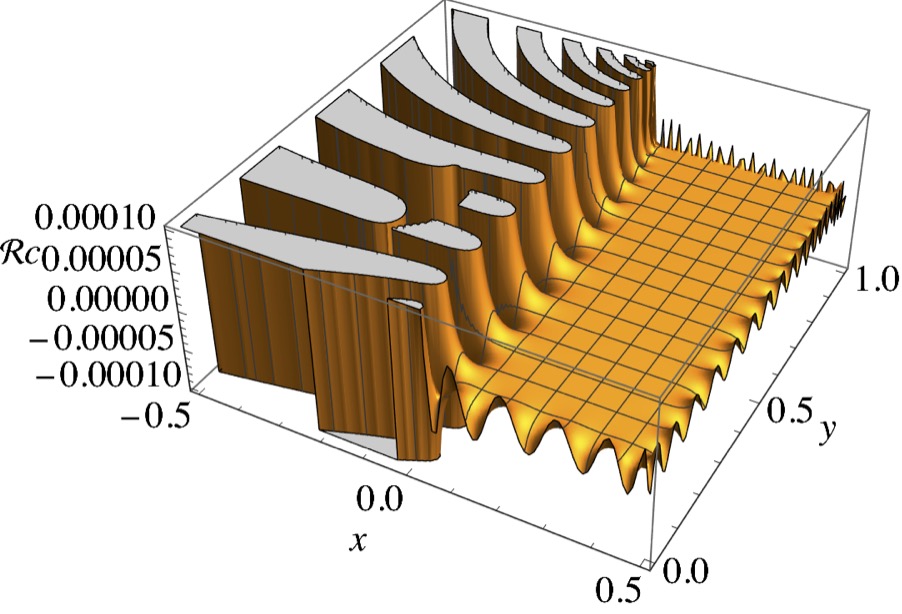}\hfill\fig{0.5}{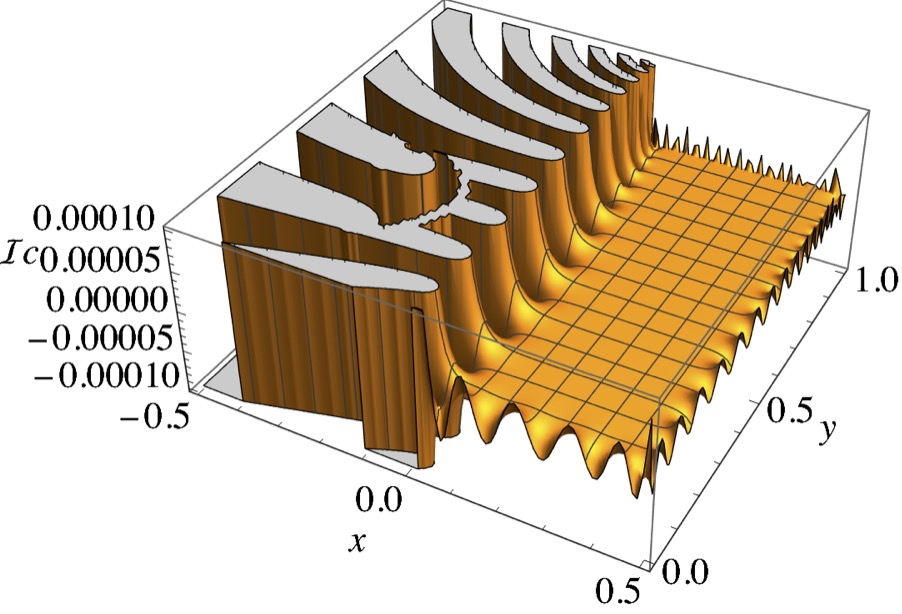}}
\caption{$c$, $L=10$, subtracted, upper half plane.}
\end{figure}
\begin{figure}[h!]
\centerline{\fig{0.5}{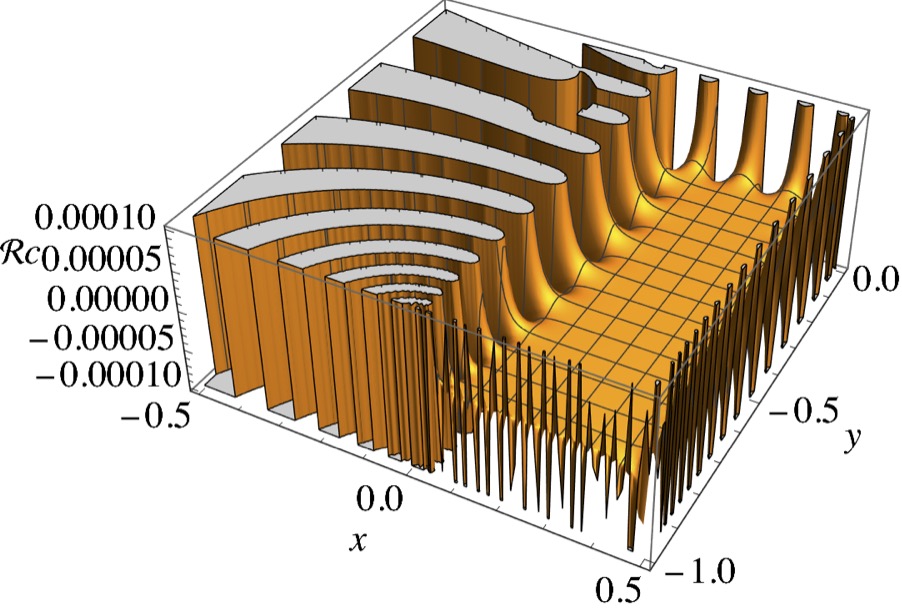}\fig{0.5}{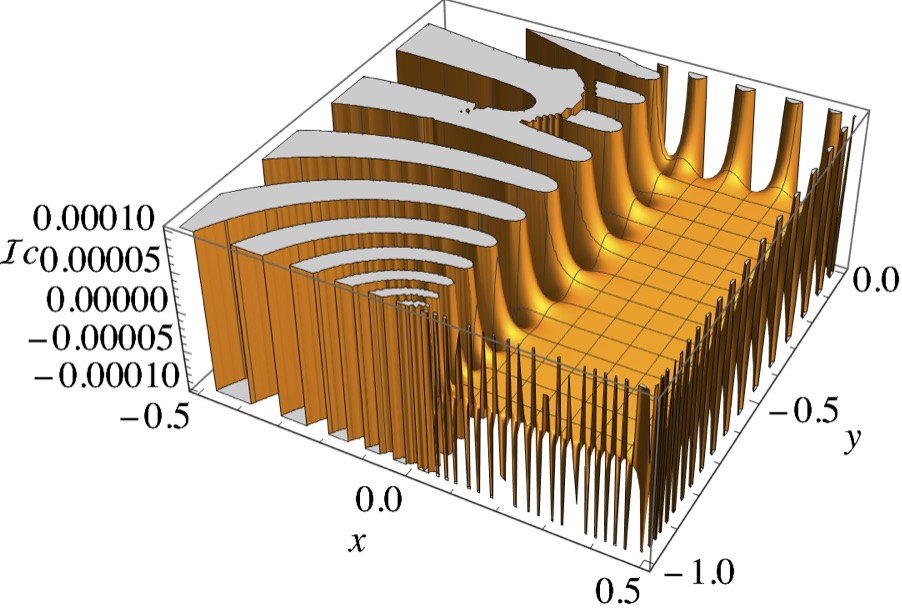}}
\caption{$c$,  $L=10$,  subtracted, lower half plane.}
\end{figure}

\begin{figure}[h!]
\centerline{\fig{0.71}{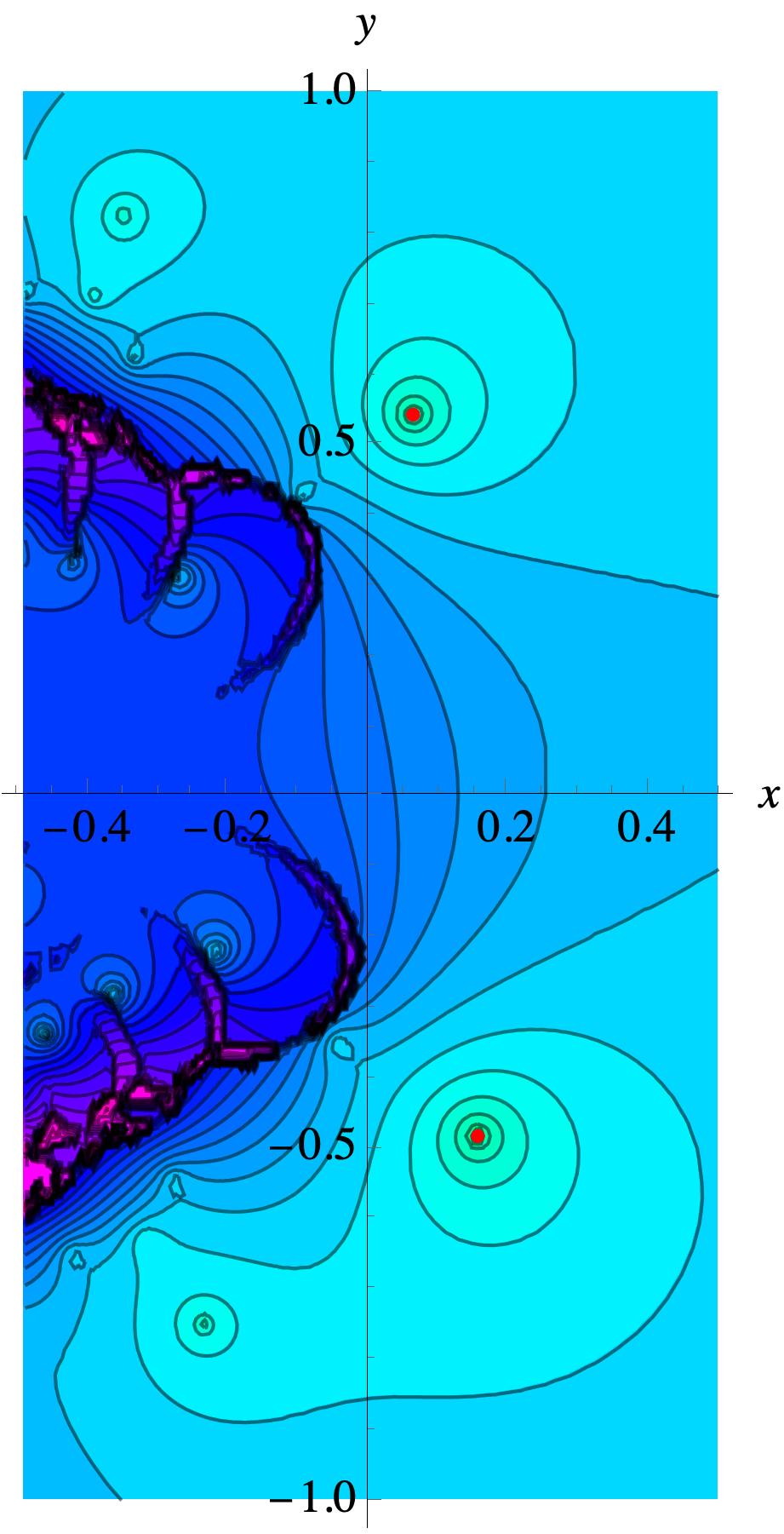}}
\caption{Contour plot for $|c(z)-c_{\rm theory}|^{0.3}$ for $Q=8+i$, $L=10$.  Saddle points $c'(z)=0$ are marked by a red dot. The level-crossing lines are clearly visible}
\label{Q=8+i-grid-c-cSP-contour-L=10.jpg}
\end{figure}

\begin{figure}[h!]
\centerline{\fig{0.6}{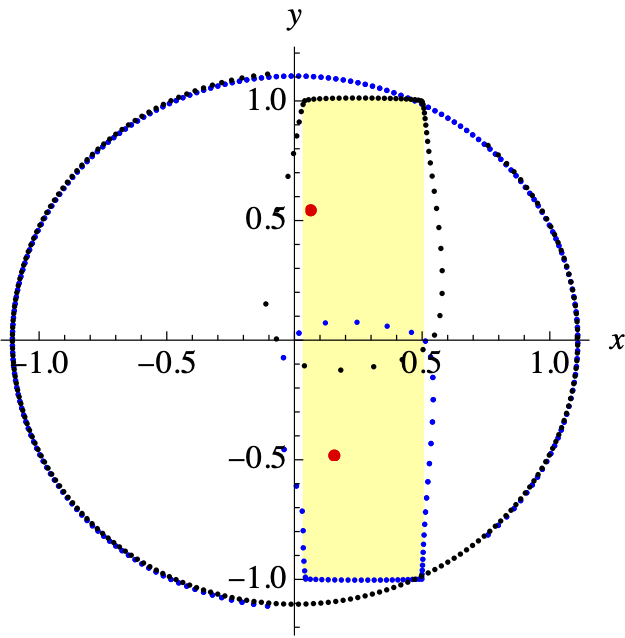}}
\caption{Zeros of $c'(z)$ at $L=10$. The yellow areas denote the points used to generate the fitting polynomials, separately  for upper and lower half plane.}
\label{Q=8+i-grid-zeros-of-c-prime-L=10}
\end{figure}

\noindent
The contour plot \ref{Q=8+i-grid-c-cSP-contour-L=10.jpg} gives a good idea about the location of the singularities. 
Note that the  two non-trivial fixed points marked by the red dots in Fig.~\ref{Q=8+i-grid-c-cSP-contour-L=10.jpg} are not complex conjugate of each other. The precise locations of $c'(z)=0$ are shown in Fig.~\ref{Q=8+i-grid-zeros-of-c-prime-L=10}; again the non-trivial ones are   in red.

\newpage

\noindent 
Extrapolations for $z_c$, $c$ and $\omega$ are shown on the next page, separately for the two fixed points.

\newpage
\enlargethispage{1cm}
\leftline{\bf Fixpoint $B_{-b}$ (upper half plane)}
\noindent Fits for upper half plane. The fitting polynomial always contains $\{1,x^2,x^3\}$, with $x:=1/L$.
\begin{figure}[h!]
\centerline{\fig{.5}{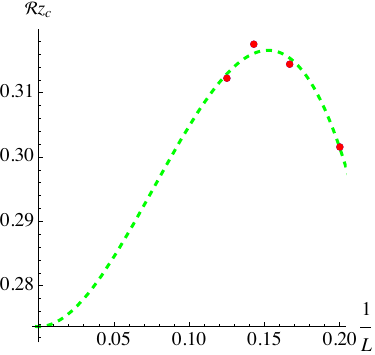}\hfill\fig{.5}{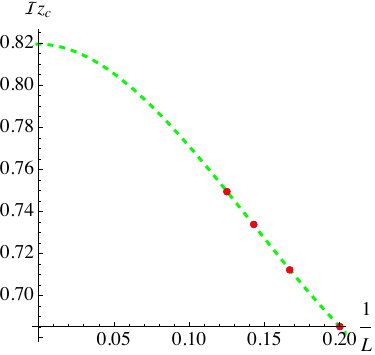}}
\caption{Extrapolation of $z_c$ in UHP.}
\label{Q=8+i-grid-Re-zc-upper}
\end{figure}
\be
  z_c^{\rm upper}({L\to \infty})=0.0546142 +0.570012 i. 
\ee
\begin{figure}[h!]
\centerline{\fig{.5}{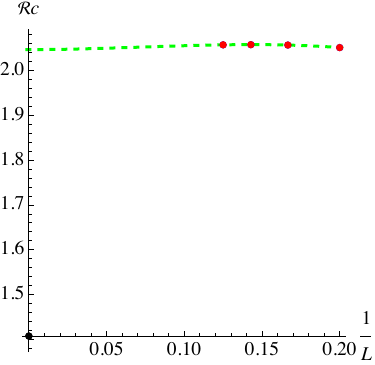}\hfill\fig{.5}{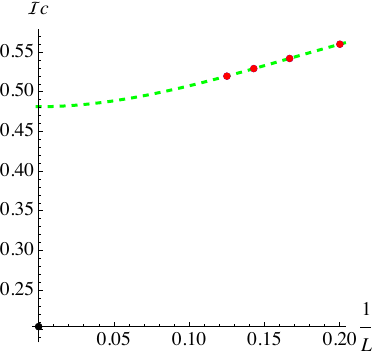}}
\caption{Extrapolation of $c$ in UHP.}
\end{figure}
\bea
  c^{\rm upper}({L\to \infty})&=&1.40773 +0.203235i ,  \\
  c^{\rm upper} _{\rm ana} &=& 1.4073 + 0.2041i.
\eea
\begin{figure}[h]
\centerline{\fig{0.5}{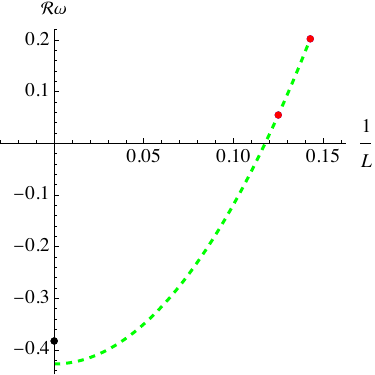}\hfill\fig{0.5}{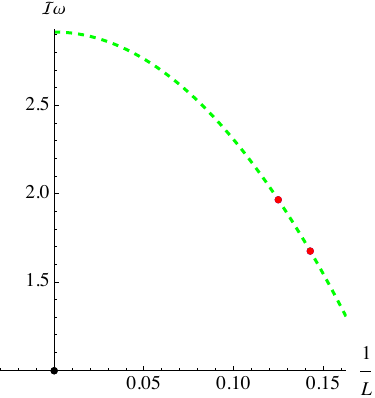}}
\caption{Extrapolation of $\omega$ in UHP.}
\label{Q=8+i-grid-Re-omega-upper}
\end{figure}
\bea
  \omega^{\rm upper}({L\to \infty}) &=& -0.534022+1.00708 i, \\
 \Delta^{\rm upper}({L\to \infty}) &=&2+ \omega^{\rm upper}({L\to \infty}) \nn\\
 &=& 1.46598 +1.00708 i \\
 { 2 h_{3,1}}&=&1.61731 +0.99649 i.
\eea

\newpage 
\enlargethispage{1cm}
\leftline{\bf Fixpoint $B_{b}$ (lower half plane)}
\enlargethispage{2cm}
\noindent Fits for lower half plane, with the same polynomial  as for the upper half plane (see left). 
\begin{figure}[h!]
\centerline{\fig{0.5}{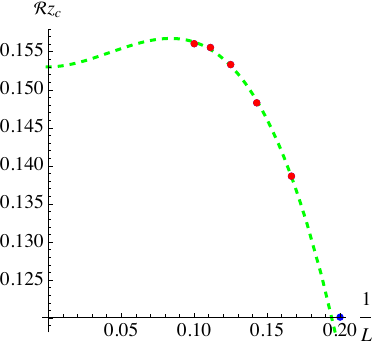}\hfill\fig{.5}{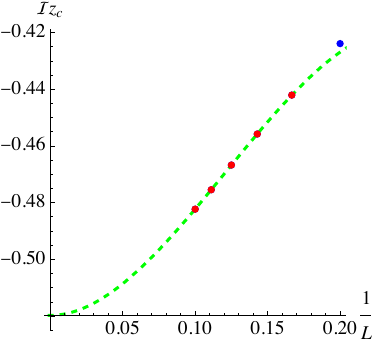}}
\caption{Extrapolation of $z_c$ in LHP.}
\end{figure}
\be
  z_c^{\rm lower}({L\to \infty})=0.153031 -0.519754  i. 
\ee
\begin{figure}[h!]
\centerline{\fig{.5}{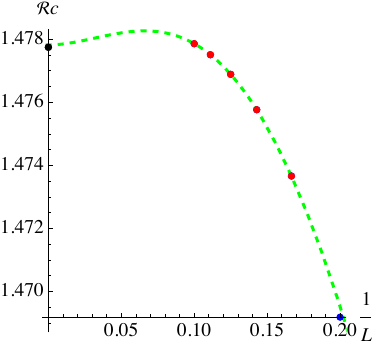}\hfill\fig{0.5}{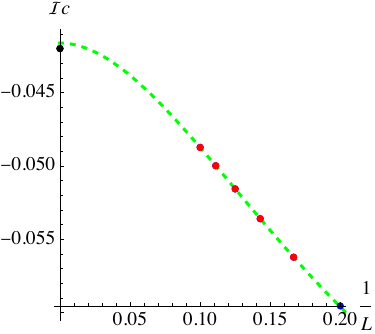}}
\caption{Extrapolation of $c$ in LHP.}
\end{figure}
\bea
  c^{\rm lower}({L\to \infty})&=&1.47782 -0.0416244 i ,\\  
  c^{\rm lower} _{\rm ana} &=& 1.4778 -0.0420 i
\eea
\begin{figure}[h!]
\centerline{\fig{.5}{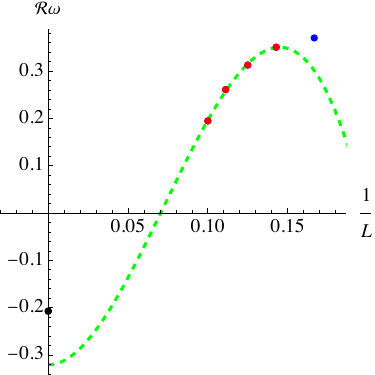}\hfill\fig{0.5}{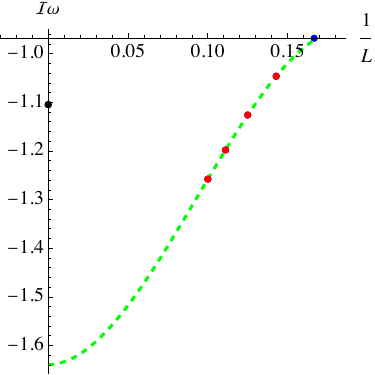}}
\caption{Extrapolation of $\omega$ in LHP.}
\label{Q=8+i-grid-Re-omega-lower}
\end{figure}
\bea
  \omega^{\rm lower}({L\to \infty})&=& -0.320701-1.63909 i\\
 \Delta^{\rm lower}({L\to \infty})&=& 2+\omega^{\rm lower}({L\to \infty}) \nn\\
 &=&  1.6793 -1.63909 i \\
 { 2 h_{3,1}} &=& 1.79269 -1.10454 i.
\eea

\newpage

\subsection{Minigrid, upper half-plane}
\enlargethispage{2cm}
\label{Q=8+i:minigid-upper}

\begin{figure}[h]
\centering
\fig{0.6}{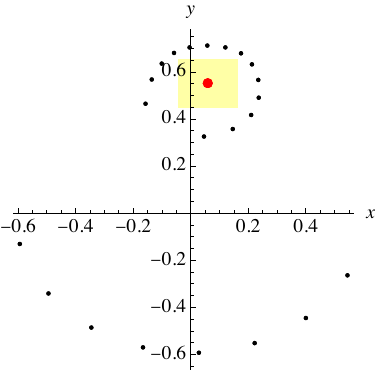}
\caption{Zeros of $c'(z)$, $L=15$, UHP.}
\label{Q=8+i-UHP-minigrid-zeros-of-c-prime-L=15}
\end{figure}

Again we use a minigrid of $5 \times 5$ points. 
As in subsection \ref{Q=5+2i on minigrid, upper half plane}, 
 we use a minigrid of $5 \times 5$ points, centered around the non-trivial fixed point identified in the preceding section. In Fig.~\ref{Q=8+i-UHP-minigrid-zeros-of-c-prime-L=15} 
we show the area covered by it, as well as the zeroes of the fitting polynomial, for our largest system sizes $L=15$. The non-trivial solution marked in red is clearly visible. 
\begin{figure}[h!]
\centerline{\fig{0.5}{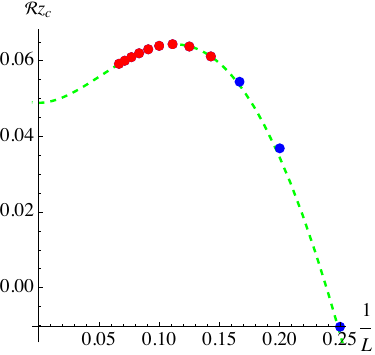}\hfill\fig{0.5}{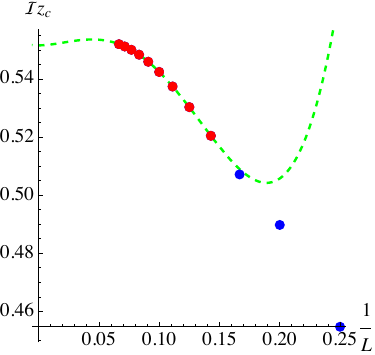}}
\caption{Extrapolation of $z_c$, UHP.}
\label{Q=8+i-minigrid-Re-zc-upper}
\end{figure}
\be
  z_c^{\rm upper}({L\to \infty})= 0.0488999 +0.551616 i  . 
\ee
\begin{figure}[h]
\centerline{\fig{0.5}{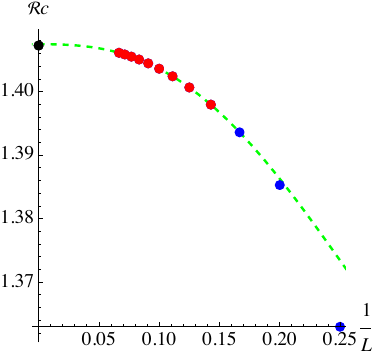}\hfill\fig{0.5}{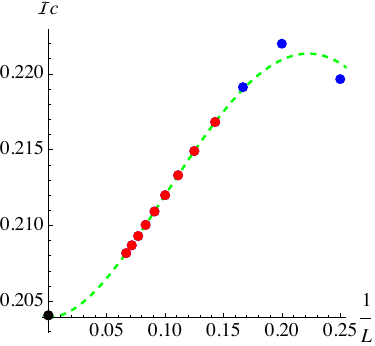}}
\caption{Extrapolation of $c$, UHP.}
\label{Q=8+i-minigrid-Re-c-upper}
\end{figure}

\newpage
\noindent
For the central charge, we find from extrapolation and analytics
\bea \nn
  c^{\rm upper}_{L\to \infty} &=& 1.40746 +0.203946 i\\
  c^{\rm upper}_{\rm ana} &=&1.40726 +0.204076 i. 
\eea
\begin{figure}[h]
\centerline{\fig{0.5}{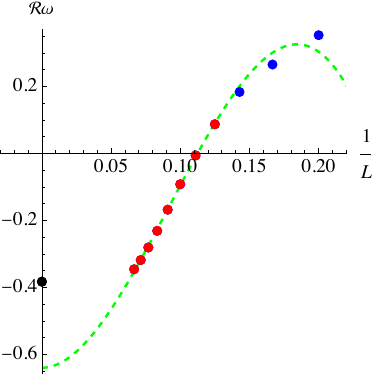}\hfill\fig{0.5}{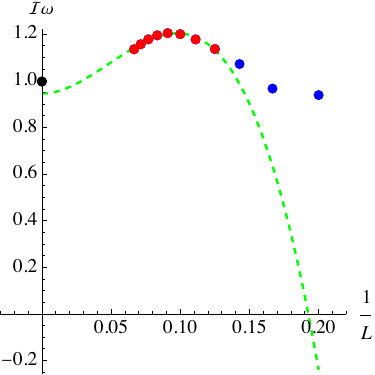}}
\caption{Extrapolation of $\omega$, UHP.}
\label{Q=8+i-minigrid-Re-omega-upper}
\end{figure}
\bea
  \omega^{\rm upper}_{L\to \infty} &=& -0.639713+0.944913 i , \\
 \Delta^{\rm upper}_{L\to \infty} &=&2+ \omega^{\rm upper}_{L\to \infty} \nn\\
 &=& 1.36029 +0.944913 i  \label{E17} \\
 {2 h_{3,1}}&=& 1.61731 +0.99649 i.
 \label{E18}
\eea
\begin{figure}[h]
\centerline{\fig{0.5}{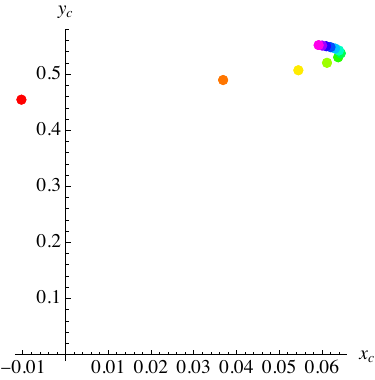}\hfill\fig{0.5}{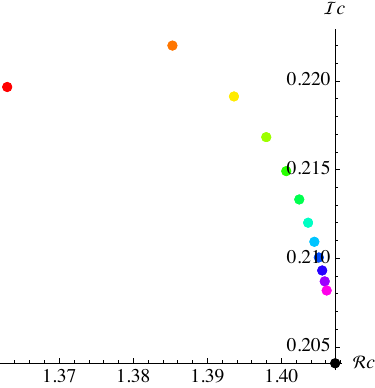}}
\caption{Movement  of $z_c$ (left) and $c$ (right) for $Q=8+i$.}
\label{Q=8+i-minigrid-zeros-of-c-prime-upper}
\end{figure}

\newpage

\subsection{Minigrid, lower half plane}
\label{a:Q=8+1: Minigrid, lower half plane}
Here we use a minigrid of $5 \times 5$ points.

\begin{figure}[h!]
\centerline{\fig{.6}{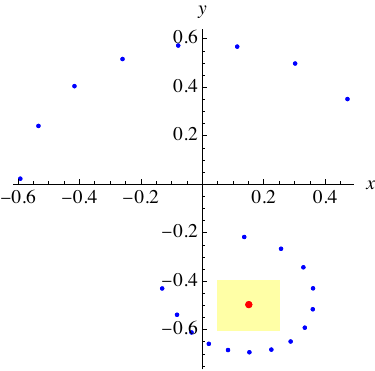}}
\caption{Zeros of $c'(z)$, $L=15$, LHP.}
\label{Q=8+i-LHP-minigrid-zeros-of-c-prime-L=14}
\end{figure}

\leftline{\bf Fixpoint $B_{-b}$}
\noindent Fits for upper half plane. If not stated otherwise, the fitting polynomial   contains $\{1,x^2,x^3\}$, with $x:=1/L$.

\begin{figure}[h!]
\centerline{\fig{.5}{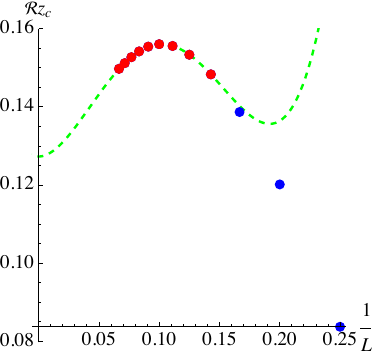}\hfill\fig{.5}{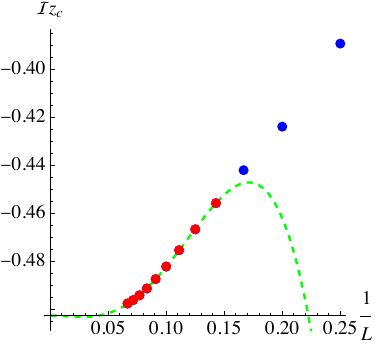}}
\caption{Extrapolation of $z_c$, LHP.}
\label{Q=8+i-minigrid-Re-zc-lower}
\end{figure}

\be
  z_c^{\rm lower}({L\to \infty})= 0.127284 -0.502652. 
\ee

\begin{figure}[h!]
\centerline{\fig{.5}{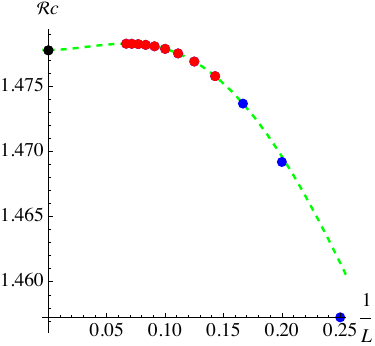}\hfill\fig{.5}{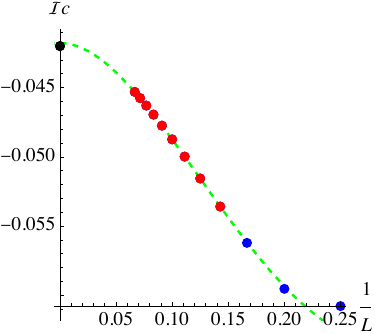}}
\caption{Extrapolation of $c$, LHP.}
\label{Q=8+i-minigrid-Re-c-lower}
\end{figure}
\bea
  c^{\rm lower}_{L\to \infty} &=& 1.47777 -0.0417788 i \nn \\
  c^{\rm lower}_{\rm ana} &=&1.47775 -0.0419944 I . 
\eea

\begin{figure}[h!]
\centerline{\fig{.5}{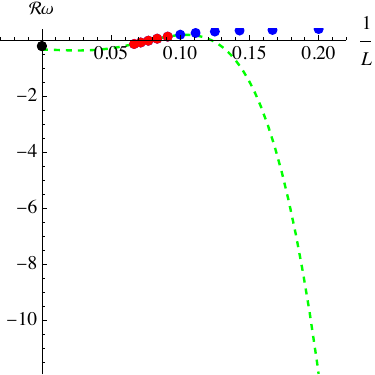}\hfill\fig{.5}{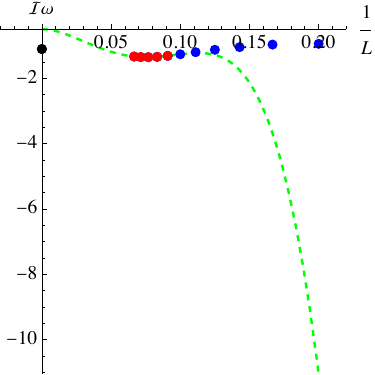}}
\caption{Extrapolation of $\omega$, LHP.}
\label{Q=8+i-minigrid-Re-omega-lower}
\end{figure}
\bea
  \omega^{\rm lower}_{L\to \infty} &=&  -0.333735-0.495648 , \\
 \Delta^{\rm lower}_{L\to \infty} &=&2+ \omega^{\rm lower}_{L\to \infty} \nn\\
 &=&  1.66627 -0.495648   \label{E23}  \\
 2 h_{3,1} &=& 1.79269 -1.10454 i.
 \label{E24}
\eea

\begin{figure}[h!]
\centerline{\fig{.52}{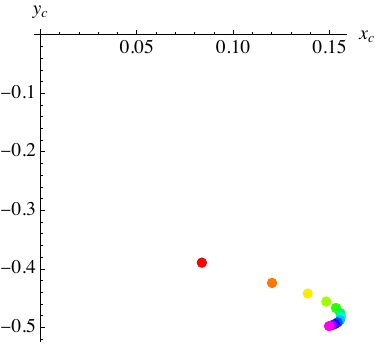}\hfill\fig{.48}{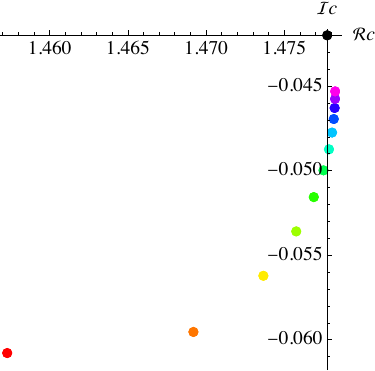}}
\caption{Movement of $z_{\rm c}$ (left)  and $c$ (right) as a function of $L$.}
\label{Q=8+i-minigrid-zeros-of-c-prime-lower}
\end{figure}

\begin{figure*}[t]

\centerline{{\fig{0.66}{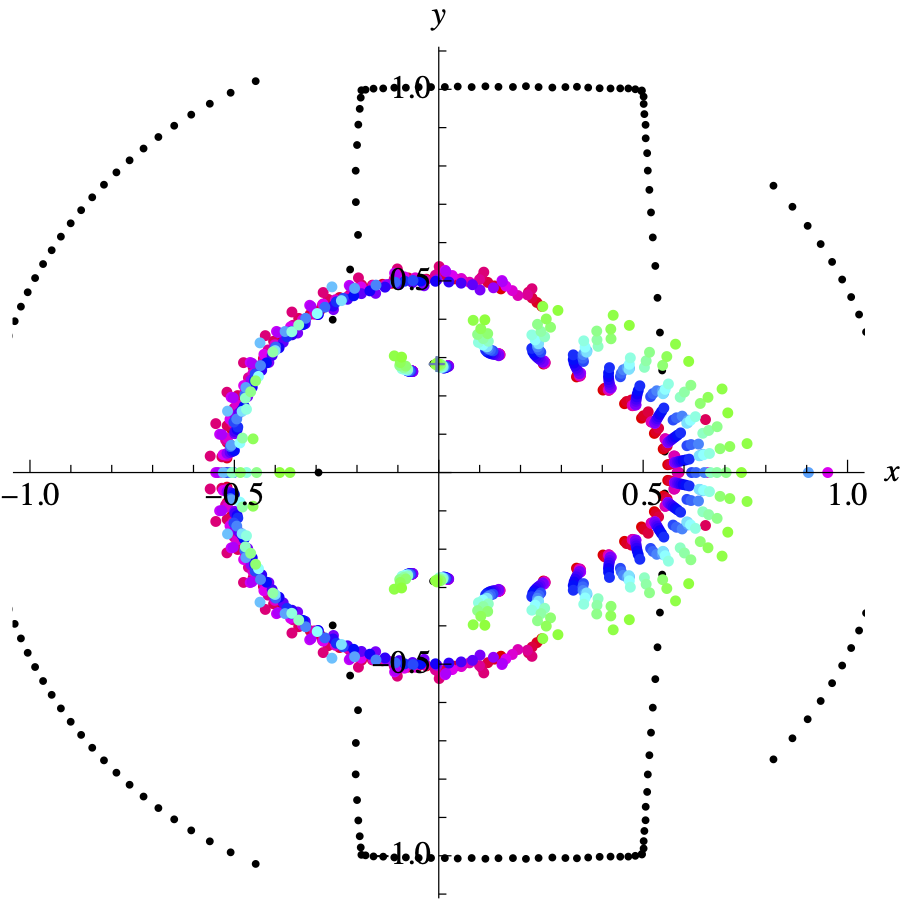}}\hfill\fig{0.66}{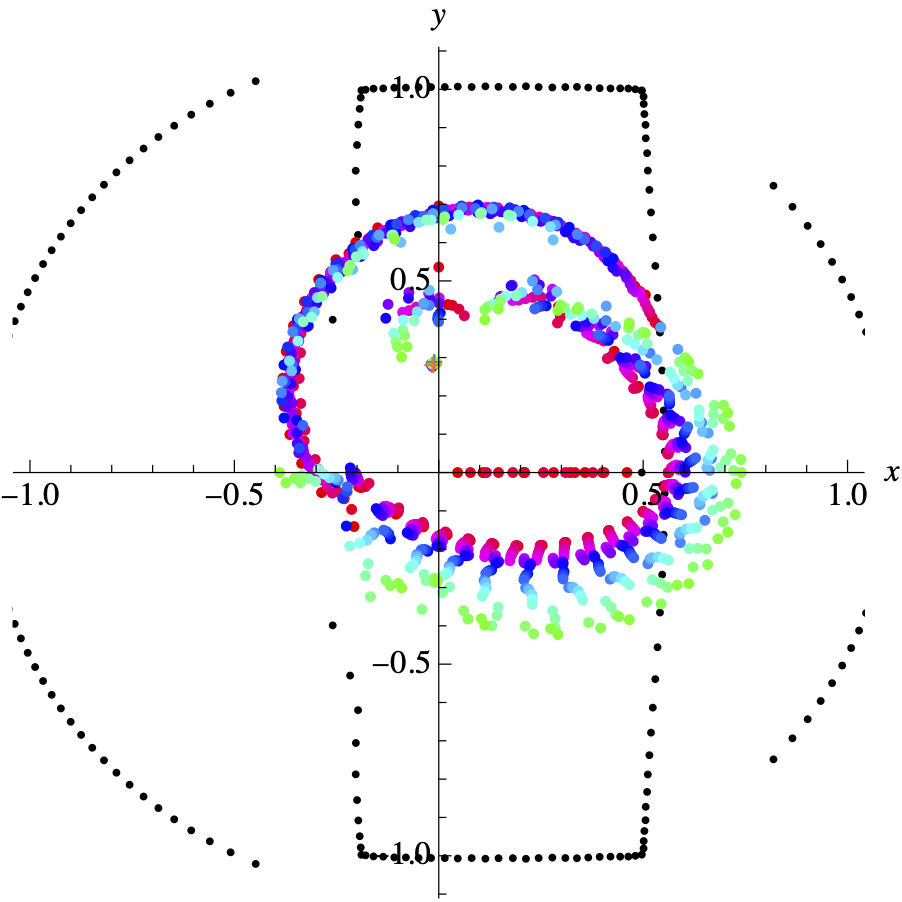}\hfill\fig{0.66}{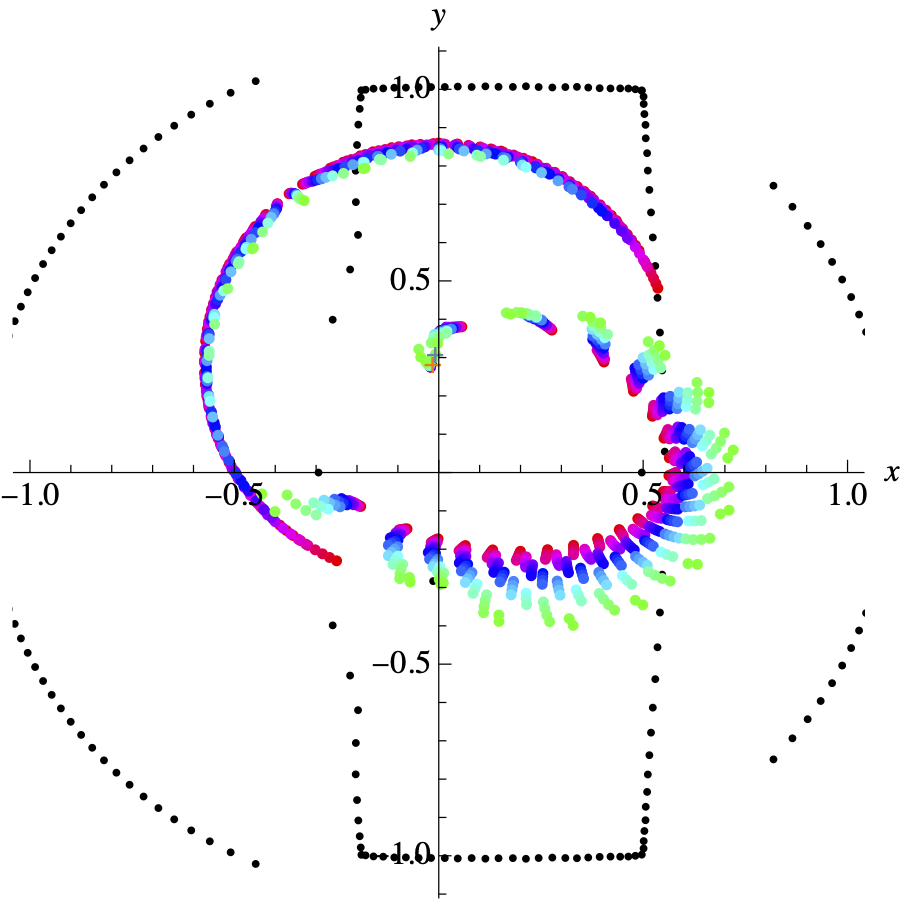}}
\centerline{{\raisebox{0mm}{\parbox{0.33\textwidth}{\fig{0.66}{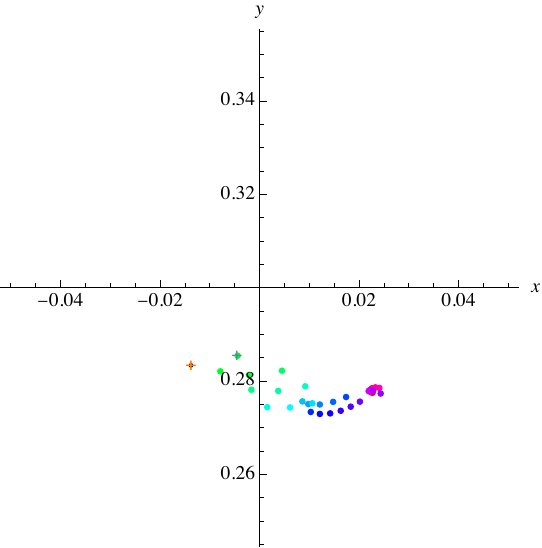}}}}\hfill\parbox{0.33\textwidth}{\fig{0.66}{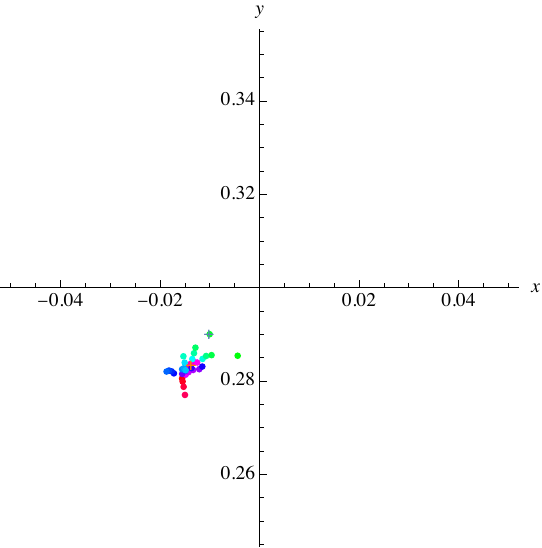}}\hfill\parbox{0.33\textwidth}{\fig{.66}{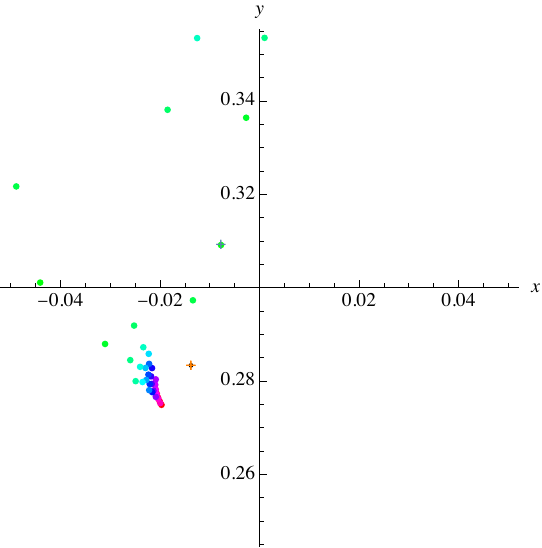}}}
\caption{Top line: solutions of $\partial_z c(z)=0$. In black from the plan-data, in color from the line data. The colors indicate the order of the fit, with green the smallest order $4$, and red the largest order $12$; larger orders are not necessarily better, as they introduce artifacts. The expansion points are $z=0$ (left), $z=0.1+0.25i$ (middle) and $z=0.3i$ (right). Bottom line: the region around the fixed point $z_{\rm c} = -0.0138+0.2833 i$.
The solution \eq{F1} is indicated by an  orange cross.}
\label{f:can one find zc from data on the real line only?}

\bigskip
\noindent\begin{minipage}{0.48\textwidth}
\section{Can one find $z_c(Q=5)$ from data on the real line only?}
\label{Can one find zc(Q=5) from data on the real line only?}
\justifying
\normalsize

An important question, relevant e.g.\ for the analysis of  the 3-dimensional 3-state Potts model \cite{YangYueTangHanZhuChen2025}, is whether one can find the location of the critical point using only the information on the real line. We show below how this can be done, and what the artifacts are. 

For this exercise we use our plane data with $L=10$. By a fit to all data points, we find the reference point for $z_{\rm c}$ and $f''(z_{\rm c})$:
\bea
\label{F1}
z_{\rm c}&=&-0.0137785+0.283345 i, \\
\label{F2}
  f''(z_{\rm c})&=&0.552501 -0.664543 i.
\eea
Let us now try to reconstruct this from   data on the real line. In order to eliminate artifacts from the singularity at $x\approx-0.48$, we retain only $x\ge 0$.
 We also have the choice to vary the degree of the polynomial fit, and its expansion point $x_0$. 
Our first example is  for $x_{\rm 0}=0$, see Fig.~\ref{f:can one find zc from data on the real line only?}, left column. 
Retaining the solution which has the lowest $f''(z_c)$ gives
\bea
z_{\rm c} &\approx& -0.0044 +0.2854 i ,\\
 f''(z_{\rm c})&\approx& 0.803 -0.456 i.
\eea
This uses 50 data points in the interval $[0,1]$; it should be compared to   \Eqs{F1}-\eq{F2}. 
Precision might increase when using more data points in the same interval.
The procedure breaks down e.g.~when using data for $x>-0.2$.  
\end{minipage}\hfill
\begin{minipage}{0.48\textwidth}
\justifying
\normalsize
In our next test, shown in the  middle column of Fig.~\ref{f:can one find zc from data on the real line only?}, 
we  use a generic expansion point somewhere in the upper half plane,  namely $x_0 = 0.1 + 0.25 i$. (The positive real part was chosen s.t. $\Re x_0 >0$, the first point kept on the real line.)
The extrapolation changes to 
\bea\label{F5}
z_{\rm c} &\approx& -0.0100 +0.2900 i  ,\\
\label{F6}
 f''(z_{\rm c})&\approx&  0.574 -0.422 i.
\eea
This  is rather close to the solution \eq{F1}-\eq{F2}, as can be seen on Fig.~ \ref{f:can one find zc from data on the real line only?} (bottom line).

Now let us assume that we know approximately where the critical point is located, and use $z_0 = 0.3i$. 
The extrapolation changes to 
\bea
z_{\rm c} &\approx& -0.0077+0.3091 i, \\
 f''(z_{\rm c})&\approx& 0.453 +0.054 i.
\eea
This is good, but no improvement over \Eqs{F5}-\eq{F6}; probably the former is working better since the expansion point is further away from the boundary of retained points, i.e.\ $x=0$. 

To conclude: it is possible to reconstruct the critical point solely from points  for real couplings. 
If need be, the  procedure could be made more systematic, and more data points could be calculated to 
increase the  precision. One should be careful to not use a  fitting polynomial of too high degree. E.g.\ the dots in the middle top plot of Fig.~\ref{f:can one find zc from data on the real line only?} are artifacts of this.

\vfill

\end{minipage}
\end{figure*}

\clearpage
 
\section{OPE-coefficients following Dotsenko-Fateev}
\label{OPE-coefficients following Dotsenko-Fateev}

The evaluation of OPE coefficients was pioneered by Dotsenko and Fateev  \cite{DotsenkoFateev1985b} (DF), and later continued in \cite{Dotsenko2016}.
While the formalism we develop in section \ref{Three-point structure constants} is able to evaluate 
more coefficients than is possible in the original framework, we keep some results here. 
There are two main reasons: First of all, it is instructive to see how this works out in \cite{DotsenkoFateev1985b}, 
with some simplifications over the original work. Secondly,  DF give explicit analytical results in terms 
of $\Gamma$-functions, while the method in section \ref{Three-point structure constants} uses the Barnes double-Gamma function,  which is   hard to evaluate numerically, and which in the cases shown can be expressed in terms of Gamma-functions.  All formulas below are  explicit enough to infer where  there are  non-analyticities in the complex $b$-plane. This also allows us to evaluate them in a series in $Q-4$.

\subsection{General formulas}
We follow Dotsenko and Fateev (DF) \cite{DotsenkoFateev1985b}. 
Their  Eqs.~(6) and (7)  read in our conventions
\bea
h_{n,n'} 
&=&\frac14 \Big[ (\alpha_- n'+\alpha_+ n)^2 -(\alpha_+ +\alpha_-)^2  \Big],\\
c&=& 1-24 \alpha_0^2 ,\\
\alpha_{\pm} &=& \alpha_0 \pm \sqrt{1+\alpha_0^2}, \\
1 &=& \alpha_+ \alpha_- .
\eea
Comparing to our formulas implies that 
\be
4 \alpha_0^2 b (b-1)=1.
\ee
Let us suppose we consider the series of minimal models with $b>1$, then the appropriate branch is
\be
 \alpha_0 >0 \quad \Longleftrightarrow \quad b>1.
\ee
This implies 
\bea
\alpha_0 &=& \frac1{2 \sqrt{b(b-1)}} ,\\
\alpha_+ &=& \sqrt{\frac{b}{b-1}}, \\
\alpha_- &=& \sqrt{\frac{b-1}{b}} .\\
\eea
DF \cite{DotsenkoFateev1985b} further use 
\bea
\rho &:=& \alpha_+^2 \equiv  \frac{b}{b-1}, \\
\rho' &:=&\alpha_-^2 \equiv  \frac{b-1}{b}.
\eea
($\rho = 1/g$ and $\rho' = g$ in the notations of Sec.~\ref{Phase diagram as a function of z}).
Note that $b \to b+1$ (from critical to tricritical point) is equivalent to $b\to -b$ {\em and } exchange of $\rho$ and $\rho'$.

 DF's expressions for the structure constants are given in their Eqs.~(15)--(18).
Below in \Eq{DFChat} we give  a slightly more compact and simplified form. 
To achieve this, define
\be
\gamma(x):= \frac{\Gamma(x)}{\Gamma(1-x)}.
\ee
The  second simplification is to shift in their Eq.~(15) the indices $i$ and $j$ by 1; this  simplifies their expressions to  
\bea
\label{DFChat}
\lefteqn{\widehat C^{(p,p')}_{(n,n'),(s,s')} =\rho ^{4 (l-1)(l'-1) } }   \nn\\
   &&\times \prod_{i=1}^{l-1}\prod_{j=1}^{l'-1}
   \frac{1}{\big[i- \rho j \big]^2 \big[s-i-\rho  (s'-j)\big]^2}\nn\\
   &&\qquad\qquad  \times  \frac1{\big[n-i-\rho  (n'-j)\big]^2 \big[p+i-\rho   (j+p')\big]^2 } \nn\\
   &&\times \prod _{i=1}^{l-1}  {\gamma  (i \rho ' )} { \gamma \big(s'-(s-i) \rho '\big)}\nn\\
&&\qquad  \times    {\gamma \big(n'-\big(n-i\big)
   \rho '\big) \gamma \big((i+p) \rho '-p'\big)} \nn\\
   && \times \prod _{j=1}^{l'-1}  {\gamma (j \rho )}  {\gamma \big(s-(s'-j)\rho \big)}\nn\\
   &&\qquad \times {\gamma \big(n-(n'-j) \rho \big)
   \gamma \big((j+p') \rho -p\big) } \\
\lefteqn{l= \frac12\big(s+n-p+1\big), \quad l'= \frac12\big(s'+n'-p'+1\big)}~~~
\label{landlprime}
\eea
Note that these formulas can only be evaluated for integer $l$ and $l'$.  For other values, one needs to continue the  products of $\Gamma$ functions analytically, in the same sprit as $n!=\prod_{i=1}^n i$ is continued analytically to $\Gamma(n+1)$. This is achieved by the Barnes double gamma function, as discussed   in section \ref{Three-point structure constants}. 
The structure constants $\omega^{(p,p')}_{(n,n'),(s,s')}$  (denoted $D$ by DF) contain an additional normalization factor $a$,  
\be
  \big[\omega^{(p,p')}_{(n,n'),(s,s')}\big]^2 =  \big[\widehat C^{(p,p')}_{(n,n'),(s,s')}\big]^2 \frac{a_{n,n'} a_{s,s'} }{a_{p,p'} }.
\ee
The latter is defined as\footnote{Attention, there is a misprint in DF in the first product in the first line after Eq~(18).}
\bea
a_{n,n'} &=&\prod_{i=1}^{n-1}  \prod_{j=1}^{n'-1} \frac{[1+i-(1+j)\rho]^2}{(i-j\rho)^2} \nn\\
&&\times \prod_{i=1}^{n-1}  \gamma(i \rho') \gamma\big(2-(i+1)\rho'\big) \nn\\
&&\times\prod_{j=1}^{n'-1} \gamma(j \rho) \gamma\big(2-(j+1)\rho\big). 
\eea

\subsection{Examples of OPE coefficients following DF, physical branch}
\label{Examples of OPE coefficients following DF, physical branch}
We now give some examples for the formulas in section \ref{OPE-coefficients following Dotsenko-Fateev}. Due to \Eq{landlprime},  $2l=s+n-p+1$ as well as $2l'=s'+n'-p'+1$ need to be even. 
For integer indices this implies that within  each group  $\{ s,n,p\}$ and $\{s',n',p'\}$,    either one index is odd and two even, or all three indices  are odd. For non-integer indices this is more complicated. 
Below we give results for the relevant physical branch; the other branch is treated in section \ref{Examples of OPE coefficients following DF, non-physical branch}. 
\bea\label{G18}
\omega_{\epsilon\epsilon\epsilon'}&=&\omega_{ (2,1)(2,1)(3,1)} \nn\\
&=& \sqrt{\frac{\Gamma  (\frac{1+2b}{b-1} ) \Gamma
    (\frac{-2}{b-1} ) \Gamma(\frac{b}{b-1} ) \Gamma (\frac{b+1}{1-b} )}{\Gamma (\frac{1}{1-b} ) \Gamma  (\frac{2 b}{b-1} ) \Gamma  (\frac{b+1}{b-1} )
   \Gamma  (\frac{b+2}{1-b} )}}. 
\\
\label{G19}
\omega_{\epsilon'\epsilon'\epsilon'}&=&\omega_{ (3,1)(3,1)(3,1)}\nn\\
&=&\frac{\Gamma  (\frac{1+3b}{b-1} )}{\Gamma  (\frac{2(b+1)}{1-b} )}
    \left[\frac{\Gamma  (\frac{b}{b-1} ) \Gamma    (\frac{b+1}{1-b} )}{\Gamma    (\frac{1}{1-b} ) \Gamma  (\frac{2
   b}{b-1} )} \right]^{3/2} \nn\\
&& \times \sqrt{\frac{\Gamma (\frac{2}{1-b} ) \Gamma(\frac{b+2}{1-b} )}{\Gamma    (\frac{2 b+1}{b-1}  ) \Gamma    (\frac{b+1}{b-1} )}}.
\\
\label{G20}
\omega_{\epsilon',\epsilon'',\epsilon''}&=&\omega_{ (3,1)(4,1)(4,1)}\nn\\
 &=&\sqrt{\frac{\Gamma  (\frac{2}{1-b} ) \Gamma
    (\frac{b+2}{1-b} ) \Gamma
    (\frac{b+1}{1-b} )}{\Gamma  (\frac{2
   b}{b-1} ) \Gamma  (\frac{b+1}{b-1} )
   \Gamma  (\frac{2 b+1}{b-1} )}} \nn\\
   && \times \frac{\Gamma  (\frac{2 b+1}{1-b} )
    \Gamma
    (\frac{4 b+1}{b-1} )}{\Gamma  (\frac{3 b+2}{1-b} ) \Gamma  (\frac{3 b}{b-1} )}
   \left[\frac{\Gamma  (\frac{b}{b-1} )}{\Gamma
    (\frac{1}{1-b} )} \right]^{\frac 32} . \qquad 
\\
%
\omega_{\epsilon,\epsilon',\epsilon''}&=& \omega_{ (2,1)(3,1)(4,1)} \nn\\
&=& \sqrt{\frac{\Gamma (\frac{2}{1-b}) \Gamma (\frac{2 b+1}{1-b}) \Gamma (\frac{b}{b-1}) 
\Gamma (\frac{3 b+1}{b-1})}{\Gamma (\frac{1}{1-b}) \Gamma (\frac{3 b}{b-1}) 
\Gamma (\frac{2 (b+1)}{1-b}) \Gamma (\frac{b+1}{b-1})}}.\qquad 
\\
%
%
\omega_{ (0, 1)(0,1 )(3,1)} &=& 
\frac{\Gamma (\frac{1}{b-1}) \Gamma (\frac{2 \
b-1}{b-1}) }{\Gamma (\frac{b-2}{b-1}) \Gamma (\frac{b}{1-b})} 
\left[\frac{\Gamma \
(\frac{b}{b-1})}{\Gamma (\frac{1}{1-b})}\right]^{\frac32} 
\nn\\ &&
\times \sqrt{\frac{\Gamma (\frac{2}{1-b}) \Gamma (\frac{b+2}{1-b}) \Gamma (\frac{b+1}{1-b})}{\Gamma (\frac{2b}{b-1}) \Gamma (\frac{b+1}{b-1}) \Gamma (\frac{2 b+1}{b-1})}}\quad 
\eea
At $Q=5$, this yields
\bea
\omega_{ (2,1)(2,1)(3,1)} &=& 0.879108 +0.140371 i\\
\omega_{ (3,1)(3,1)(3,1)} &=& 2.26873 +1.1967  i
\\
\omega_{ (3,1)(4,1)(4,1)} &=&  3.92606 +3.32608  i \\
\omega_{ (2,1)(3,1)(4,1)} &=& 0.831751 +0.202731 i \\
\omega_{ (0, 1)(0,1 )(3,1)} &=& - 0.245254+ 0.131105  i 
\eea

\subsection{Examples of OPE coefficients following DF, non-physical branch}
\label{Examples of OPE coefficients following DF, non-physical branch}
Below are OPE coefficients for operators not realized in the Potts model. They are relevant for other models; some of them appear in  \cite{GorbenkoRychkovZan2018b}. Note that this list contains some OPE coefficients with non-integer entries, see \Eqs{G34}-\eq{G35}.

To compare to the perturbative expansion of \cite{GorbenkoRychkovZan2018b}, we also give its $(4-Q)$ expansion. This is achieved by first establishing 
the $1/b$ expansion, and then  expressing it in powers of $(4-Q)$. 
\bea
\lefteqn{ \omega_{(1,2)(1,2)(1,3)}=}  
\nn\\
&=&\frac{ \Gamma
    (\frac{2}{b} ) \Gamma  (\frac{b-1}{b} )}{\Gamma \left(\frac{1}{b}\right) \Gamma
    (\frac{b-2}{b} )} \sqrt{\frac{b^3 \Gamma  (2-\frac{3}{b} ) \Gamma  (\frac{1}{b} )}{(b-2)^2
   \Gamma  (\frac{3}{b}-1 ) \Gamma  (-\frac{1}{b} )}}\qquad \nn\\
&=& \frac{\sqrt{3}}{2}-\frac{\sqrt{3}  }{2b}-\frac{\sqrt{3}}{b^2}  +\frac{ 2\sqrt{3}  [\zeta
   (3)-1]}{b^3} \nn\\
&& -\frac{2   \sqrt{3}[\zeta (3)+2] }{b^4}  +\ca O (b^{-5}) \nn\\
&=&  \frac{\sqrt{3}}{2} 
-\frac{\sqrt{3}   \sqrt{{4{-}Q}}}{4 \pi }
-\frac{\sqrt{3} ({4{-}Q})}{4 \pi ^2} \nn\\
&& 
-\frac{({4{-}Q})^{3/2} \left[24+\pi ^2-24 \zeta (3)\right]}{32 \sqrt{3} \pi ^3} \nn\\
&&  -\frac{({4{-}Q})^2 \left[6 \zeta (3)+12+\pi ^2\right]}{16 \sqrt{3} \pi ^4}+ \ca O({4{-}Q})^{5/2}.\qquad 
\eea
The first two terms of this OPE-coefficient   agree with \cite{GorbenkoRychkovZan2018b} Eq.~(4.19), who obtained their results via a perturbative expansion in $Q-4$.
\bea
\lefteqn{\omega_{(1,3)(1,3)(1,3)}  
}\nn\\
& =& \frac{2^{2-\frac{4}{b}} \Gamma \left(\frac{3}{2}-\frac{2}{b}\right) \Gamma
   \left(\frac{1}{b}-\frac{1}{2}\right) }{\Gamma \left(\frac{3}{2}-\frac{1}{b}\right) \Gamma
   \left(\frac{2}{b}-\frac{1}{2}\right)} \sqrt{\frac{(b-2)^2 \Gamma \left(\frac{3}{b}-1\right) \Gamma
   \left(-\frac{1}{b}\right)}{b^3 \Gamma \left(2-\frac{3}{b}\right) \Gamma
   \left(\frac{1}{b}\right)}} \nn\\
   &=& \frac{4}{\sqrt{3}}-\frac{4 \sqrt{3}}{b}-\frac{20 }{\sqrt{3}b^2}+\frac{4 [24 \zeta (3)-7]}{\sqrt{3} b^3}
   - \frac{4[5+72 \zeta(3)]}{\sqrt3 b^4} \nn\\
&&   +\ca O(b^{-5})\nn\\
   &=& \frac{4}{\sqrt{3}}-\frac{2 \sqrt{3} \sqrt{{4-Q}}}{\pi }-\frac{5 ({4-Q})}{\sqrt{3}\pi ^2} \nn\\
&&-\frac{({4-Q})^{3/2}  (\pi ^2-48 \zeta (3)+14 )}{4 \sqrt{3} \pi ^3} \nn\\
&&  - \frac{(4-Q)^2[15+ 5 \pi^2 + 216 \zeta (3)]}{12 \sqrt3 \pi^4}+ \ca O({4-Q})^{5/2}.
\eea
The first two terms of the OPE-coefficients above agree with Eq.~(4.11) in \cite{GorbenkoRychkovZan2018b}. 
\bea
\lefteqn{ \omega_{(1,2)(1,3)(1,4)}  
}\nn\\
& =& \sqrt{\frac{2^{\frac{4}{b}-1} \Gamma \left(3-\frac{4}{b}\right)
   \Gamma \left(\frac{1}{2}+\frac{1}{b}\right) \Gamma
   \left(\frac{3}{b}-2\right)}{\Gamma
   \left(3-\frac{3}{b}\right) \Gamma
   \left(\frac{1}{2}-\frac{1}{b}\right) \Gamma
   \left(\frac{4}{b}-2\right)}} \nn\\\
   &=& \sqrt{\frac{2}{3}}-\frac{\sqrt{\frac{3}{2}}}{b}-\frac{13
   }{2
   \sqrt{6} b^2}+\frac{ (80 \zeta
   (3)-63)}{4 \sqrt{6} b^3} \nn\\
   && -  \sqrt{\frac{3}{2}} \frac{    80 \zeta
   (3)+111 }{8 b^4} 
+\ca O (b^{-5}) \nn
\eea
\bea
&=& 
\sqrt{\frac{2}{3}}
-\sqrt{\frac{3}{2}}  \frac{\sqrt{4-Q}}{2 \pi}
-\frac{13 (4-Q)}{8    \sqrt{6} \pi ^2} 
\nn\\ &&
-\frac{(4-Q)^{3/2}  [63+2 \pi ^2 -80 \zeta (3) ]}{32 \sqrt{6} \pi ^3}
\nn\\ &&
-\frac{(4-Q)^2  [999+52 \pi ^2+720 \zeta (3) ]}{384 \sqrt{6} \pi ^4} \nn\\
&& + \ca O(4-Q)^{5/2}. \\
\lefteqn{ \omega_{(1,3)(1,4)(1,4)} }\nn\\
& =& \frac{2^{ \frac4b -2} \Gamma (4- \frac{5}b ) \Gamma \left(\frac1b
   +\frac{1}{2}\right) \Gamma ( \frac3b -2) }{\Gamma (3- \frac3b ) \Gamma
   \left(\frac{3}{2}-\frac1b \right) \Gamma ( \frac 5b -3)} \sqrt{\frac{  
   \Gamma (-\frac1b ) \Gamma ( \frac3b -1)}{b\Gamma (2- \frac3b )
   \Gamma (\frac1b )}} \nn \\
&=& 
\frac{5 \sqrt{3}}{2}-\frac{65   }{2 \sqrt{3}b}-\frac{155 }{3 \sqrt{3}b^2}
+\frac{65  [72 \zeta (3)-11]}{12\sqrt{3}b^3} \nn\\
&&+\frac{5 [127-4056 \zeta (3)]}{12\sqrt{3}b^4}
+\ca O (b^{-5}) \nn\\
&=& 
\frac{5 \sqrt{3}}{2}
-\frac{65 \sqrt{4-Q}}{4 \sqrt{3} \pi  }
-\frac{155 (4-Q)}{12 \sqrt{3} \pi ^2 }\nn\\
&&
-\frac{65 (4-Q)^{3/2}[11+\pi ^2-72 \zeta (3)  ]}{96 \sqrt{3} \pi^3 }
\nn\\
&&
-\frac{5 (4-Q)^2 [12168 \zeta(3)-381+124 \pi ^2 ]}{576  \sqrt{3} \pi^4 }
\nn\\
&&
+\ca O (4-Q)^{5/2} .
\eea

\newpage
\bea
\label{G34}
\omega_{ (0, \frac12 )(0,\frac12 )(3,1)} &=& 
\sqrt{\frac{\Gamma  (\frac{b+2}{1-b} ) \Gamma (\frac{b+1}{1-b} ) \Gamma   (\frac{b}{b-1} ) \Gamma  (\frac{2
   b}{b-1} )}{\Gamma  (\frac{1}{1-b} ) \Gamma (\frac{2}{1-b} ) \Gamma (\frac{b+1}{b-1} ) \Gamma  (\frac{2
   b+1}{b-1} )}} \nn\\
   && \times  4^{\frac{b+1}{1-b}} \frac{ \Gamma  (\frac{b+1}{2 (b-1)} ) }{\Gamma  (\frac{b+1}{2
   (1-b)} )} \qquad 
\\   
\omega_{ (0, \frac32)(0,\frac32 )(3,1)} &=&  \sqrt{\frac{\Gamma (\frac{b+2}{1-b}) \Gamma (\frac{b-3}{2 (b-1)}) \Gamma (\frac{b}{b-1}) \Gamma (\frac{2 b}{b-1})}{\Gamma (\frac{b+1}{1-b}) \Gamma (\frac{b+1}{b-1}) \Gamma (\frac{2 b+1}{b-1})}}
\nn\\
&& \times
\frac{2^{-\frac{5}{b-1}-\frac{11}{2}} (1-3 b)^2 \Gamma (\frac{3-b}{2 (b-1)}) }{\sqrt[4]{\pi } (b-1)^2 \Gamma (\frac{3 b-5}{2 (b-1)})}  
\label{G35}
\eea
Numerically we find 
\bea
\!\!\!\!\!\!  \omega_{(1,2)(1,2)(1,3)}(Q=5)&=& 0.90131 + 0.134446 i ,\\
\!\!\!\!\!\!  \omega_{(1,2)(1,2)(1,3)}(Q=5)
 &=&2.48214 + 1.20601 i, \\
\!\!\!\!\!\!  \omega_{(1,2)(1,3)(1,4)}(Q=5)
&=&0.864836 + 0.196605 i, \\
\!\!\!\!\!\!  \omega_{(1,3)(1,4)(1,4)}(Q=5)
&=& 4.53186 + 3.48543i \qquad  \\
\!\!\!\!\!\! \omega_{ (0, \frac12 )(0,\frac12 )(3,1)} &=& 0.0438596 -0.0665729 i, \\
\omega_{ (0, \frac32)(0,\frac32 )(3,1)} &=&   0.578473 -0.134703 i .\qquad 
\eea

\vfill

%
%

\ifx\doi\undefined
\providecommand{\doi}[2]{\href{http://dx.doi.org/#1}{#2}}
\else
\renewcommand{\doi}[2]{\href{http://dx.doi.org/#1}{#2}}
\fi
\providecommand{\link}[2]{\href{#1}{#2}}
\providecommand{\arxiv}[1]{\href{http://arxiv.org/abs/#1}{#1}}
\providecommand{\hal}[1]{\href{https://hal.archives-ouvertes.fr/hal-#1}{hal-#1}}
\providecommand{\mrnumber}[1]{\href{https://mathscinet.ams.org/mathscinet/search/publdoc.html?pg1=MR&s1=#1&loc=fromreflist}{MR#1}}

\tableofcontents
\label{table-of-contents}

\vfill

\end{document}